\documentclass[final,5p,times,twocolumn]{elsarticle}

\usepackage{stmaryrd}
\usepackage[export]{adjustbox}
\usepackage{amsmath}
\usepackage{amssymb}
\usepackage{bm}
\usepackage{dsfont}
\usepackage{multicol}
\usepackage{subfig}
\usepackage{array, makecell}
\usepackage{graphicx, adjustbox}
\usepackage{mathtools}
\usepackage{setspace}
\usepackage[section]{placeins}
\usepackage{placeins}
\usepackage{tikz}
\usetikzlibrary{positioning,shapes.geometric}
\usepackage{hyperref}
\usepackage{lineno}
\usepackage{enumitem}

\usepackage{xcolor}
\definecolor{C0}{RGB}{31,119,180}
\definecolor{C1}{RGB}{255, 127, 14}
\definecolor{C2}{RGB}{44, 160, 44}
\definecolor{C3}{RGB}{214, 39, 40}
\definecolor{C4}{RGB}{148, 103, 189}
\definecolor{C5}{RGB}{140, 86, 75}
\definecolor{C6}{RGB}{227, 119, 194}
\definecolor{C7}{RGB}{127, 127, 127}
\definecolor{C8}{RGB}{188, 189, 34}
\definecolor{C9}{RGB}{23, 190, 207}

\newcommand{\bra}[1]{\langle #1 \rvert}
\newcommand{\ket}[1]{\lvert #1 \rangle}
\newcommand{\braket}[2]{\langle #1 \vert #2 \rangle}

\biboptions{sort&compress}

\newcounter{bla}

\journal{Computer Physics Communications}

\newcommand{\etal}{\textit{et al}.}
\DeclarePairedDelimiter\onenorm{\lVert}{\rVert_{1}}

\begin{document}

\begin{frontmatter}



\title{QSW\_MPI: a framework for parallel simulation of quantum
  stochastic walks}


\author[a]{Edric Matwiejew\corref{author}}
\author[a]{Jingbo Wang}

\cortext[author]{Corresponding author.\\\textit{E-mail address:} Edric.Matwiejew@research.uwa.edu.au}
\address[a]{Department of Physics, The University of Western Australia, Perth, Australia}

\begin{abstract} 
QSW\_MPI is a python package developed for time-series simulation of continuous-time quantum stochastic walks. This model allows for the study of Markovian open quantum systems in the Lindblad formalism, including a generalisation of the continuous-time random walk and continuous-time quantum walk. Consisting of a python interface accessing parallelised Fortran libraries utilising sparse data structures, QSW\_MPI is scalable to massively parallel computers, which makes possible the simulation of a wide range of walk dynamics on directed and undirected graphs of arbitrary complexity. 
\end{abstract}


\end{frontmatter}



{\bf PROGRAM SUMMARY}

\begin{small}
\noindent
{\em Program Title:} QSW\_MPI                                          \\
{\em Licensing provisions:} GPLv3
\\
{\em Programming language:} Python 3 + Fortran 2003                                   \\
{\em Computer and operating system:} Systems supporting python 3 and an MPI implementation.                                         \\
{\em RAM:} 16 GB minimum recommended, memory requirements scale with digraph size and connectivity.                                              \\
{\em Number of processors used:} Arbitrary number of processors supported via MPI.                           \\
{\em Classification:} 4.15 Quantum Computing, 6.5 Software including Parallel Algorithms                                     \\
{\em External routines/libraries:} NumPy [1], SciPy [2], mpi4py [3], h5py[4]                         \\
{\em Nature of problem:}\\
 QSW\_MPI provides a framework for the simulation of quantum stochastic walks
 on arbitrary graphs (directed/undirected, weighted/unweighted).
   \\
{\em Solution method:}\\
A parallel distributed-memory implementation of the matrix exponential via a truncated Taylor series expansion with scaling and squaring [5].

\noindent {\em Restrictions:}\\
QSW\_MPI will provide support for the simulation of multiple quantum walkers in a future version.

\end{small}

\section{Introduction}

The exploration of nanoscale systems and quantum information is at the core of a new generation of technologies, including quantum computation and molecular-scale electronics. This necessitates the development of efficient simulation software capable of modelling the dynamics of these systems in their practical application, namely when subject to an external environment. To this effort, we have developed QSW\_MPI \cite{edric_qsw_mpi}, a package designed for the efficient time-series simulation of continuous-time quantum stochastic walks on both workstations and massively parallel computers.  

Walk based models describe the time evolution of a system consisting of discrete sites linked by a coupling potential, which together can be represented as a graph. These include the continuous-time random walk (CTRW) and its quantum analogue, the continuous-time quantum walk (CTQW). Such models have been applied to a wide range of physical and informatic systems. For example, their natural correspondence to the tight-binding model in solid-state physics has seen their application to the study of quantum and classical energy transport in molecular systems \cite{zhang_forster_2016, mulken_continuous-time_2011}. Elsewhere in the rapidly developing field of quantum computing, quantum-walk based algorithms have been developed which are exponentially faster than their classical counterparts \cite{childs_exponential_2003}. Other quantum walk based algorithms promise near-term quantum advantage in noisy intermediate-scale quantum computation \cite{bravyi_quantum_2018, marsh_combinatorial_2020}. 

Developed by Whitfield \etal~in 2010, quantum stochastic walks describe a quantum walk weakly coupled to an external environment, whose effects are derived axiomatically from an underlying directed or undirected graph \cite{whitfield_quantum_2010}. The continuous-time form obeys the Gorini-Kossakowski-Sudarshan-Lindblad (GKSL) equation\footnote{A \textit{stochastic} equation of motion.} and encompasses as special cases a generalisation of the CTRW and CTQW. Quantum stochastic walks are defined for both continuous and discrete-time, however, their initial introduction and subsequent research have focused primarily on the continuous-time quantum stochastic walk (QSW). Note, in this paper, QSW refers only to continuous-time quantum stochastic walks. \cite{whitfield_quantum_2010,sanchez-burillo_quantum_2012,liu_steady_2017,govia_quantum_2017,tang_experimental_2019,schijven_modeling_2012,domino_properties_nodate,domino_superdiffusive_nodate,glos_limiting_2018,loke_comparing_2016}.

Studies of the propagation and steady states of QSWs have demonstrated significant variability with graph type and environmental interaction.
For example, a `local-interaction' QSW (L-QSW) reduces to diffusive propagation \cite{domino_properties_nodate}, while a `global-interaction' QSW (G-QSW) retains superdiffusive propagation that does not reflect the structure of the originating graph \cite{domino_properties_nodate,domino_superdiffusive_nodate}.       

Significantly, QSWs have been employed to describe quantum assisted transport through the light-harvesting complexes of photosynthetic bacteria \cite{mohseni_environment-assisted_2008}, which extended the QSW model to include the non-unitary processes of absorption into and emission from the graph. These processes have since been formally introduced to the L-QSW model and applied to the study of transport in monomers, dimers and topologically disordered graphs \cite{schijven_modeling_2012}. 

L-QSWs have also been explored as a basis for developing quantum algorithms. For example, an L-QSW based variant of the PageRank search engine algorithm exhibited final distributions which broke degeneracies present in the classical version, while also speeding up the rate of convergence \cite{sanchez-burillo_quantum_2012, loke_comparing_2016}. Experimentally, L-QSWs have been implemented as associative memory for potential use in artificial neurons \cite{tang_experimental_2019}. 

The established expressiveness and flexibility of the QSW model thus motivate making possible its application to systems of greater complexity. To this end, software packages for the simulation of QSWs have been developed using the Wolfram programming language (QSWalk.m) and the Julia programming language (QSWalk.jl) \cite{falloon_qswalk:_2017, glos_qswalk.jl:_2019,falloon_reply_2019}. With QSWalk.jl notably introducing support for the `demoralisation correction scheme', which makes possible superdiffusive G-QSWs which are consistent with the underlying digraph structure. They both offer a user-friendly interface but are limited in the simulated graph size due to memory constraints and their reliance on single-process linear algebra libraries. Additionally, they do not provide support for efficient time-series calculation. 

QSW\_MPI addresses these limitations by taking a distributed memory approach to the construction of the QSW superoperator and the calculation of system evolution via matrix exponentiation. This takes the form of subroutines contained in Fortran libraries with which the user interacts using a python interface; taking advantage of the ubiquity of the interpreted python language and the speed afforded by highly optimised Fortran compilers. With QSW\_MPI, a user can write simulations appropriate for execution on massively parallel computers with minimal background in programming or parallel computation. Support for L-QSWs and G-QSWs is provided, including graph demoralisation. For L-QSWs, absorption and emission process may be easily modelled through modifications to the structure of the originating directed graph.  

This paper thus proceeds as follows. In Section \ref{sec:theory} the mathematical framework of QSWs is introduced. Section \ref{sec:computation} provides a summary of the computational methods utilised in QSW\_MPI. This is followed by an overview of the software package and usage examples. Validation and performance of QSW\_MPI is discussed in Section \ref{sec:bench}, and concluding statements given in Section \ref{sec:conclusion}. 

\section{Theory} \label{sec:theory}

This section provides an overview of the mathematical formalism underpinning QSWs. The starting point is a discussion of graph theory terminology, which draws primarily from Refs. \cite{domino_superdiffusive_nodate} and \cite{falloon_qswalk:_2017}, followed definition of CTRWs on digraphs and CTQWs on graphs. An overview of the master equation approach to the description of Markovian open systems is then provided. From this, the locally-interacting quantum stochastic walk (L-QSW) master equation is introduced, which unifies the CTRW and CTQW models under a density theoretic framework. The practical extension of this equation to the inclusion of non-Hermitian absorption and emission process is then presented. Next, the globally interacting quantum stochastic walk (G-QSW) is introduced, along with the demoralisation correction scheme. We conclude by discussing vectorisation of the QSW master equations, the numerical approximation of which is the primary task at hand.

\subsection{Digraphs and Graphs} \label{sec:graphs}

A weighted digraph is defined as an object $\mathcal{G} = (V,E)$ comprised of vertex set $V = \{v_1, ...,v_N\}$ connected by arc set $E = \{(v_i, v_j), (v_k, v_l),...\}$. Associated with each edge is a positive non-zero weight, $\text{w}(v_i,v_k) \in \mathbb{R}$, describing the magnitude of connection between $v_i$ and $v_j$. $\mathcal{G}$ is represented by an $N \times N$ adjacency matrix, $G$:
\begin{equation}
    \label{eq:adjacency_graph}
    G_{ij} =
    \begin{cases}
        \text{w}(v_i,v_j), & (v_i, v_j) \in E \\
        0, & \text{otherwise}.
    \end{cases}
\end{equation}
Vertices of form $(v_i, v_i)$ are known as self-loops, with a graph containing no self-loops being referred to as a simple graph. QSW\_MPI considers only the case of simple graphs where $\text{Tr}(G) = 0$.

Associated with $\mathcal{G}$ is the weighted but undirected graph $\mathcal{G}^u = (V,E^u)$, where $E^u$ is a set of edges. This is represented by a symmetric adjacency matrix, $G^u$, with weightings $\text{w}^u(v_i,v_j)= \text{max}(\text{w}(v_j,v_i),\text{w}(v_i,v_j))$ in Equation (\ref{eq:adjacency_graph}). 
 A digraph is weakly connected if there exists a path between all $v_i \in \mathcal{G}^u$. Additionally, a digraph which satisfies the further condition of having a path between all $v_i \in \mathcal{G}$ is strongly connected. 

The sum total of the outgoing edge weights from vertex $v_j$, 
        
\begin{equation} \label{eq:out_degree}
    \text{OutDeg}(v_j) = \sum_{i \neq j}\text{w}(v_i,v_j)
\end{equation}
        
\noindent is termed the vertex out-degree. A connected vertex, $v_j$, in $\mathcal{G}$ for which $\text{OutDeg}(v_j) = 0$ is refereed as a `sink'. Similarly, the total of the incoming edge weights at vertex $v_i$,

\begin{equation} \label{eq:in_degree}
  \text{InDeg}(v_i) = \sum_{i \neq j}\text{w}(v_i,v_j)
\end{equation}

\noindent is termed the vertex in-degree. A connected vertex, $v_i$, in $\mathcal{G}$ for which $\text{InDeg}(v_i) = 0$ is referred to as a `source'. A regular digraph or graph has equal in-degree and out-degree for all vertices. 

\subsection{Continuous-Time Classical Random Walks}

A continuous-time random walk (CTRW) describes the probabilistic evolution of a system (walker) though a parameter space as a continuous function of time. Most typically, CTRWs refer to a type of Markov process. This describes a scenario where the future state of a system depends only on its current state. Heuristically, one might describe such systems as having a `short memory'. Under this condition, a CTRW over a digraph is described by a system of first-order ordinary differential equations,

\begin{equation}
    \label{eq:CTRW}
    \frac{d \vec{p}(t)}{dt} = -M \vec{p}(t)
\end{equation}

\noindent where element $p_i \geq 0$ of $\vec{p}(t)$ is the probability of the walker being found at vertex $i$ of the digraph, and $\vec{p}(t)$ has the solution $\vec{p}(t) = \exp(-tM)\vec{p}(0)$  which satisfies $\sum\vec{p}(t) = 1$ \cite{kampen_n._g._stochastic_2007,whitfield_quantum_2010}. $M$ is the transition matrix derived from $G$,

\begin{equation}
    \label{eq:generator_matrix}
    M_{ij} =
    \begin{cases}
    -\gamma \ G_{ij}, & i \neq j \\ 
    \gamma \ \text{OutDeg}(j), & i = j
    \end{cases}
\end{equation}

\noindent where the off-diagonal elements $M_{ij}$ represent the probability flow along an edge from vertex $j$ to vertex $i$, while the diagonal elements $M_{jj}$ account for the total outflow from vertex $j$ per unit time. Scalar $\gamma \in \mathbb{R}$ is the system wide transition rate \cite{falloon_qswalk:_2017}.

\subsection{Continuous-Time Quantum Walks}

A continuous-time quantum walk (CTQW) is constructed by mapping $\mathcal{G}$ to an $N$-dimensional Hilbert space where the set of its vertices $\{\ket{v_1}, ..., \ket{v_N}\}$ form an orthonormal basis. The matrix elements of the system Hamiltonian $H$ are then equal to the classical transition matrix ($\bra{v_j}H\ket{v_i} = M_{ij}$). In place of $\vec{p}(t)$, the evolution of the state vector $\ket{\Psi(t)} = \sum_{i=1}^{N} \ket{v_i}\braket{v_i}{\Psi(t)}$ is considered, the dynamics of which are governed by the Schr\"odinger equation \cite{falloon_qswalk:_2017},

\begin{equation}
    \label{eq:CTQW}
    \frac{d \ket{\Psi(t)}}{dt} = %
    -\frac{\mathrm{i}}{\hbar} H \ket{\Psi(t)}
\end{equation}

\noindent which has the formal solution $\ket{\Psi(t)} = \exp(-i tH)\ket{\Psi(0)}$ when $H$ is time-independent\footnote{In atomic units where $\hbar = 1 \ \text{a.u} =  1.054 \ 571 \times 10^{-34} \text{J.s}$ and $t = 2.418 884 \times 10^{-17} s = 24.188 \ 84 \ \text{fs}$.}. The probability associated with vertex $v_i$ at time $t$ is $|\braket{v_i}{\Psi(t)}|^2$.

While Equations (\ref{eq:CTRW}) and (\ref{eq:CTQW}) appear superficially similar, there are several fundamental differences between the two processes. Firstly, $\ket{\Psi(t)}$ describes a complex probability amplitude, meaning that its possible paths may interfere. Secondly, the Hermiticity requirement on $H$ needed to maintain unitary evolution of the system dictates that M be derived from $\mathcal{G}^u$ \cite{whitfield_quantum_2010}. 

\subsection{Markovian Open Quantum Systems}

A density matrix,

\begin{equation}
    \label{eq:density matrix}
    \rho(t) = \sum_k p_k \ket{\Psi_k(t)}\bra{\Psi_k(t)} \text{,}
\end{equation}

\noindent describes a statistical ensemble of quantum states, $\ket{\Psi_k(t)}$, each with an associated probability $p_k \geq 0$ and $\sum_k p_k = 1$. The case where $p_k$ is non-zero for more than one $k$ is termed a mixed state while the case of only one non-zero $p_k$ is termed a pure state. Diagonal elements $\rho_{ii}$ represent the probability density at a given vertex and are termed `populations', while off-diagonal elements $\rho_{ij}$ describe phase coherence between vertices $i$ and $j$  \cite{falloon_qswalk:_2017}. 

Density matrices satisfy:

\begin{itemize}
\item $\rho(t)^\dagger = \rho(t)$.
\item $\text{Tr}(\rho(t)) = 1$.
\item $\text{Tr}(\rho(t)^2) \leq 1$, with equality holding for only pure states.
\item $\langle A \rangle = \text{Tr}(\rho(t)A)$, where $A$ is a quantum operator.
\end{itemize}

The dynamics of $\rho(t)$ are given by the Liouville-von Neumann equation, 

\begin{equation}
    \label{eq:liouville-von-neumann}
    \frac{d\rho(t)}{dt} = -\text{i}[H, \rho(t)],
\end{equation}

\noindent which is the density theoretic equivalent of the Schr\"odinger equation (Equation (\ref{eq:CTQW})) \cite{breuer_theory_2009}.

Consider a system, $S$, coupled to an external reservoir (or `bath'), $B$. The Hilbert space of $S + B$ is given by \cite{breuer_theory_2009},

\begin{equation}
    \label{eq:open_hilbert_space}
    \mathcal{H} = \mathcal{H}_S \otimes \mathcal{H}_B,
\end{equation}

\noindent where $\mathcal{H}_S$ and $\mathcal{H}_B$ are the Hilbert spaces of $S$ and $B$. $S$ is referred to as an `open' system, while $S + B$ is closed in the sense that its dynamics can be described unitarily. Under the conditions that the evolution of S is Markovian with no correlation between S and B at t = 0, and given $\mathcal{H}_S$ of finite dimensions $N$. The dynamics of S are described by a generalization of Equation (\ref{eq:liouville-von-neumann}): the GKSL quantum master equation \cite{breuer_theory_2009},

\begin{equation}
  \label{eq:gksl}
  \frac{d\rho_S(t)}{dt} = -\frac{\text{i}}{\hbar}[H, \rho_S(t)] + \sum_k \mathcal{D}_k[\rho_S(t)]
\end{equation}

\noindent with

\begin{equation} \label{eq:KL_eq}
    \mathcal{D}_k[\rho_S(t)] = \tau_k(L_k\rho_S(t)L_{k}^{\dagger} %
  - \frac{1}{2}\{L_{k}^{\dagger}L_k,\rho_S(t)\}),
\end{equation}

\noindent where $H$ is the Hamiltonian describing the unitary dynamics of $\mathcal{H}_s$ and $\mathcal{H}_B$, the Lindblad operators $L_k$ span the Liouville space and the scalars $\tau_k \geq 0$. The reduced density operator $\rho_s(t)$ is formed by tracing out the degrees of freedom associated with B. Equation (\ref{eq:gksl}) is invariant under unitary transformations of the Lindblad operators, allowing for the construction of a wide range of phenomenological models. 

\subsection{Quantum Stochastic Walks} \label{sec:qsw}

\subsubsection{Local Environment Interaction} \label{sec:l_qsw}

An local-interaction quantum stochastic Walk (L-QSW) on an arbitrary simple $\mathcal{G}$ is derived from Equation (\ref{eq:KL_eq}) by defining $\rho_s(t)$ in the basis of vertex states, $\{\ket{v_1},...,\ket{v_N}\}$, setting $H$ equal to the transition matrix of $G^u$, and deriving the local interaction Lindblad operators from the transition matrix of $G$, 

\begin{equation}
    \label{eq:lindblad}
    L_{k}=\sqrt{|M_{ij}|}\ket{v_i}\bra{v_j}.
\end{equation}

\noindent where $k=N(j-1)+i$. Each $L_k$ describes an incoherent scattering channel along an arc of $\mathcal{G}$ when $i \neq j$ and dephasing at $v_i$ when $i = j$ \cite{whitfield_quantum_2010,falloon_qswalk:_2017}.       

A scalar decoherence parameter $0 \leq \omega \leq 1$ is introduced \cite{whitfield_quantum_2010}. This allows for the model to be easily tuned to explore a continuum of mixed quantum and classical dynamics. The standard form of a QSW is then,

        \begin{equation}
            \label{eq:qsw}
            \frac{d\rho(t)}{dt} = -\text{i}(1-\omega)[H, \rho(t)] %
            + \omega \sum_{k=1}^{N^2} \mathcal{D}_k[\rho(t)] 
        \end{equation}

\noindent with $\rho_s(t)$ denoted as $\rho(t)$ and $\tau_k = 1$ for all dissipator terms. At $\omega = 0$, Equation (\ref{eq:qsw}) reduces to a CTQW obeying the Liouville-von Neumann equation (Equation (\ref{eq:liouville-von-neumann})) and, at $\omega = 1$, the density-matrix equivalent of the CTRW equation (Equation (\ref{eq:CTRW})) is obtained. 

It is worth noting that QSWs are defined elsewhere directly from $G$ and $G^u$, such that $\bra{v_j}L_k\ket{v_i} = G_{ij}$ and $\bra{v_j}H\ket{v_i} = G^u_{ij}$ \cite{glos_qswalk.jl:_2019}. Additionally, the continuous-time open quantum walk (CTOQW) \cite{liu_steady_2017} defines quantum walks on undirected graphs obeying Equation (\ref{eq:gksl}), where $H$ is defined by Equation (\ref{eq:generator_matrix}) and, in place of $\sqrt{M_{ij}}$ in Equation (\ref{eq:lindblad}), is the canonical Markov chain transition matrix,

\begin{equation}
    \label{eq:markov_chain}
    C_{ij} =  
    \begin{cases}
    \frac{1}{\text{OutDeg}(v_j)}, & (v_i, v_j) \in E \\ 
    0, & \text{otherwise}. 
    \end{cases}
\end{equation}

\noindent In each case, these walks are consistent with the generalised definition of QSWs with locally-interacting Lindblad operators \cite{whitfield_quantum_2010}. 

\begin{figure} 
    \centering
    \begin{tikzpicture}
    \node (img) [anchor=south west,inner sep=0] at (0,0) {\includegraphics[width=6cm, valign=c]{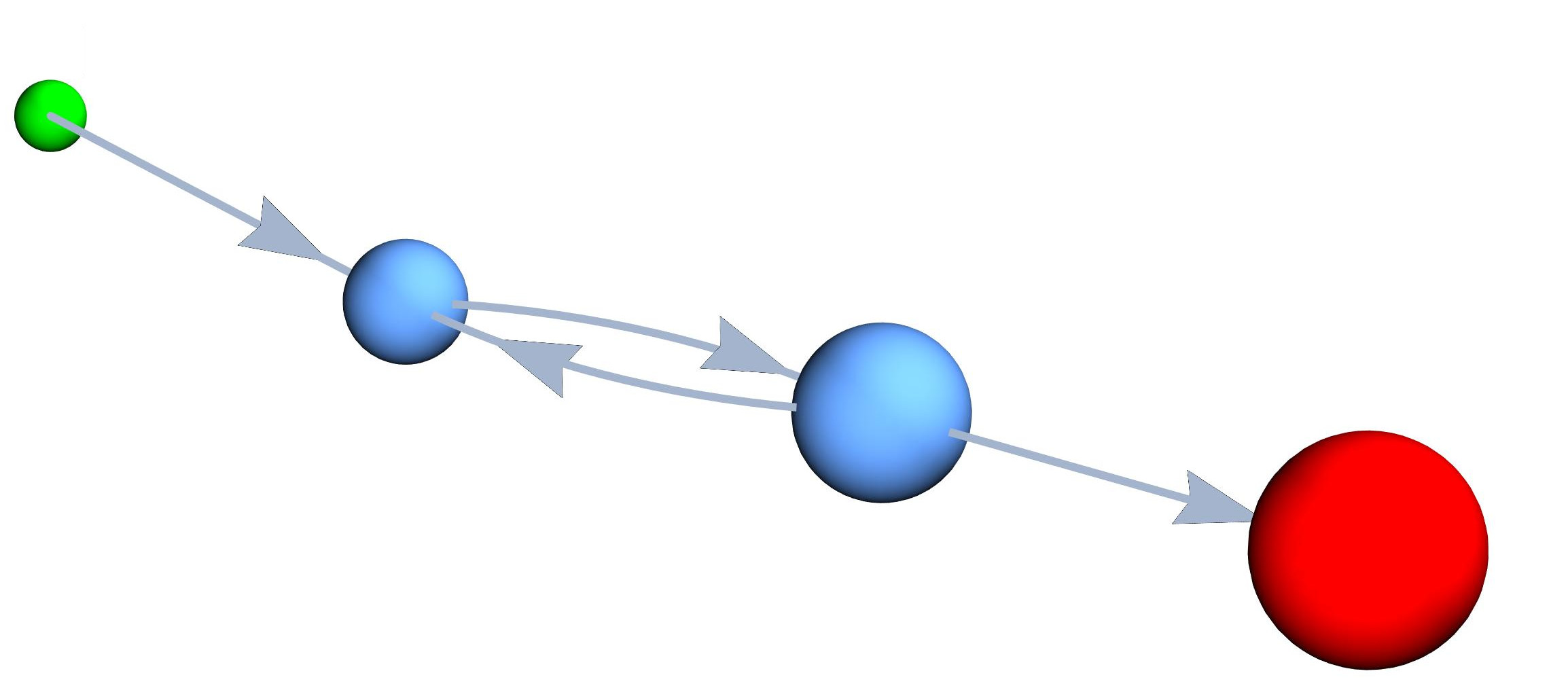}};
    \node[] at (0.20,2.51) {$v_3$};
    \node[] at (0.60,1.73) {$\Gamma_3$};
    \node[] at (1.58,1.9) {$v_1$};
    \node[] at (3.42,1.55) {$v_2$};
    \node[] at (4.1,0.65) {$\Theta_{14}$};
    \node[] at (5.25,1.15) {$v_4$};
\end{tikzpicture}
    \caption{A dimer graph with a source, $\Gamma_3 = 2 $, attached to $v_1$ and a sink, $\Theta_{14} = 3$, attached to $v_2$ (see Equations (\ref{eq:dimer_aug}) and (\ref{eq:qsw_ss})). Note that the absorption and emission channels are unidirectional.}
    \label{fig:dimer}
\end{figure}

The local interaction QSW model naturally facilitates the modelling of non-Hermitian transport through connected $\mathcal{G}$. This is achieved by introducing a source vertex set, $V^\Gamma$, and a sink vertex set, $V^\Theta$, which are connected unidirectionaly to $\mathcal{G}$ by arc sets $E^\Gamma$ and $E^\Theta$. Together with $\mathcal{G}$, these form the augmented digraph, $\mathcal{G}^{\text{aug}}$. For example, consider the dimer graph shown in Figure \ref{fig:dimer} on which absorption is modeled at $v_1$ and emission at $v_2$. In QSW\_MPI, $G^u$ and $G^{\text{aug}} = G + G^\Gamma + G^\Theta$ are represented as,

\begin{align}
\label{eq:dimer_aug}
G^u = \begin{bmatrix}
0 & 1 & 0 &0 \\ 
1 & 0 & 0 & 0\\ 
0 & 0 & 0 & 0\\ 
0 & 0 & 0 & 0
\end{bmatrix}, &&
G^{\text{aug}}=\begin{bmatrix}
0 & 1 & 2 &0 \\ 
1 & 0 & 0 & 0\\ 
0 & 0 & 0 & 0\\ 
0 & 3 & 0 & 0
\end{bmatrix}.
\end{align}

The walk Hamiltonian is then derived from $G^u$ and the $L_k$ corresponding to scattering and dephasing on $\mathcal{G}$ from $G$. Finally, $L_k$ originating from $\mathcal{G}^\Gamma$ and $\mathcal{G}^\Theta$ are formed as $\bra{v_j}L_k\ket{v_i} = G^{\Gamma}_{ij}$ and $\bra{v_j}L_k\ket{v_i} = G^{\Theta}_{ij}$ respectively, appearing in additional terms appended to Equation (\ref{eq:qsw}) outside the scope of $\omega$. An L-QSW incorporating both absorptive and emissive processes is then succinctly expressed as,

\begin{align}
  \begin{split}
    \label{eq:qsw_ss}
    & \frac{d\rho(t)}{dt} = -\text{i}(1-\omega)[H, \rho(t)] 
     + \omega \sum_{k = 1}^{\tilde{N}^2} \mathcal{D}_k[\rho(t)] \\ 
    & + \sum_{k = 1}^{\tilde{N}^2}\mathcal{D}^{\Gamma}_k[\rho(t)] 
   + \sum_{k = 1} ^{\tilde{N}^2}\mathcal{D}^{\Theta}_k[\rho(t)] \\ 
  \end{split}
\end{align}

\noindent where $k = \tilde{N}(j-1) + i$ with $\tilde{N}$ equal to $N$ plus the total vertices in $V^\Gamma$ and $V^\Theta$, and $\rho(t)$ is of dimensions $\tilde{N} \times \tilde{N}$. Terms $\mathcal{D}^{\Gamma}_k[\rho(t)]$ are defined as per Equation (\ref{eq:KL_eq}) with $\tau_k = \Gamma_k$ where $\Gamma_k$ is the absorption rate from source $v_j \in \mathcal{G}^\Gamma$ to vertex $v_i \in \mathcal{G}$. Similarly, for $\mathcal{D}^{\Theta}_k[\rho(t)]$, $\tau_k = \Theta_k$ where $\Theta_k$ is the emission rate from vertex $v_j \in \mathcal{G}$ to sink $v_i \in \mathcal{G}^{\Theta}$. 

\subsubsection{Global Environment Interaction} \label{sec:g_qsw}

A global-interaction quantum stochastic walk (G-QSW) differs from a L-QSW in that it utilizes a single Lindblad operator derived from the digraph adjacency matrix,

\begin{equation}
  \label{eq:L_global}
L_{\text{global}} = \sum_{i,j=1}^{N}G_{ij}\ket{v_i}\bra{v_j}.
\end{equation}

However, a Lindblad operator of this form has the potentially undesirable effect of inducing transitions between vertices whose arcs connect to a common outgoing vertex, a phenomena termed spontaneous moralisation. A demoralisation correction scheme can be applied to arrive at a non-moralising G-QSW (NM-G-QSW), which respects the connectivity of the originating digraph. This proceeds by a homomorphic mapping of $\mathcal{G}$ and $\mathcal{G}^u$ to an expanded vertex space\cite{domino_superdiffusive_nodate}. First supported by QSWalk.jl  \cite{glos_qswalk.jl:_2019}. Provided here is a practical overview of the process, which is implemented in QSW\_MPI with respect to weighted digraphs. 

\paragraph{Graph Demoralisation} \label{par:demoralisation}

\begin{enumerate} \item From $\mathcal{G}^u = (V, E^u)$, construct a set of vertex subspaces $V^D = \{V^D_i\}$ with $V^D_i = \{v^0_i,...,v^{\text{InDeg(i)-1}}_i\}$ for each $v_i \in V$. Associated with $V^D$ is edge set $E^{uD} = \{(v^i_j,v^k_l), (v^m_n,v^o_p),...\}$, where ($v^l_i,v^k_j) \in E^{uD} \iff (v_i,v_k) \in E^u$. These have weightings,
    \begin{equation}
        \label{eq:nm_weight} 
    \text{w}^{D}(V_i^D,V_k^D) =  \left(\text{SubDeg}(V_i^D,V_k^D)\text{w}(v_i,v_k)\right)^{-\frac{1}{2}} 
    \end{equation}
  
    \noindent where $\text{SubDeg}(V^D_i,V^D_k) = \dim(\{(v_i^l,v_k^j) : (v_i^l,v_k^j) \in E^{D}\})$ and, for $G^u$, $E^D = E^{uD}$. This forms the demoralised graph, $\mathcal{G}^{uD} = (V^D,E^{uD})$.
    
\item Construct the demoralised digraph, $\mathcal{G}^D = (V^D,E^D)$ where $(v_i^j,v_k^l) \in E^D \iff (v_i,v_k) \in E$ and the arc weights, are given by Equation (\ref{eq:nm_weight}).

\item  Form the Lindblad operator form orthogonal matrices, $\{F_i\}\in \mathbb{C}^{\dim(V^D_i) \times \dim(V^D_i)}$, such that,

  \begin{equation}
    \label{eq:dm_lind}
    L^D = (F_i)_{l(k+1)}\text{G}^{D}_{v_i^l,v_k^j}\ket{v^j_i} \bra{v^l_k},
  \end{equation}

\noindent and QSW\_MPI follows the convention of choosing for $\{F_i\}$ the Fourier matrices \cite{glos_qswalk.jl:_2019}.

\item Construct the rotating Hamiltonian, 

  \begin{equation}
    \label{eq:H_rot}
    \bra{v^k_l} H^D_{\text{rot}} \ket{v^i_j} =
    \begin{cases}
      \text{i}, & i=j \text{ and } l = k + 1 \mod \text{InDeg}(v_i) \\
      -\text{i}, & i=j \text{ and } l = k - 1 \mod \text{InDeg}(v_i) \\
      0, & \text{otherwise}
    \end{cases}
  \end{equation}

\noindent which changes the state within subspaces of V in order to prevent occurrence of stationary states dependant only on the expanded vertex set of $\mathcal{G}^D$.  

\end{enumerate}

Through formation of $L^D$, the spontaneous moralisation is destroyed, but, the induced dynamics may not correspond with symmetries present in $\mathcal{G}$. In this case, symmetry may be reintroduced by constructing additional $L^D$ formed using unique permutations of $\{F_i\}$. However, the generality of this symmetrisation process has not been confirmed \cite{domino_superdiffusive_nodate}. The master equation of a NM-G-QSW is then,

\begin{equation}
\begin{split}
  \label{eq:nm_gqsw}
\frac{d\rho^D(t)}{dt} = & -\text{i}(1-\omega)[H^D, \rho^D(t)] \\ 
& + \omega \left( \text{i}[H^D_{\text{rot}}, \rho^D(t)] + \sum_{\{L^D\}} \mathcal{D}_k[\rho^D(t)] \right).
\end{split}
\end{equation}

\noindent where $H^D$ is formed from $\mathcal{G}^{uD}$ as per Equation (\ref{eq:generator_matrix}). The probabilities of the demoralised density operator, $\rho^{D}(t)$, are related to the probability of measuring the state in vertex $v_i$ at time $t$ by

\begin{equation}
  \label{eq:nm_rho_map}
  p(v_i, t) = \sum_{v^k_i \in V_i^D}\bra{v^k_i}\rho^{D}(t)\ket{v^k_i}.
\end{equation}

\subsubsection{Vectorization of the Master Equation}

Equations \ref{eq:qsw}, \ref{eq:qsw_ss} and \ref{eq:nm_gqsw} may be recast as a system of first order differential equations through their representation in an $\tilde{N}^2 \times \tilde{N}^2$ Liouville space \cite{breuer_theory_2009}, where $\tilde{N}$ is the dimension of the system. This process, termed `vectorization', makes use of the identity $\text{vec}(XYZ) = (Z^T \otimes X)\text{vec}(Y)$ \cite{banerjee_linear_2014} to obtain the mappings,

\begin{align}
  \label{eq:vec_mappings}
  & [X,Y] \leftrightarrow (I \otimes X - X^T \otimes I)\text{vec}(Y), \\
  & \{X,Y\} \leftrightarrow (I \otimes X + X^T \otimes I)\text{vec}(Y), \\
  & X.B.X^{\dagger} \leftrightarrow (X^* \otimes X)\text{vec}(Y)
\end{align}

\noindent where $X, Y, Z \in \mathbb{C}^{\tilde{N} \times \tilde{N}}$. Such that, for each QSW variant, its equation of motion has the solution,

\begin{equation} \label{eq:qsw_vec_sol}
   \tilde{\rho}(t) = \exp(t\tilde{\mathcal{L}})\tilde{\rho}(0),
\end{equation}

\noindent where $\rho(t)$ is related to the vectorised density matrix, $\tilde{\rho}(t)$, by the mapping $\tilde{\rho}_k \leftrightarrow \rho_{ij}$ and $\tilde{\mathcal{L}}$ is the vectorized superoperator.

\section{Computational Methods} \label{sec:computation}

QSW\_MPI has been developed primarily for use on distributed memory systems, computational clusters consisting of multiple networked CPUs with local memory. This strategy affords a much higher degree of parallelisation than is possible with a threaded model whereby multiple tasks run on the same CPU with shared memory access. Parallelisation was achieved using the Message Passing Interface (MPI) protocol, a well established and highly portable standard for distributed memory computation. Essentially, MPI runs multiple copies of a given program over an \textit{MPI communicator}, in which each process (or \textit{node}) is identified sequentially by its \textit{rank}. Communication occurs between these isolated nodes via the passing of messages as instructed by directives placed in code using the MPI API. A finer layer of shared-memory parallelism using OpenMP is included as a compile-time option. 

 Overall, the Fortran libraries developed for QSW\_MPI consist of approximately 45 subroutines comprised of over 3000 lines of code, most of which are optimised for parallel execution. This section thus provides a high-level overview of approaches and design considerations used in the development of the package. We being with a discussion of the selected matrix exponentiation methods which, by Equation (\ref{eq:qsw_vec_sol}), is the primary task at hand. This is followed by an outline of the data structures and parallelisation scheme used to achieve fast and memory-efficient QSW simulation.
 
\subsection{Matrix Exponentiation}

The matrix exponential is defined by a converging Taylor series

\begin{equation}
\exp(A) := \sum_{j=0}^\infty \frac{(A)^j}{j!}  
\end{equation}

\noindent where $A \in \mathbb{C}^{n \times n}$. Direct application of this formula is not generally practical as its rate of convergence can vary wildly. An additional challenge is that, for  $A = \mathcal{L}$, $n$ grows exponentially with $\tilde{N}$, meaning that multiple powers of $A$ may not be stored easily in computer memory. For this reason, and in light of Equation (\ref{eq:qsw_vec_sol}), it is instead preferable to directly approximate $\exp(A)\vec{u}$ where $\vec{u} \in \mathbb C^n$  \cite{moler_nineteen_2003}. 

Of the algorithms used to compute $\exp(A)\vec{u}$, perhaps the most common for sparse matrices are the Krylov subspace techniques. These proceed by approximating the $n$-dimensional problem in a smaller $m$-dimensional subspace of $\text{span}\{\vec{u}, A\vec{u},...,A^{m-1}\vec{u}\}$,  on which efficient dense matrix exponentiation methods may then be used \cite{sheehan_computing_2010}.

Other popular techniques are based on polynomial expansions of $\exp(A)$. For example, the Chebyshev approximation method is based on the Chebyshev expansion of the matrix exponential about the point $\left[ \lambda_{min}, \lambda_{max} \right] \subset \mathbb{C}$, where $\lambda_{min}$ and $\lambda_{max}$ are the eigenvalues of $A$ with the smallest and largest real values \cite{WangScholz1998, MidgleyWang2000, izaac_computational_2018}. 

\begin{figure} 
    \centering
    \subfloat{\includegraphics[width=6cm, valign=c]{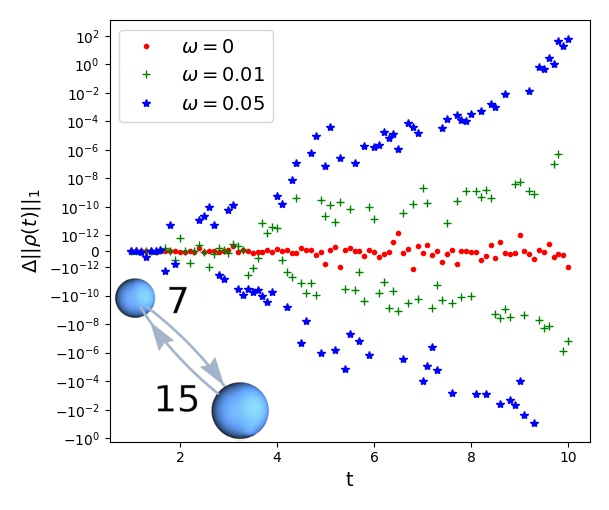}}
    \caption{Norm-wise error resulting from application of the Chebyshev approximation method as given in Ref. \cite{izaac_computational_2018} to simulate an L-QSW on a directed dimer graph with target numerical error less than $10^{-18}$. $\Delta \onenorm{\rho(t)} = \onenorm{\rho_C(t)} - \onenorm{\rho_M(t)}$, where $\onenorm{.}$ is the matrix 1-norm, $\rho_C(t)$ is the result given by the Chebyshev approximation, and $\rho_M(t)$ by the Mathematica \texttt{MatrixExp} function.}
    \label{cheb_error}
\end{figure}

This method is popular in quantum simulation as it offers fast and reliable convergence for Hermitian matrices \cite{moler_nineteen_2003, izaac_computational_2018, izaac_pyctqw:_2015}. It also has a lower memory overhead than the Krylov methods, not requiring storage of basis vectors and ancillary matrices \cite{sheehan_computing_2010}. However, if $A$ is non-Hermitian with eigenvalues off the negative real axis of the complex plane, the expansion can produce a poor approximation of $\exp(A)$ \cite{moler_nineteen_2003}. This is shown in Figure \ref{cheb_error}, where complex eigenvalues resulting from inclusion of the non-unitary Lindblad operators in $\tilde{\mathcal{L}}$ result in poor numerical stability proportional to $\omega$ and $t$.

A third approach, `scaling and squaring', takes advantage of the relationship, 

\begin{equation} \label{scale}
    \exp(A) \vec{u} = \exp\left(s^{-1} A\right) ^s \vec{u} = \left( \prod_{\text{s times}} \exp(s^{-1} A) \right) \vec{u} = \vec{v}
\end{equation}

\noindent to reduce the number of terms in a series expansion of $\exp(A)$ needed to satisfy a numerical error less than $\epsilon$ \cite{al-mohy_computing_2011}. Let $T_m(A)$ denote a Taylor series expansion to $m$ terms. Then,

\begin{equation} \label{eq:ss_tay}
    \exp(A)\vec{u} \approx \left( T_m( s^{-1} A) \right)^s \vec{u}  = \left( \sum_{j = 0}^{m} \frac{s^{-1} A^j}{j!} \right)^s \vec{u}.
\end{equation}

Historically, this method has not been favoured due to difficulties in the selection of the optimal $m$ and $s$ parameters \cite{moler_nineteen_2003}. However, the scaling and squaring algorithm developed by Al-Mohy and Higham \cite{al-mohy_computing_2011} achieves this reliably via backwards error analysis while additionally allowing though efficient time series calculation. This method shares the advantages of the Chebyshev approximation over the Krylov methods. It has also been experimentally demonstrated as being numerically stable for both Hermitian and non-Hermitian matrices, with computational performance comparable to the Chebyshev approximation in both cases \cite{al-mohy_computing_2011}. This method is also well tested; it forms the basis for sparse matrix exponentiation in the widely used SciPy Python library \cite{jones_scipy:_2001}. Despite its perceived success, there has been little explicit discussion of its use in the literature of quantum simulation. This may be due in part to it not being currently implemented in widely used parallel numerical libraries such as PetSc \cite{balay_petsc_2019}. As such, this method has been selected for its purported advantages, stability with non-Hermitian matrices and novelty in the context of computational quantum physics. Specifically, QSW\_MPI implements Algorithm 3.2 (named \texttt{step} in QSW\_MPI) for single time point calculations and Algorithm 5.2 for time-series calculations (named \texttt{series} in QSW\_MPI) as described in Ref. \cite{al-mohy_computing_2011}, without the optional balancing or minimisation of the Frobenius norm.  

\subsection{Sparse operator representation}
QSW\_MPI increases the scope of possible simulations by representing $\tilde{\mathcal{L}}$ using a sparse matrix format. For this, the Compressed Sparse Rows (CSR) format has been selected. An advantageous property of the CSR datatype is that it provides for the efficient access of matrix rows, as each row has its non-zero column indices and values stored as a contiguous sub-array. This quality distinguishes CSR from other standard sparse matrix formats and allows for the efficient computation of matrix-vector products: the fundamental algebraic operation in the action of matrix exponentiation. 

For the case of an L-QSW, as all of the terms in $\tilde{\mathcal{L}}$ involve a Kronecker product with an identity matrix or sparse Lindblad operator, $\tilde{\mathcal{L}}$ is highly sparse. While the total number of matrix elements in $\mathcal{L}$ is $\tilde{N}^4$, the number of non-zero entries is of the order $\tilde{N}^{3}$. This results in a sparsity of over 90\% for digraphs with 20 vertices or greater, meaning that the CSR format provides a memory-efficient representation of the L-QSW for all but the smallest of systems. 

Given a G-QSW or NM-QSW,  Potential non-sparsity in the $L_{\text{global}}$ or $L^D_k$ results in an $\tilde{\mathcal{L}}$ sparsity proportional to the square of the non-zeros in $G$. Practically, this limits the scope of memory-efficient simulations using the CSR representation to $G$ with $\geq \sim 76 \%$ non-zeros, assuming single-precision CSR indexing arrays and complex non-zero entries stored in double precision. 
 
\subsection{Parallel Partitioning Scheme}  

\begin{figure}%
    \centering
    \subfloat[]{{\includegraphics[width=3.5cm, valign=c]{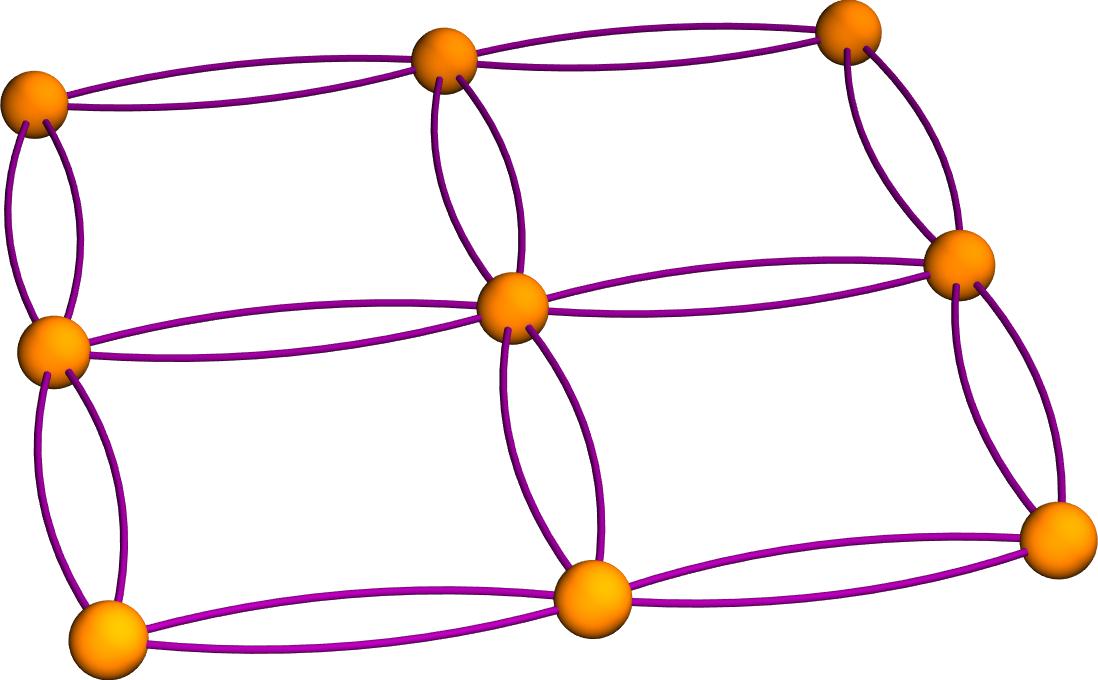} }}%
    \qquad
    \subfloat[]{{\includegraphics[width=3.5cm, valign=c]{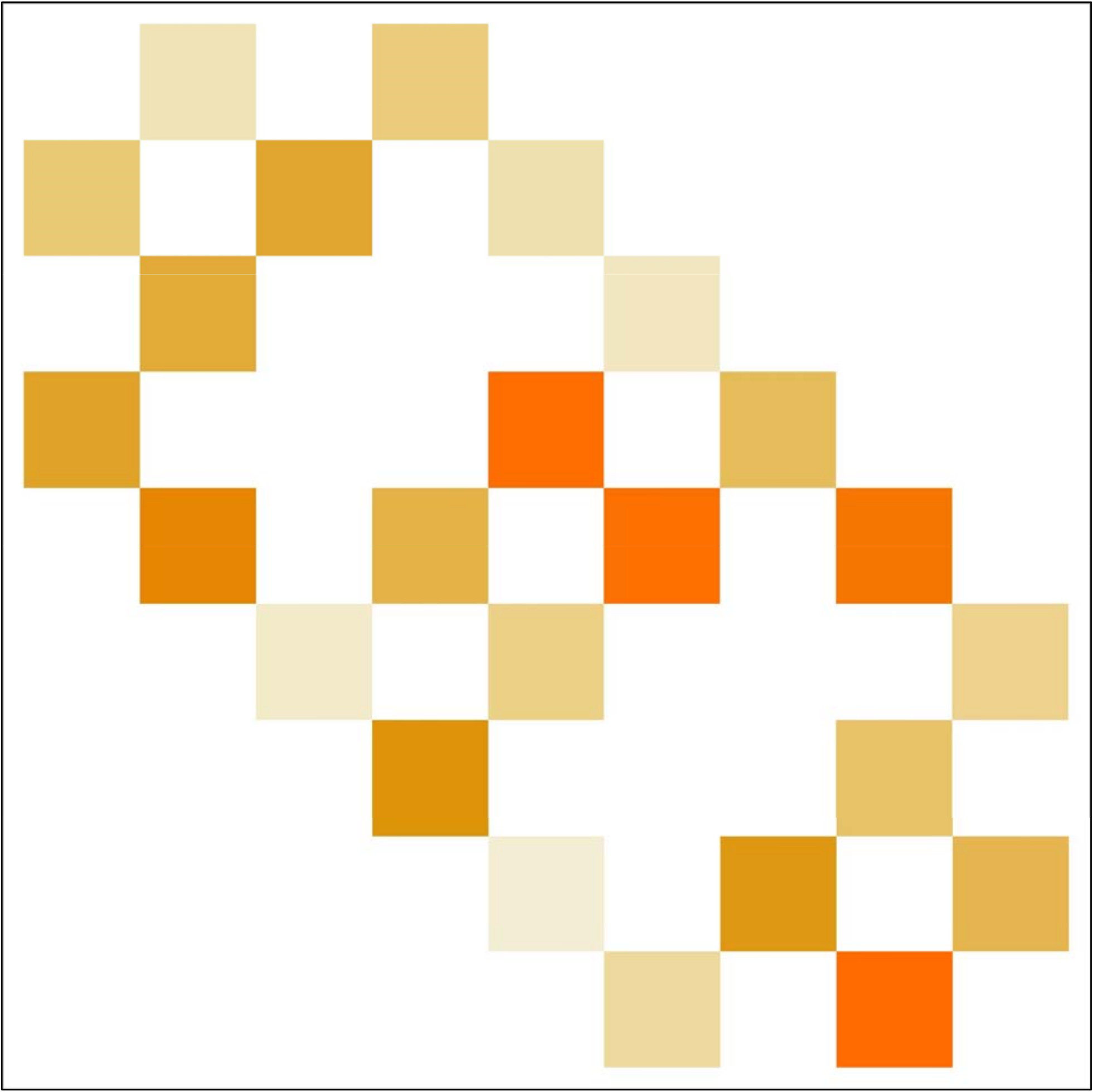} }}%
    \newline
    \subfloat[]{{\includegraphics[width=3.9cm]{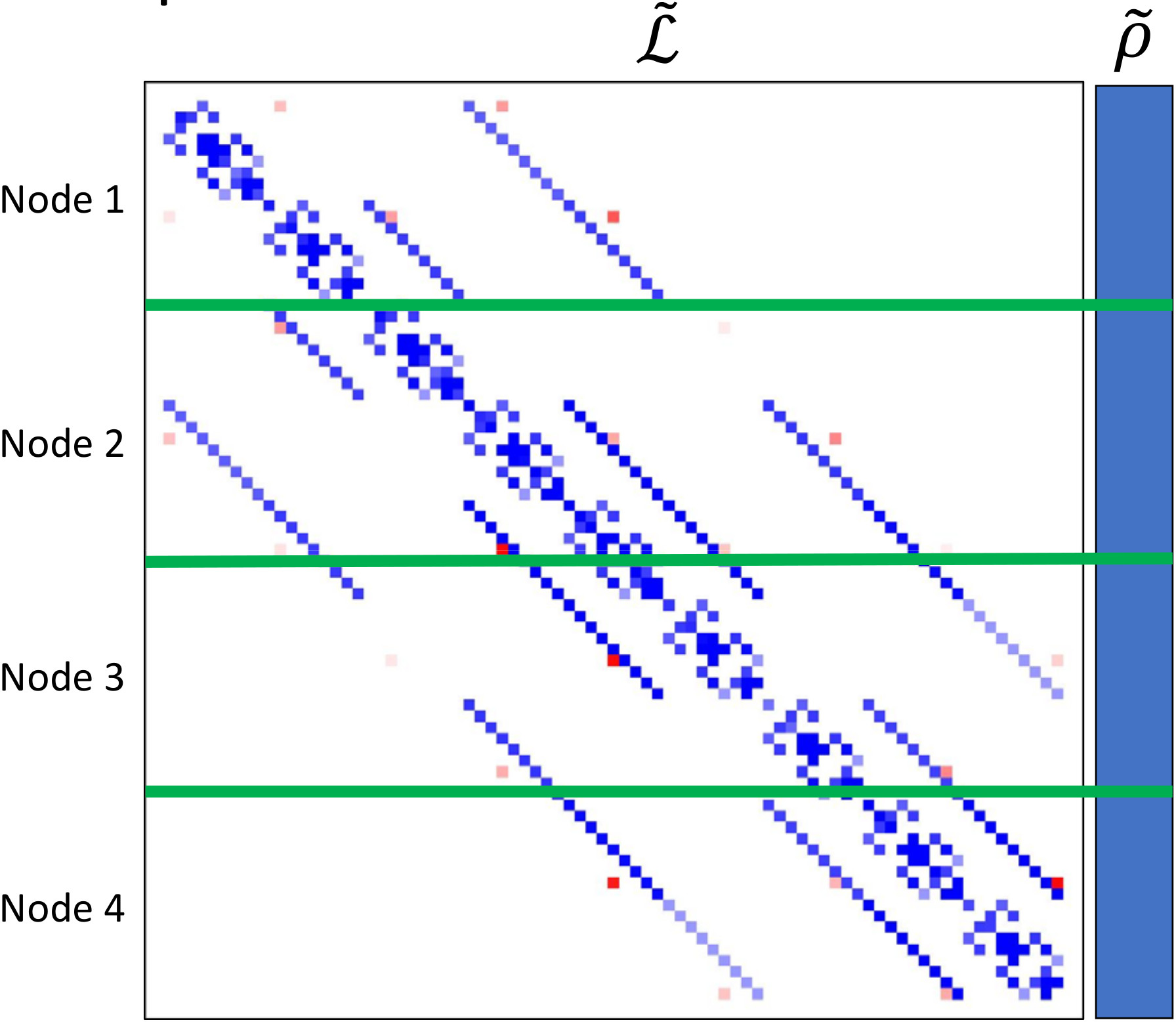} }}%
  \qquad
    \subfloat[]{{\includegraphics[width=3.9cm]{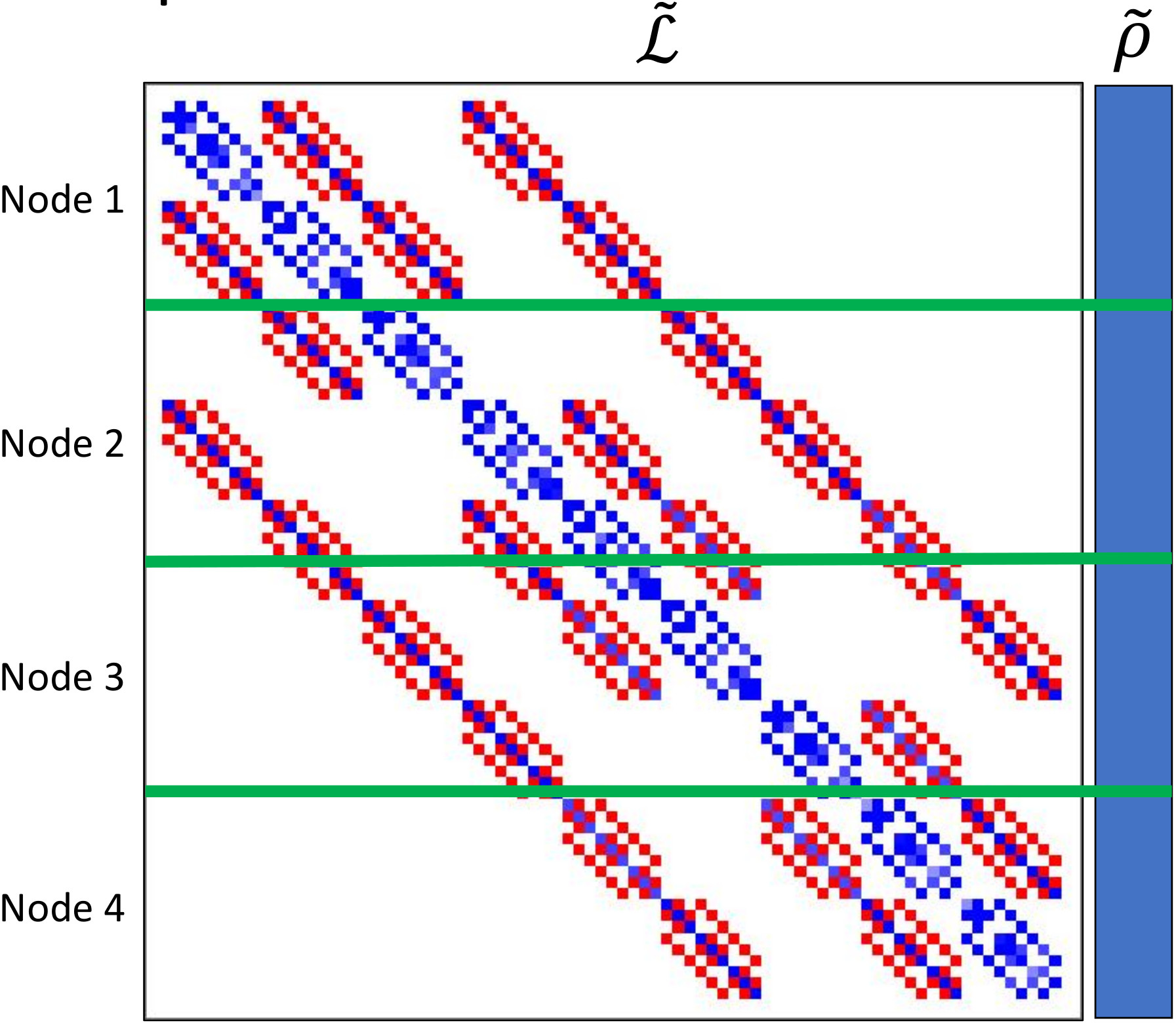} }}%
    \caption{(a) A directed square lattice graph with edge weights $(0,1)$ and
    adjacency matrix, $G$, of non-zero structure as shown in (b). The orange
    squares denote the relative magnitude of each value. The structure of
    $\tilde{\mathcal{L}}$ is depicted for a (c) L-QSW and (d) G-QSW for $\omega = 0.5$. Red and blue intensity depicts the relative magnitude of the non-zero values, with red denoting a real valued entry and blue denoting a complex valued entry. Green vertical lines depict the parallel row-wise partitioning of $\mathcal{L}$ and the vectorized density operator, $\tilde{\rho}$, for a MPI communicator consisting of four nodes.}
    \label{fig:struct}
\end{figure}
    
The structure of $\tilde{\mathcal{L}}$ resulting from a directed lattice graph
is shown in Figure \ref{fig:struct} for an L-QSW and G-QSW. In each case, it
displays a high degree of structural symmetry, with a block-circulant structure
along its diagonal flanked by diagonal striping. However, it is important to
note that, aside from the case of $\omega=0$ and $\{L_\Gamma\} = \{L_\Theta\} = \emptyset$, $\tilde{\mathcal{L}}$ is not block circulant as generally $L^T_k \neq L_k$. Hence, efficient eigendecomposition techniques, which take advantage of analytical solutions for the eigenvalues and eigenvectors of circulant block matrices, do not provide a general method for obtaining $\tilde{\rho(t)}$ \cite{rjasanow_effective_1994}. 

Despite this, these structural properties are easily exploited. Firstly, the diagonal structure results in each row being primarily dependant during matrix-vector multiplication on vector elements within the `vicinity' of its row index. This motivates the adoption of the row-wise partitioning scheme depicted in Figure \ref{fig:struct}, which ensures that the parallel processes communicate primarily with nodes containing vector elements adjacent to their local partition, thus limiting inter-process communication. 

The block structure of $\tilde{\mathcal{L}}$ means that the L-QSW
superoperator can be efficiently constructed directly from the CSR
representations of $H$ and a single-matrix representation of the
local-interaction Lindblad operators,

\begin{equation}
  \label{eq:condensed_lindblads}
M_L = \sqrt{|M_{ij}|},
\end{equation}

\noindent thus avoiding the need to form
intermediate Kronecker products or store each Lindblad operator separately. Generality in the structure of $L_k$ for G-QSWs necessitates
the explicit formation of $L_k$ and the vectorisation terms given by Equation \ref{eq:vec_mappings}. However, for both cases, construction of $\tilde{\mathcal{L}}$ is independent at each node following receipt of $G$, and $\tilde{\mathcal{L}}$ is consistently stored in distributed memory.

As $\exp(t \tilde{\mathcal{L}})\vec{u}$ is found through repeated action of $\tilde{\mathcal{L}}$ on a $\tilde{\rho}(0)$, MPI inter-process communication in each multiplication cycle is reduced by the determination of the specific vector elements that need to be sent and received with each multiplication on formation of $\tilde{\mathcal{L}}$. This has the additional benefit of ensuring that the sparse matrix multiplication algorithm implemented in QSW\_MPI acts on contiguous row sub-arrays of the CSR values and column index arrays, as is typically the case for non-parallel sparse matrix multiplication \cite{kepner_graph_2011}. 

If enabled, sparse matrix multiplication, and other intensive exponentiation-related do-loops, are further parallelised within each MPI process using OpenMP. For digraphs with a high degree of connectivity or distributed systems with slow networks, this can produce superior performance at a smaller number of MPI processes through a reduction in inter-process communication overhead, and a corresponding increase in cache-locality for the sparse matrix multiplication do-loops.  

\section{Package Overview} \label{chap:QSW}

The QSW\_MPI software package brings together the computational methods described in Section \ref{sec:computation} through a Python interface, thus providing a user friendly means of high performance QSW simulation. This primarily occurs through use of the \texttt{qsw\_mpi.MPI} submodule which provides for the creation of distributed $\tilde{\mathcal{L}}$, vectorization of $\rho(0)$, and evolution of the system dynamics. In particular, the user creates and calls methods from one of the following \texttt{walk} classes: 

\begin{itemize}
  \item \texttt{LQSW}: L-QSWs (Section \ref{sec:l_qsw}).
  \item \texttt{GQSW}: G-QSWs (Section \ref{sec:g_qsw}).
\end{itemize}

A \texttt{walk} object is in instantiated by passing to it the relevant operators, coefficients and MPI-communicator. On doing so the distributed $\tilde{\mathcal{L}}$ is generated and its 1-norm series calculated\footnote{Selection of optimal series expansion terms ($m$) and scaling and squaring parameters ($s$) is achieved through backwards error analysis dependant on $A_\text{norms} = \{\onenorm{A^n}\}$, where $n = 1,...,9$ and $\onenorm{.}$ is the matrix 1-norm \cite{al-mohy_computing_2011}. As $\onenorm{tA^n} = t \onenorm{A^n}$, $A_\text{norms}$ is reusable for all exponentiation at the same $\omega$. It is thus included as part of the $\tilde{\mathcal{L}}$ construction phase.}. After this the user provides defines $\rho(0)$ and generates the distributed $\tilde{\rho}(0)$ via the \texttt{initial\_state} method. 

Simulations are carried out for a single time point with the \texttt{step} method or for a number of equally spaced points using the \texttt{series} method. These return $\tilde{\rho}(t)$ (or $\tilde{\vec{\rho}}(t)$) as a distributed vectorized matrix which can be reshaped gathered at a specified MPI process via \texttt{gather\_result}, or measured via \texttt{gather\_populations}. Otherwise, results may be reshaped and saved directly to disk using \texttt{save\_result} or \texttt{save\_population}. File I/O is carried out using h5py \cite{collette_python_2013},  a python interface to the HDF5 libraries, and will default to MPI parallel-I/O methods contained in the non-user accessible \texttt{qsw\_mpi.parallel\_io} module  if such operations are supported by the host system. Finally, a second user accessible module \texttt{qsw\_mpi.operators} provides for creation of L-QSW and NM-G-QSW operators from $\mathcal{G}$ stored in the SciPy CSR matrix format \cite{jones_scipy:_2001}. 

The following provides an overview of QSW\_MPI workflows using examples drawn from prior studies - which correspond to files included in `QSW\_MPI/examples'. In addition to the program dependencies of QSW\_MPI, the example programs make use of the python packages Networkx \cite{hagberg_exploring_2008} for graph generation, and Matplotlib \cite{hunter_matplotlib:_2007} for visualisation. Note that a complete accounting of the methods contained in QSW\_MPI exceeds the scope of this document. Comprehensive documentation and installation instructions are included with the package and are additionally hosted on Read the Docs \cite{Matwiejew}.

\subsection{Usage Examples} \label{sec:usage}

\subsubsection{Execution}

QSW\_MPI programs, and other python 3 programs utilising MPI, are executed with the command,

\begin{verbatim}
mpirun -N <n> python3 <program_file.py>
\end{verbatim}

\noindent where \texttt{<n>} is a user specified parameter equal to the number of MPI processes.

\subsubsection{Graph Demoralisation} 

Here we provide an example of the typical workflow of QSW\_MPI through an exploration of the graph demoralisation process. This begins by loading the required modules and external methods.

\begin{verbatim}
import qsw_mpi as qsw 
import numpy as np
from scipy.sparse import csr_matrix as csr
from mpi4py import MPI
\end{verbatim}

As the system explored in this example is small, its simulation will not benefit from multiple MPI processes. However, initialisation of an MPI communicator is required to use the \texttt{QSW\_MPI.MPI} module.

\begin{verbatim}
comm = MPI.COMM_WORLD
\end{verbatim}

\begin{figure}[h]
  \centering
    \subfloat[]{{\includegraphics[width=3.5cm]{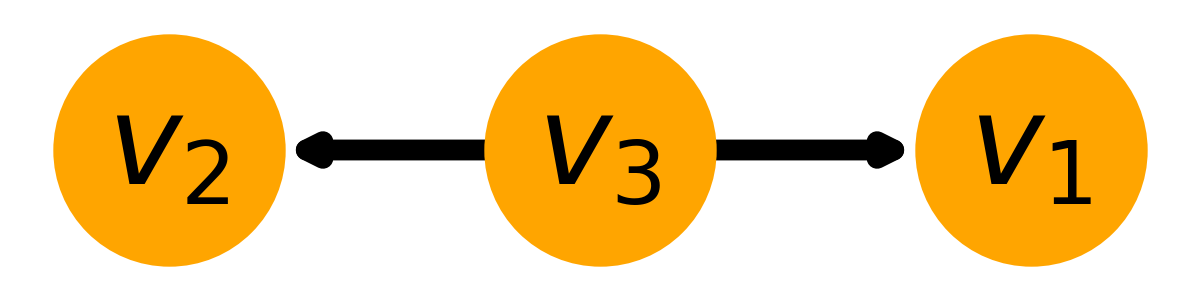} }}
    \qquad
    \subfloat[]{{\includegraphics[width=3.5cm]{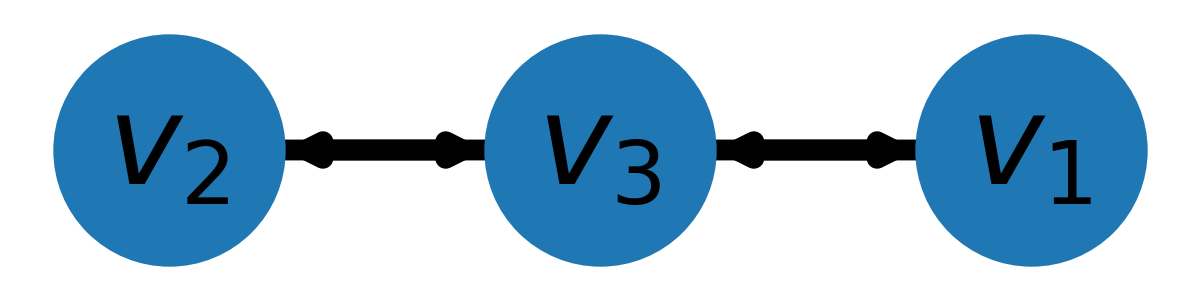} }}
    \caption{A three vertex (a) digraph consisting of two source verticies connected to a common sink and its (b) undirected counterpoint.}
    \label{fig:1_example}%
\end{figure}

Adjacency matrices $G$ and $G^u$ are defined here by writing them directly into the CSR format, where the arguments of \texttt{csr} are an ordered array of non-zero values, a corresponding tuple containing the row indices and column indices, and the dimensions of the adjacency matrix. The structure of the directed graph and its undirected counterpart is shown in Figure \ref{fig:1_example}.

{
\small
\begin{verbatim}
G = csr(([1,1],([2,2],[0,1])),(3,3))
GU = csr(([1,1,1,1],([0,1,2,2],[2,2,0,1])),(3,3))
\end{verbatim}
}

First, we examine the behaviour of a G-QSW. The Lindblad operator and Hamiltonian are created as per Equations \ref{eq:generator_matrix} and \ref{eq:L_global}. Note that the Lindblad operator is contained within an array.

\begin{verbatim}
gamma = 1.0
L = [G]
H = qsw.operators.trans(gamma, GU)
\end{verbatim}

Next, the starting state of the system is specified as a pure state at $v_1$. This may be achieved by either specifying $\rho(0)$ completely or by giving a list of probabilities, in which case its off-diagonal entries are assumed to be $0$. Here, the latter approach is employed.

\begin{verbatim}
rho_0 = np.array([1,0,0])
\end{verbatim}

A \texttt{GQSW} walk object is now initialised with $\omega = 1$, such that the dynamics induced by $L_{\text{global}}$ can be examined in isolation. The initial state of the system is then passed to the walk object.

\begin{verbatim}
omega = 1.0
GQSW = qsw.MPI.GQSW(omega, H, Ls, comm)
GQSW.initial_state(rho_0)
\end{verbatim}

Using the \texttt{step} method the state of the system at $t = 100$ is examined. Note that the result is gathered to a single MPI process. As such, commands acting on the gathered array should be contained within a conditional statement which first checks for the correct MPI process rank. 

\begin{verbatim}
GQSW.step(t = 100)
rhot = GQSW.gather_result(root = 0)

if comm.Get_rank() == 0:
    print(np.real(rhot.diagonal()))
\end{verbatim}

After the period of evolution, we find that there is a non-zero probability of there being a walker at $v_2$, despite it having an in-degree of 0.

\begin{verbatim}
>> [0.25 0.25 0.5]
\end{verbatim}

This is an example of spontaneous moralisation, a non-zero transition probability between $v_1$ and $v_3$  occurs due to them having a common `child' node.

We will now demonstrate how to use QSW\_MPI to apply the demoralisation correction scheme. First, we create a set of vertex subspaces, $V^D$.

\begin{verbatim}
vsets = qsw.operators.nm_vsets(GU)
\end{verbatim}

These are then used with adjacency matrices G and GU to create the Hamiltonian, Lindblad operators and rotating Hamiltonian which capture the structure of the demoralised graph and demoralised digraph.

\begin{verbatim}
H_nm = qsw.operators.nm_H(gamma, GU,vsets)
L_nm = [qsw.operators.nm_L(gamma, G,vsets)]
H_loc = qsw.operators.nm_H_loc(vsets)
\end{verbatim}

When creating the \texttt{GQSW} walk object, it is initialised with additional arguments specifying the vertex subspaces and rotating Hamiltonian.

\begin{verbatim}
nm_GQSW = qsw.MPI.QSWG(omega, H_nm, L_nm, 
                       comm, H_loc = H_loc, 
                       vsets = vsets)
\end{verbatim}

The initial system state is then mapped to the moralised graph as per Equation (\ref{eq:nm_rho_map}),

\begin{verbatim}
rho_0_nm = qsw.operators.nm_rho_map(rho_0, vsets)
\end{verbatim}

\noindent and passed to the walk object via \texttt{nm\_GQSW.initial\_state}. System propagation and measurement proceeds as previously described. At $t = 100$ the system is now in a pure state at the sink node, as expected by the originating graph topology.

\begin{verbatim}
>> [3.72007598e-44 0.00000000 1.00000000]
\end{verbatim}

As a further point of consideration, we will now compare the dynamics of the NM-G-QSW to an L-QSW on the same digraph, with $H$ and $M_L$ defined as the adjacency matrices \texttt{GU} and \texttt{G}. Note that $M_L$ is provided as a single CSR matrix.

\begin{verbatim}
LQSW = qsw.MPI.LQSW(omega, GU, G, comm)
LQSW.initial_state(rho_0)
\end{verbatim}

\noindent Evolving the state to $\rho(100)$ with \texttt{LQSW.step} yields,

\begin{verbatim}
>> [-9.52705648e-18  0.00000000  1.00000000].
\end{verbatim}

\noindent Which corresponds to the state of the NM-G-QSW. 

The coherent evolution of the two systems is examined by first rebuilding $\tilde{\mathcal{L}}$ at $\omega = 0$.

\begin{verbatim}
GQSW.set_omega(0)
LQSW.set_omega(0)
\end{verbatim}

\begin{figure}[h]
  \centering
  \includegraphics[width=5cm]{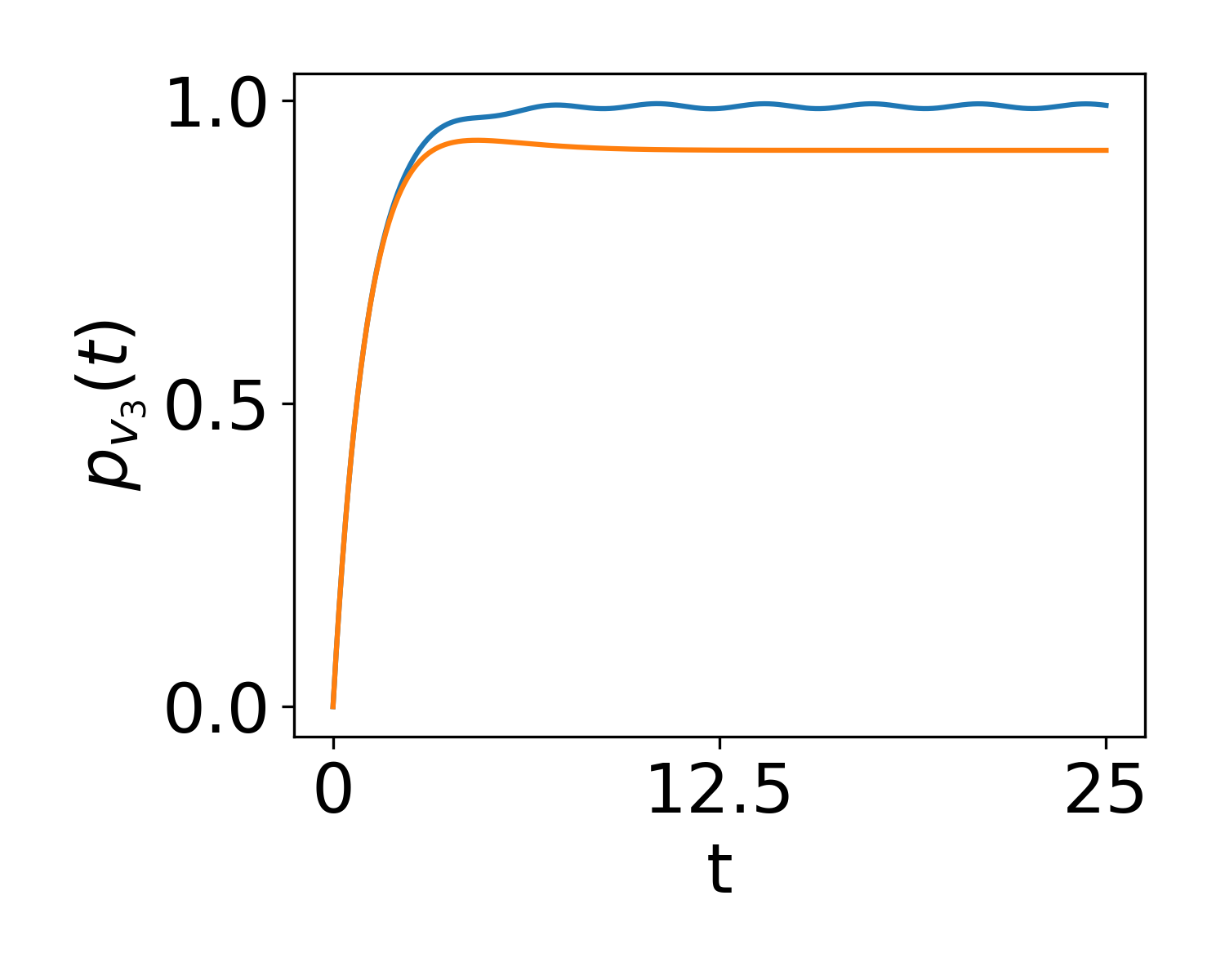}
  \caption*{{\color{C1} \textbf{---}} L-QSW, {\color{C0} \textbf{---}} NM-G-QSW}
  \caption{Probability at $v_3$ for an L-QSW and NM-G-QSW defined on the digraph and graph depicted in Figure \ref{fig:1_example} at $\omega = 0.9$.}
  \label{fig:1_sink_dynamics}
\end{figure}

\noindent After which a \texttt{step} to $t = 100$ yields,
{
  \small
\begin{verbatim}
>> [3.80773381e-07 9.98766244e-01 1.23337485e-03]
\end{verbatim}
 }

\noindent for the NM-G-QSW and,

{
\small
\begin{verbatim}
>> [3.80773217e-07 9.98766244e-01 1.23337485e-03]
\end{verbatim}
}

\noindent for the L-QSW. In fact, for this particular system, the limiting dynamics of a NM-G-QSW correspond to that of a CTQW and CTRW, as is the case for the L-QSW. However, if we examine the time evolution of the two systems  at $\omega = 0.9$ using the \texttt{series} method,

\begin{verbatim}
nm_GQSW.series(t1=0,tq=25,steps=500)
LQSW.series(t1=0,tq=25,steps=500)
\end{verbatim}

\noindent notably different dynamics are observed as shown in Figure \ref{fig:1_sink_dynamics}. the NM-G-QSW results in a higher transfer of probability to the sink vertex and does not as readily decay to a quasi-stationary state.

\subsubsection{Graph Dependant Coherence}

Here the steady-state solutions for an L-QSW on a 2-branching tree graph and a cycle graph are examined with respect to support for coherence. The graphs were generated and converted to sparse adjacency matrices using NetworkX and L-QSWs defined as per Equation (\ref{eq:qsw}) using the \texttt{LQSW} subclass.

\begin{figure}[h!]
  \centering
  \subfloat[]{{\includegraphics[width=3.5cm]{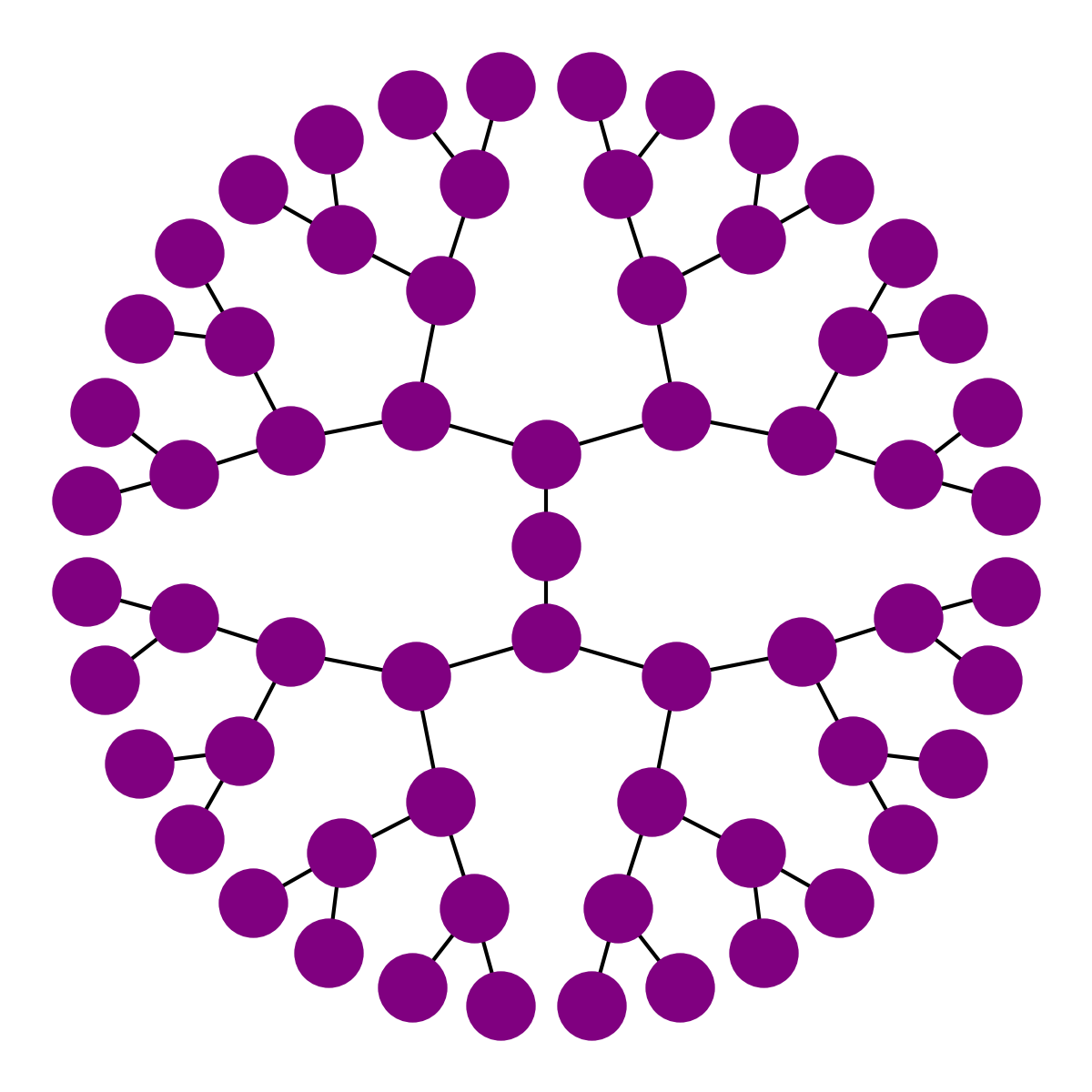} }}
  \qquad
  \subfloat[]{{\includegraphics[width=3.5cm]{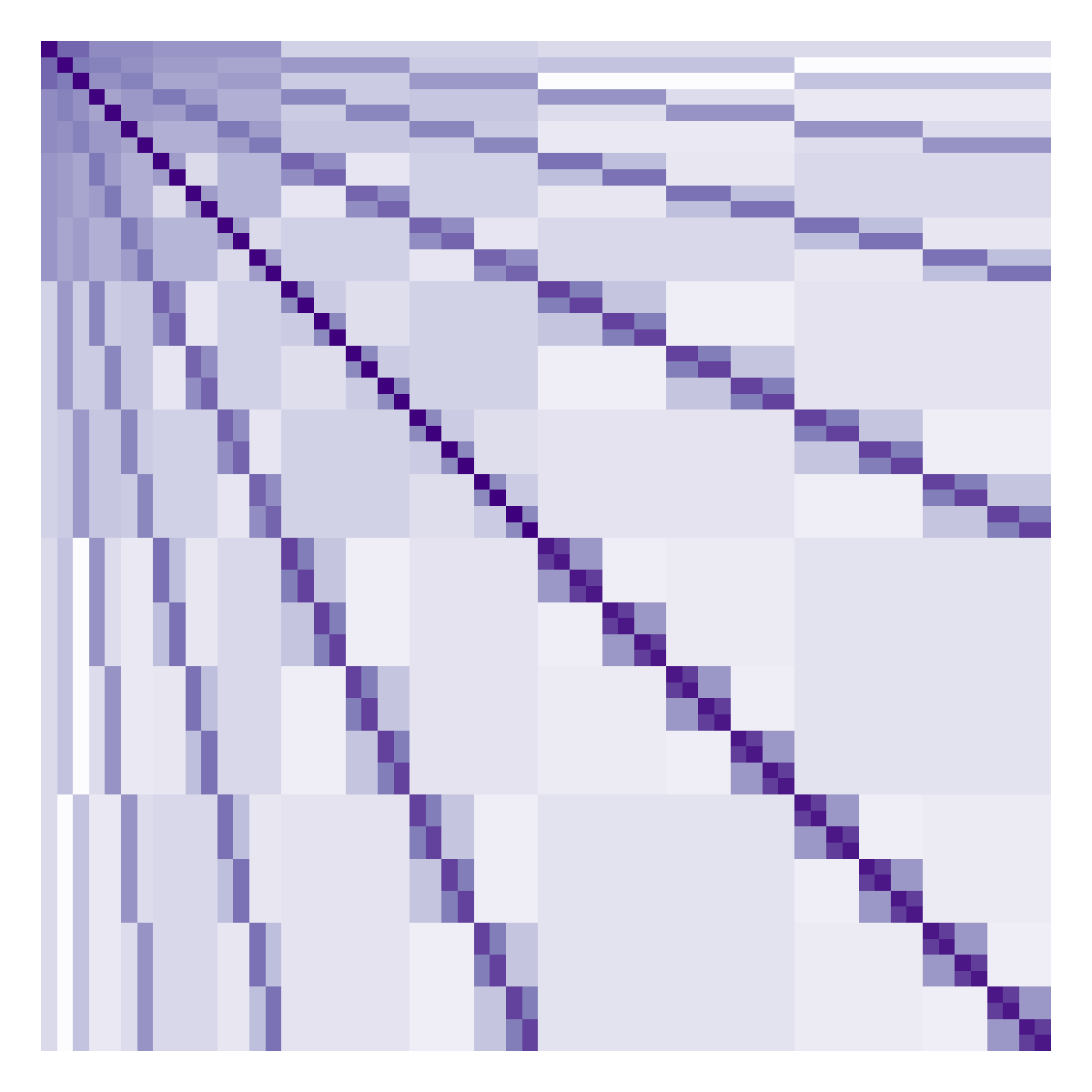} }}
  \newline
  \subfloat[]{{\includegraphics[width=3.5cm]{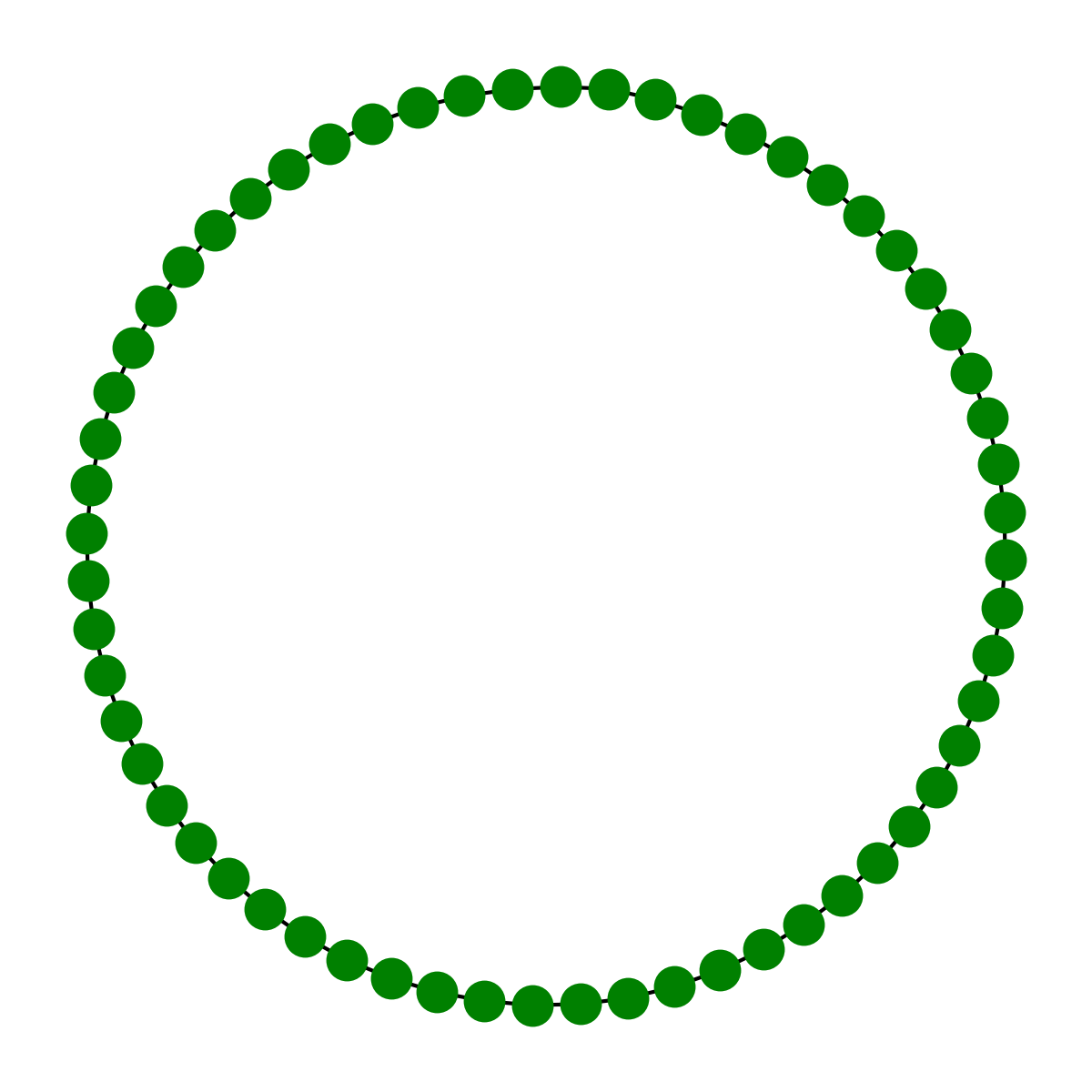} }}
  \qquad
  \subfloat[]{{\includegraphics[width=3.5cm]{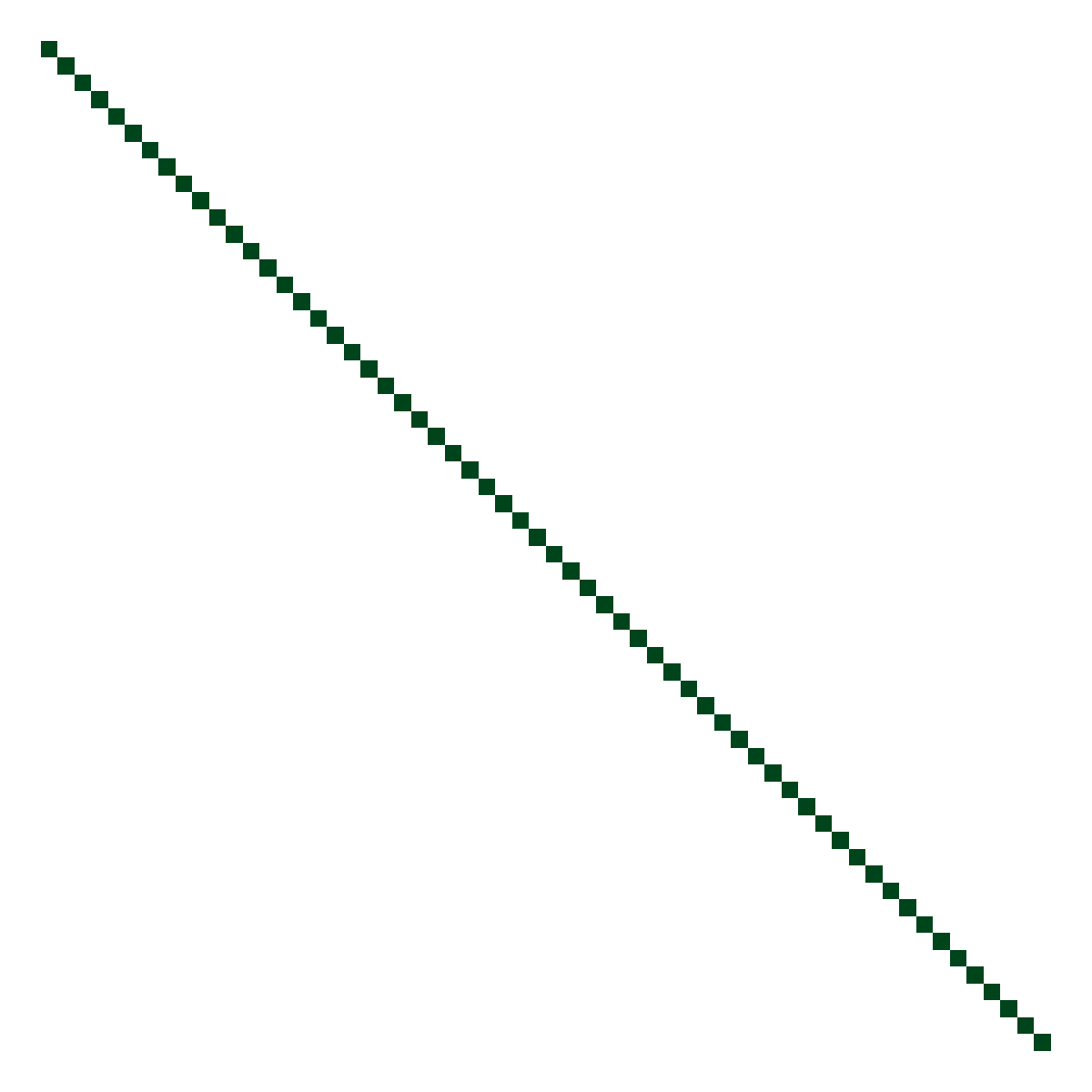} }}
  \caption{(a) A 2-branching tree graph of depth 5 (60 vertices), (b) its L-QSW steady state, (c) a cycle graph of 60 verticies, and (d) its L-QSW steady state. Darker colours in (b) and (d) correspond to a higher value of $|\rho_{ij}(t_\infty)|$ with white entries representing $0$.}
  \label{fig:2_example}%
\end{figure}

Starting in a maximally mixed state, $\rho(0)$, was evolved via the \texttt{step} method to the steady-state, $\rho(t_\infty)$, by choosing a sufficiently large time ($t = 100$). This is visualised in Figure \ref{fig:2_example}, where it is apparent that $\rho(t_\infty)$ for the balanced tree exhibits significant coherence, as opposed to the cycle graph which exhibits none. In fact, it has been established that, for regular graphs, $\rho(t_\infty)$ will always exhibit no coherence \cite{liu_steady_2017}. 

\subsubsection{Transport Through a Disordered Network}

\begin{figure}[t]
  \centering
 \includegraphics[width=6cm]{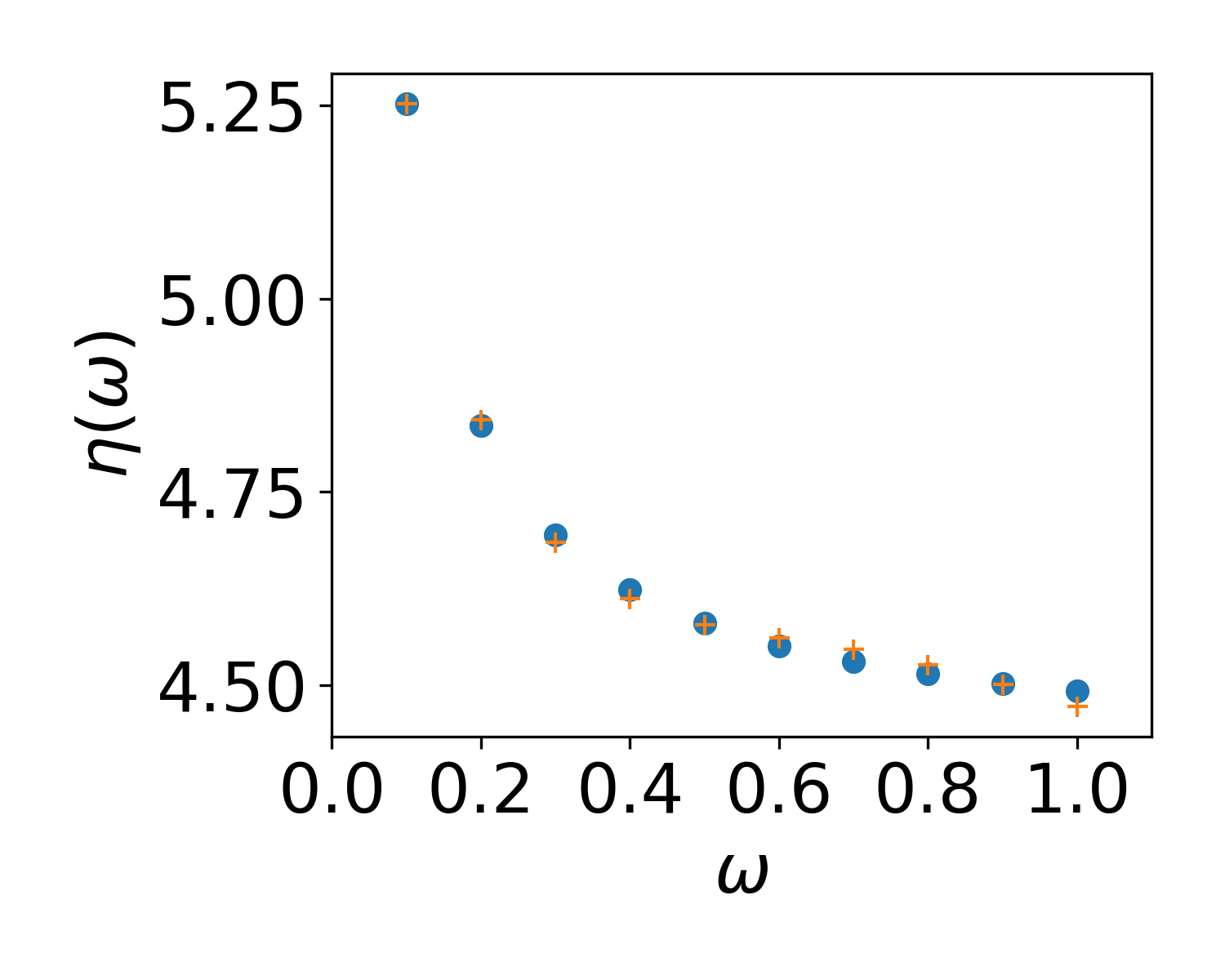}
  \caption{Expected survival time of the network and optimised dimer ($\delta = 1.5$) after 43 evaluations of the objective function. Starting parameters of the dimer were $V =\Gamma_D = \gamma_d = 0.5$.}
  \label{fig:3_dimer_fit}
\end{figure}

This example makes use of time series calculations to illustrate that the efficiency of transport through a disordered network, as modelled by an L-QSW, can be closely approximated as transport through an energetically disordered dimer\cite{schijven_modeling_2012}. A system of $N$ points randomly distributed in a unit sphere undergoing dipole-dipole is considered, leading to the potential,

\begin{equation}
  V_{ij} =
         \begin{cases}
                 -d^{-3}_{ij}, & i \neq j \\
                 0, & i = j \\
         \end{cases}
       \end{equation}

\noindent which is set equal to $G$. A source with $\Gamma = 0.5$ is attached to $v_1$ and a sink with $\gamma = 0.5$ to $v_N$.

The efficiency of transport is quantified through the Expected Survival Time (EST),

\begin{equation}
  \eta(\omega) = \int^\infty_0 \text{dt} (1-p_\gamma(t,\omega))
\end{equation}

\noindent where $p_\gamma$ is the accumulated probability at the sink vertex. Numerically this is approximated by making use of the \texttt{series} method to calculate $1 - p_\gamma(t)$ at $q$ evenly spaced intervals between $t_1 = 0$ and some time $t_q$ where $p_\gamma(t) \approx 1$. The resulting vector is then numerically integrated using the Simpson's Rule method provided by SciPy. By repeating this for a series of omega values where $0 < \omega \leq 1$, the response of $\eta(\omega)$ is specified for the network.

An energetically disordered dimer is described by the Hamiltonian,

\begin{equation}
  \label{eq:disorered_dimer}
  H =
  \begin{bmatrix}
    0 & -V \\
    -V & \delta \\
  \end{bmatrix}
\end{equation}

\noindent where $V$ represents the hopping rates between the vertices and $\delta$  is the energetic disorder. To this, a source of rate $\Gamma_D$ is attached to the first vertex and a sink of rate $\gamma_D$ to the second. The response of $\eta(\omega)$ between $0 < \omega \leq 1$ is then determined as previously described. 

To arrive at values of $V$, $\Gamma_D$ and $\gamma_D$ which produce a similar $\eta(\omega)$ response, the problem is formulated as an optimisation task with the objective function being minimisation of the vector $\Delta \vec{\eta}(\omega)$, the difference in EST between the disordered network and dimer at corresponding $\omega$ values. For this, the SciPy \texttt{least\_squares} optimisation algorithm was used. The result of the fitting process is shown in Figure \ref{fig:3_dimer_fit} for a network with N = 10. Despite being a much simpler system, the dimer closely approximates $\eta(\omega)$ of the disordered network.

\section{Validation and Performance} \label{sec:bench}

Analysis of the accuracy and scalability of QSW\_MPI was carried out on four digraph types generated via Networkx. This included line digraphs, square lattices, Erd\H{o}s-R\'{e}nyi digraphs with $\sim N \log(N)$ edges, and complete digraphs. These were assigned random arc weights between $(0,1)$ and stored in the matrix market sparse matrix format which is supported by the Python, Julia and Wolfram programming languages. As all tests study the performance of sparse matrix operations, matrix size is considered in terms of the number of $\tilde{\mathcal{L}}$ non-zeros. In each case, $\rho(0)$ was initialised as a maximally mixed state and propagated to $t = 100$. All depicted results are for a single run with unless otherwise stated, OpenMP threading set to one.

Sparse matrix exponentiation in QSWalk.jl is provided by \texttt{expmv} from the Julia implementation of Expokit \cite{glos_qswalk.jl:_2019,sidje_expokit_1998}. This has a user-definable error tolerance, which defaults to $10^{-8}$ (single precision). Here this was set to $2^{-53}$ in order to match the double precision target of QSW\_MPI \cite{al-mohy_computing_2011}. QSWalk.m utilises the Mathematica function \texttt{MatrixExp} \cite{noauthor_matrixexpwolfram_nodate}, which appears to also target double precision accuracy.

\begin{figure*}[hbt!]
  \centering
    \subfloat[]{{\includegraphics[width=3.9cm]{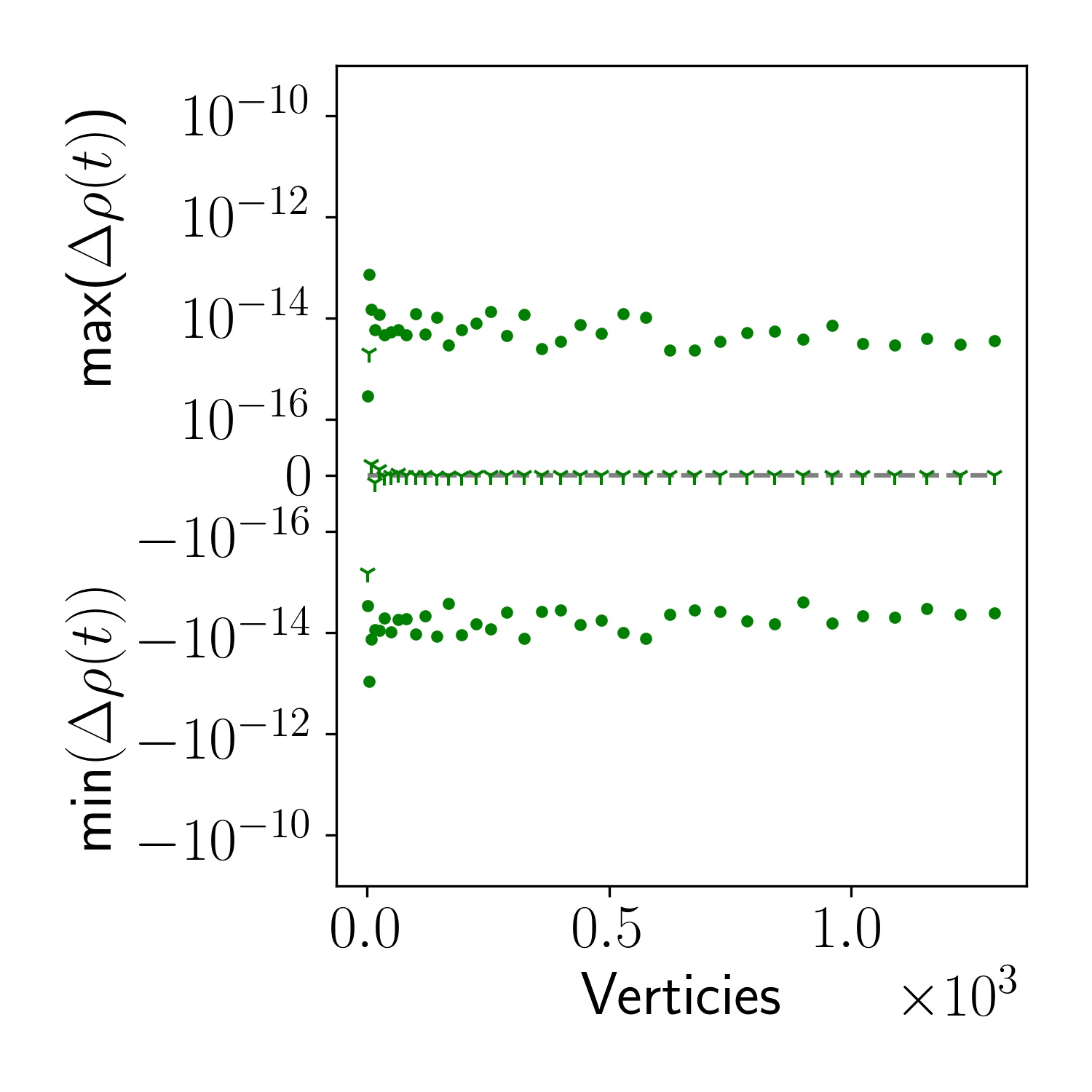} }}%
    \qquad
    \subfloat[]{{\includegraphics[width=3.9cm]{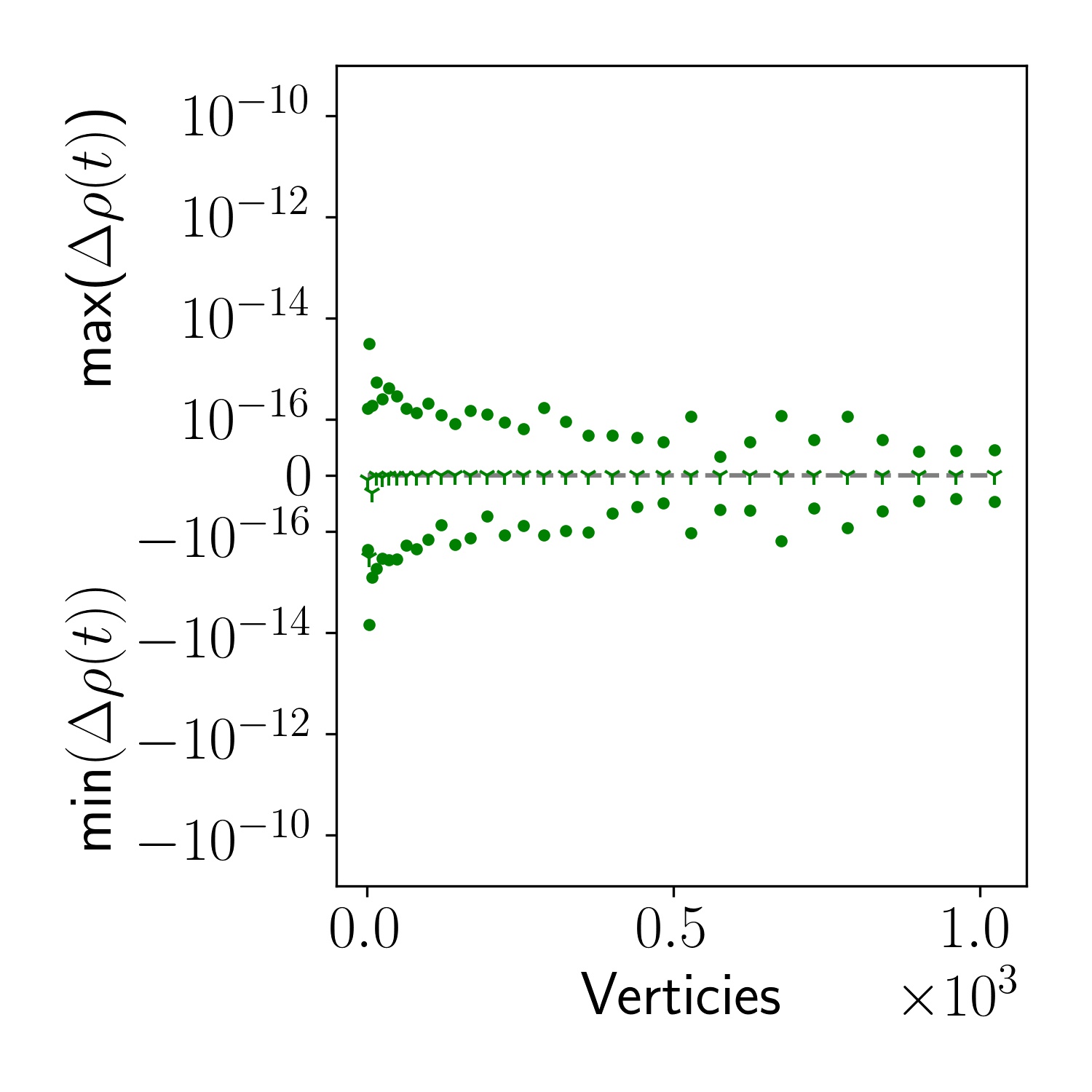} }}%
    \qquad
    \subfloat[]{{\includegraphics[width=3.9cm]{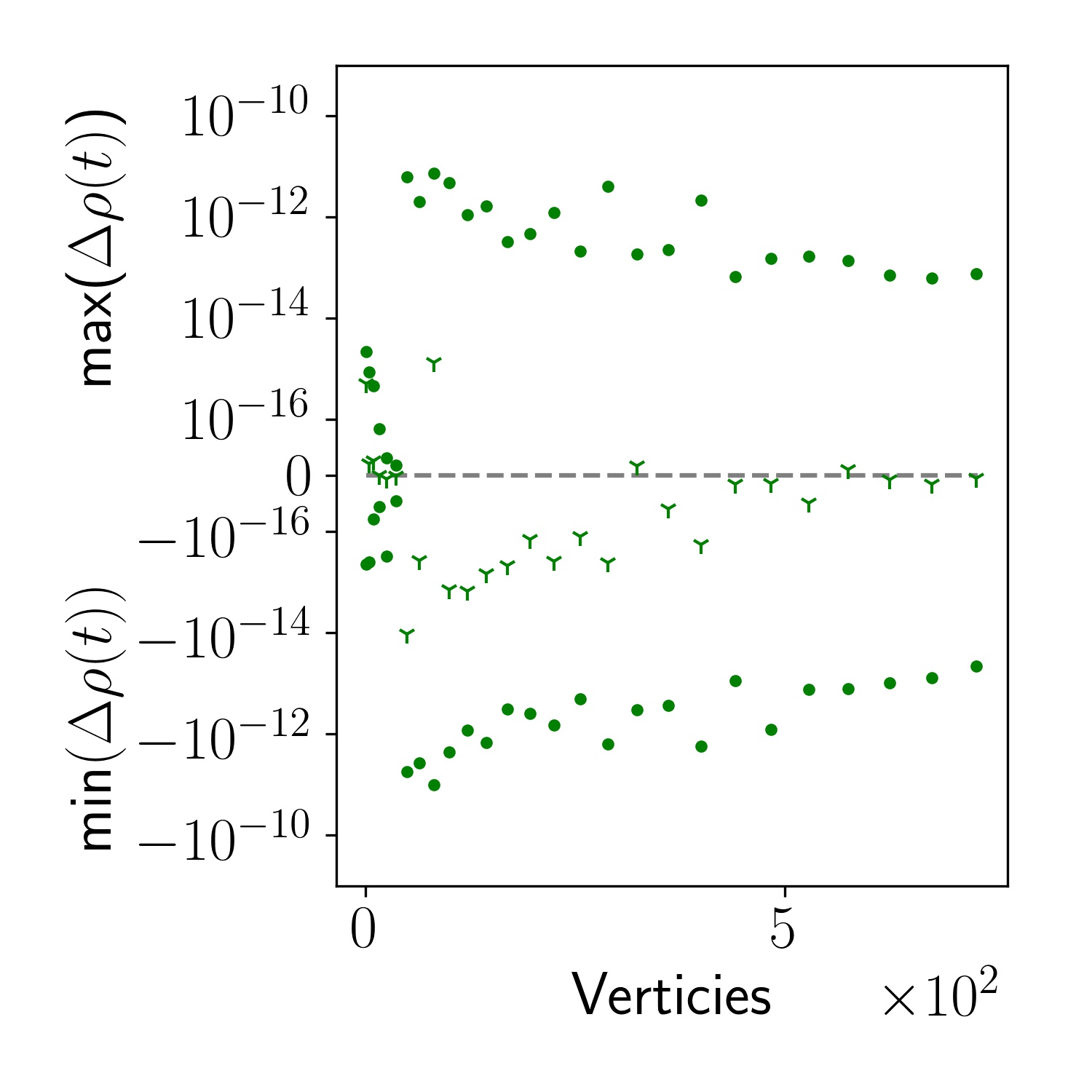} }}%
    \qquad
    \subfloat[]{{\includegraphics[width=3.9cm]{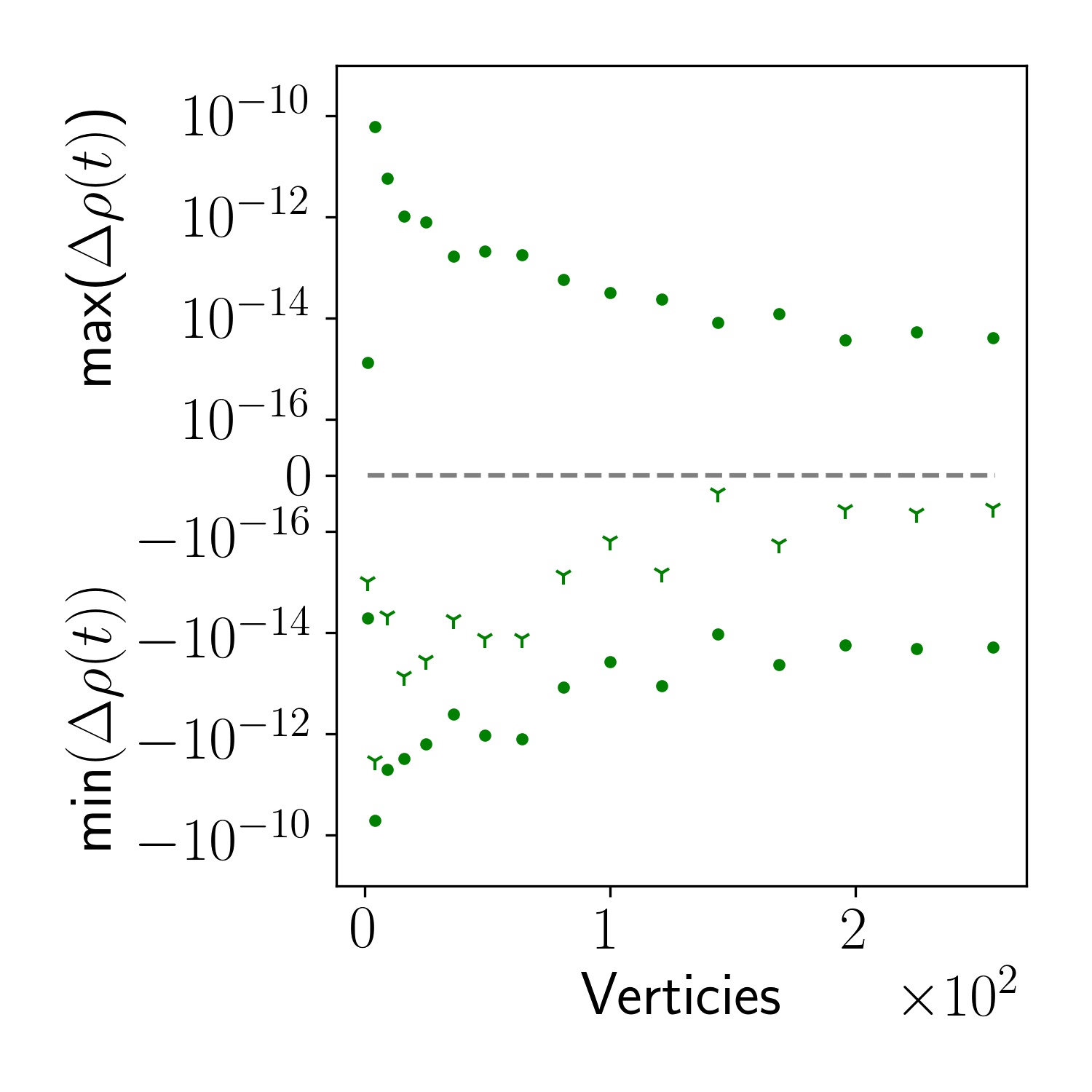} }}%
    \caption*{\footnotesize $\color{C2} \bullet$: $\min(\text{Re}(\Delta\rho(t)))$ or
      $\max(\text{Re}(\Delta\rho(t)))$, $\color{C2} \Ydown$: $\text{mean}(\text{Re}(\Delta\rho(t)))$}
    \subfloat[]{{\includegraphics[width=3.9cm]{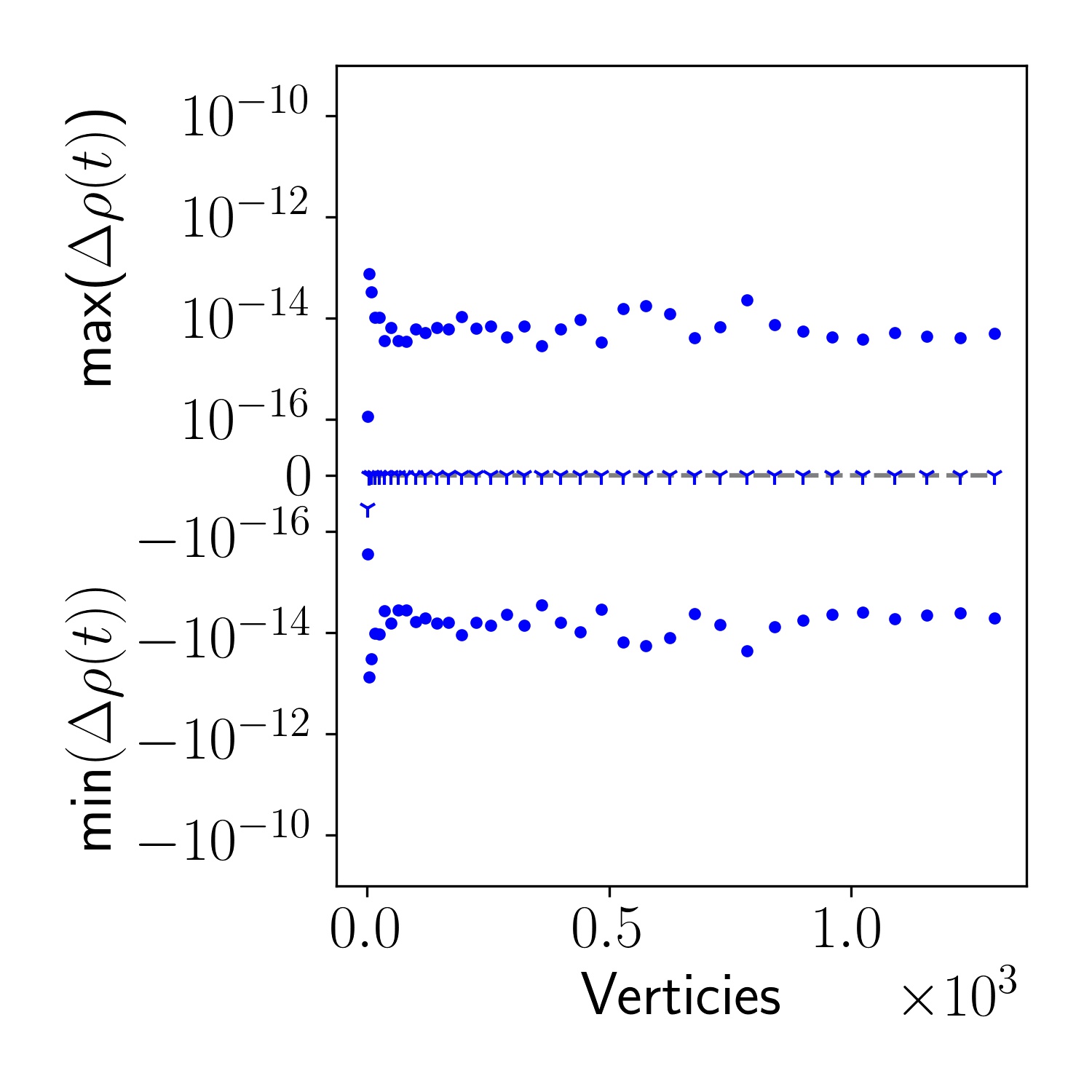} }}%
    \qquad
    \subfloat[]{{\includegraphics[width=3.9cm]{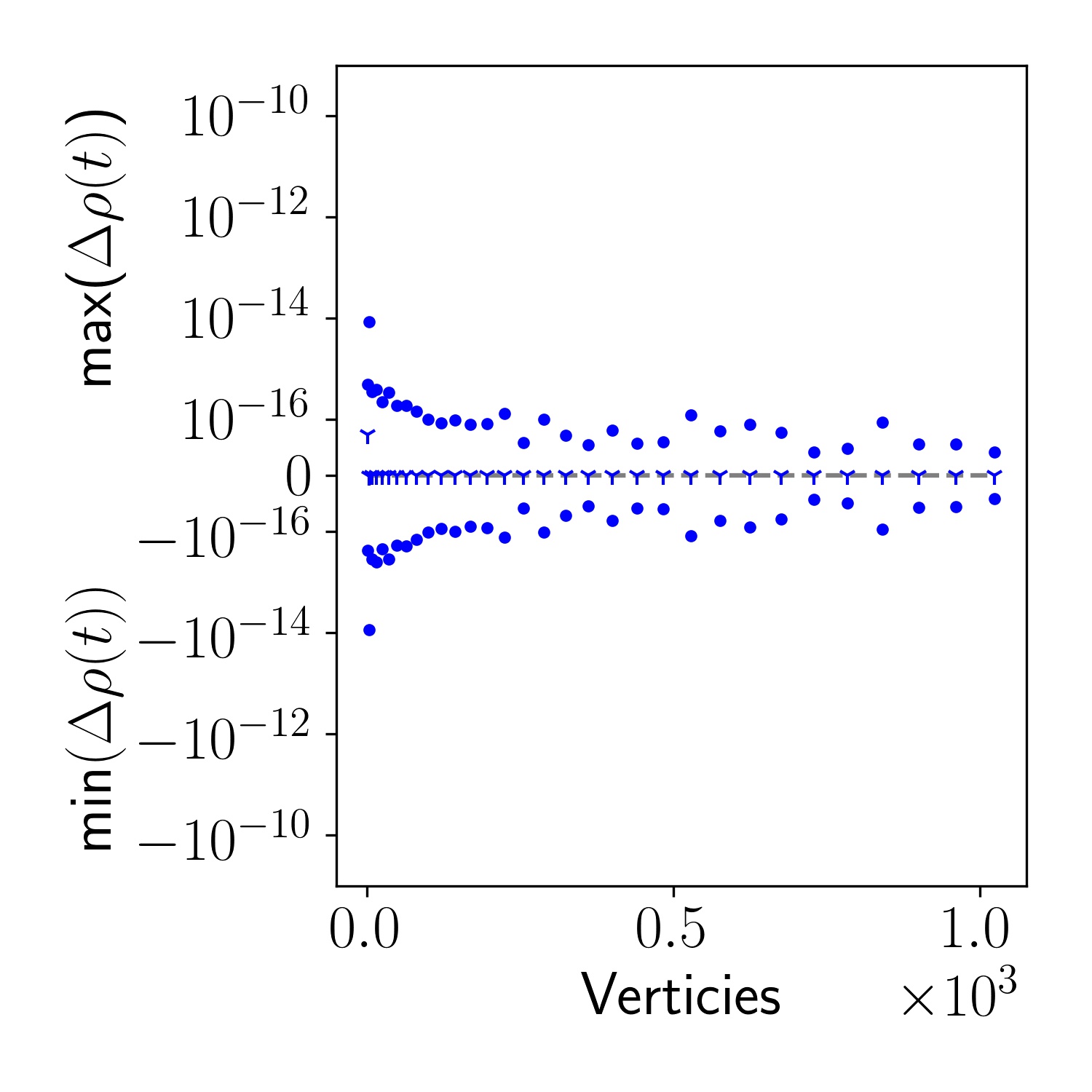} }}%
    \qquad
    \subfloat[]{{\includegraphics[width=3.9cm]{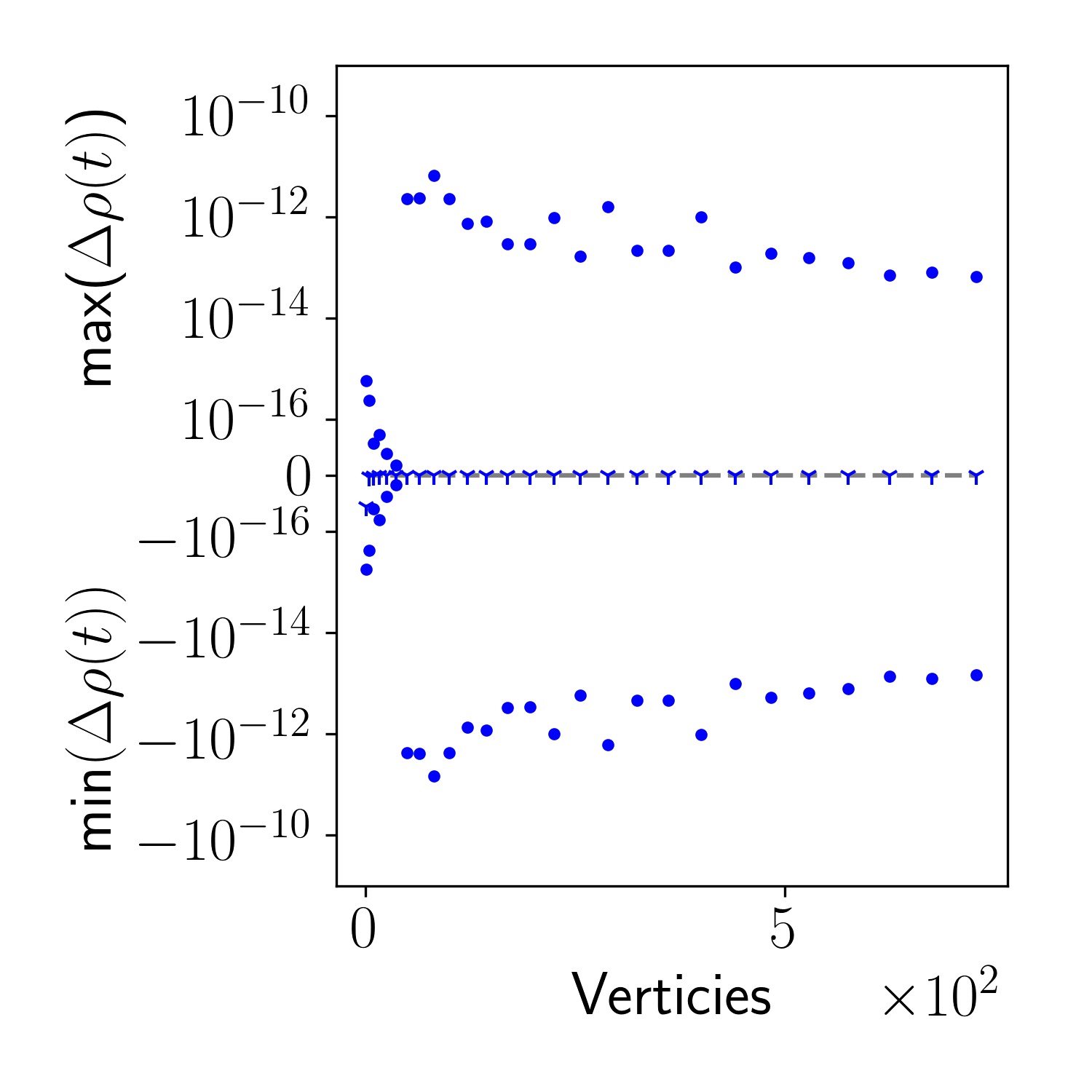} }}%
    \qquad
    \subfloat[]{{\includegraphics[width=3.9cm]{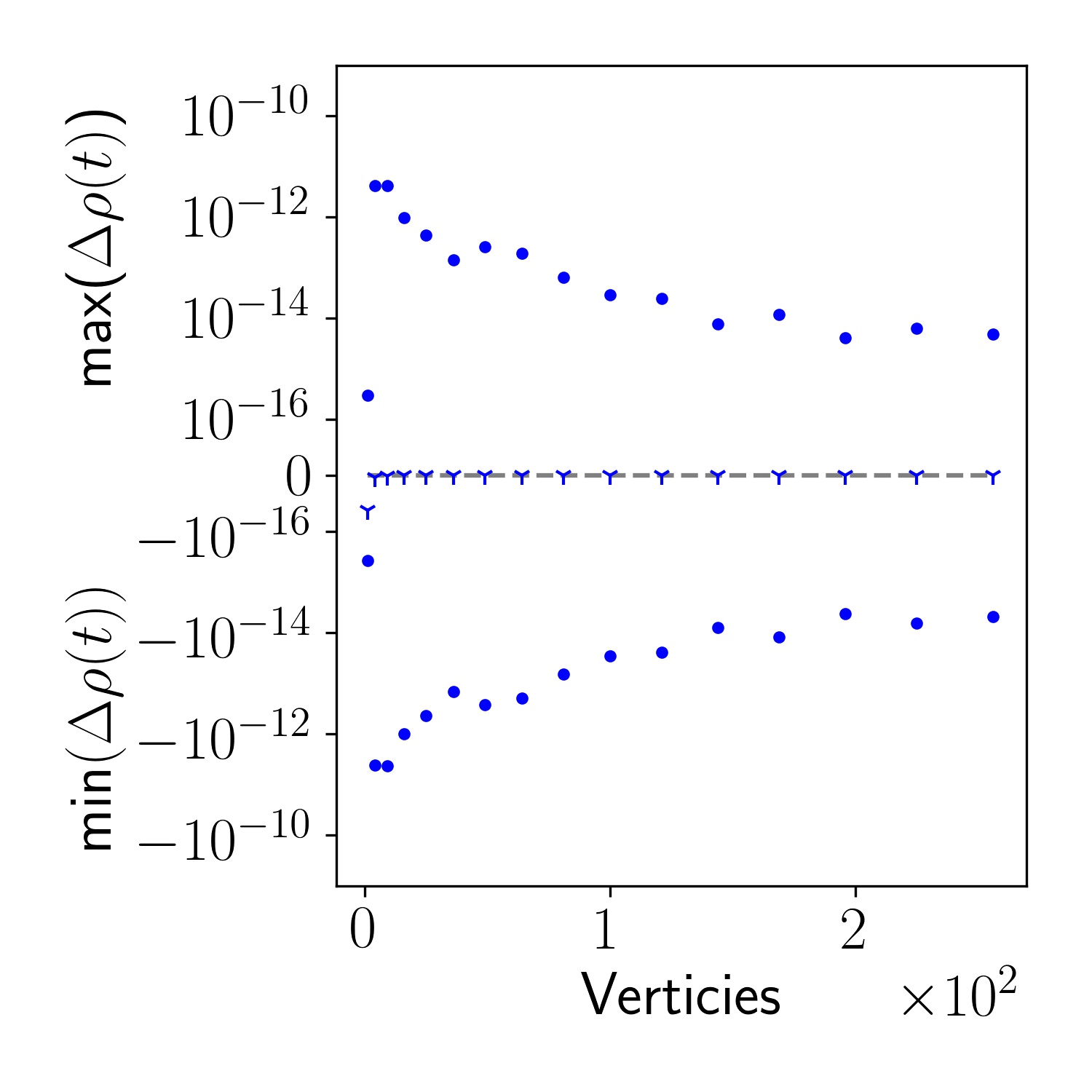} }}%
    \caption*{\footnotesize $\color{blue} \bullet$: $\min(\text{Im}(\Delta\rho(t)))$ or
      $\max(\text{Im}(\Delta\rho(t)))$, $\color{blue} \Ydown$: $\text{mean}(\text{Im}(\Delta\rho(t)))$}
    \caption{Difference between L-QSW  results calculated to double precision using the QSW\_MPI \texttt{step} method and QSWalk.m for (a) line digraphs, (b) square lattices and (c) Erd\H{o}s-R\'{e}nyi digraphs, and (d) complete digraphs. $\Delta \rho(t) = \rho_L(t) - \rho_M(t)$ where $\rho_L(t)$ and $\rho_M(t)$ are the results obtained with QSW\_MPI and QSWalk.m.}
    \label{fig:QSW_MPI_qswalk_delta}%
\end{figure*}

\begin{figure}[p]
  \centering
    \subfloat[]{{\includegraphics[width=3.9cm]{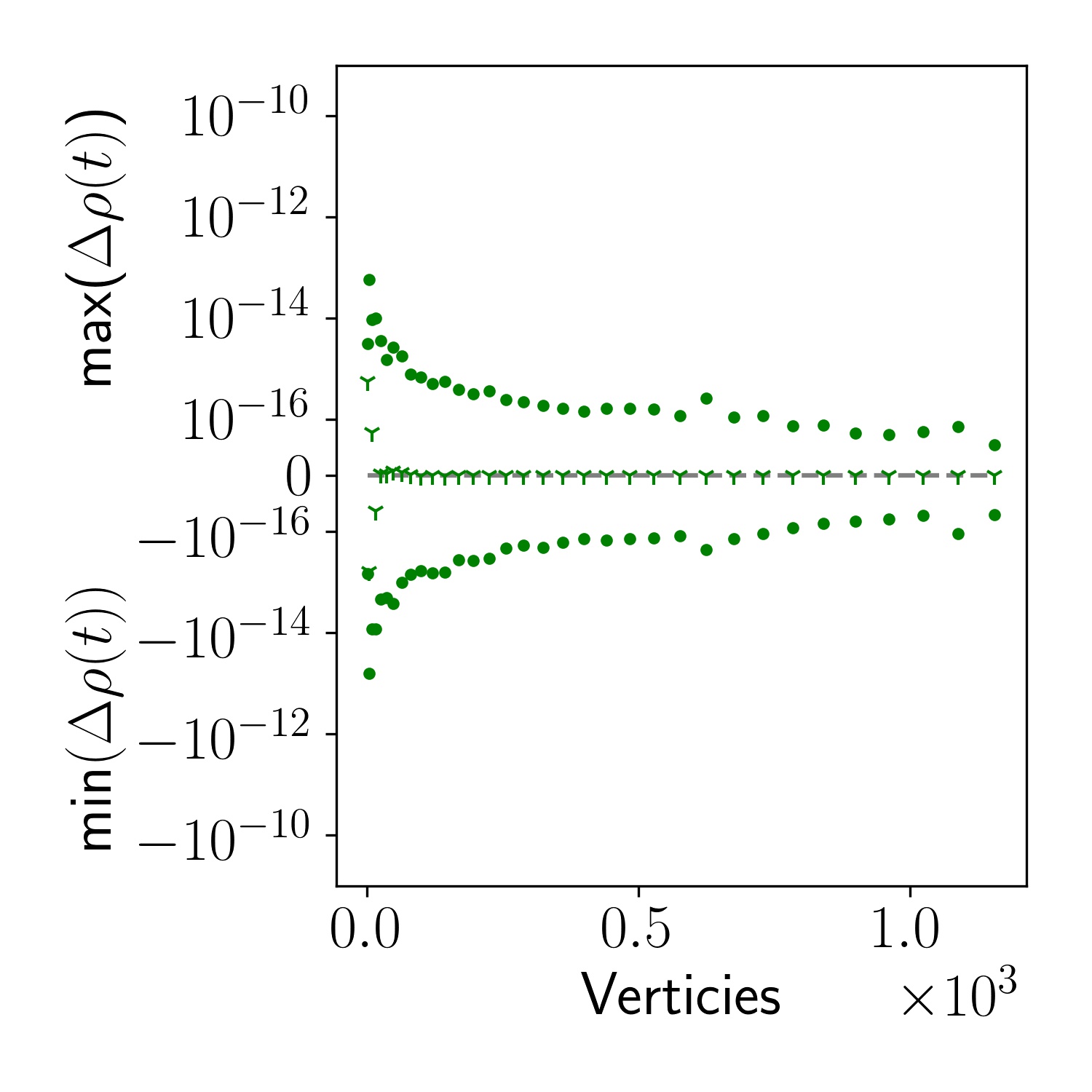} }}%
    \qquad
    \subfloat[]{{\includegraphics[width=3.9cm]{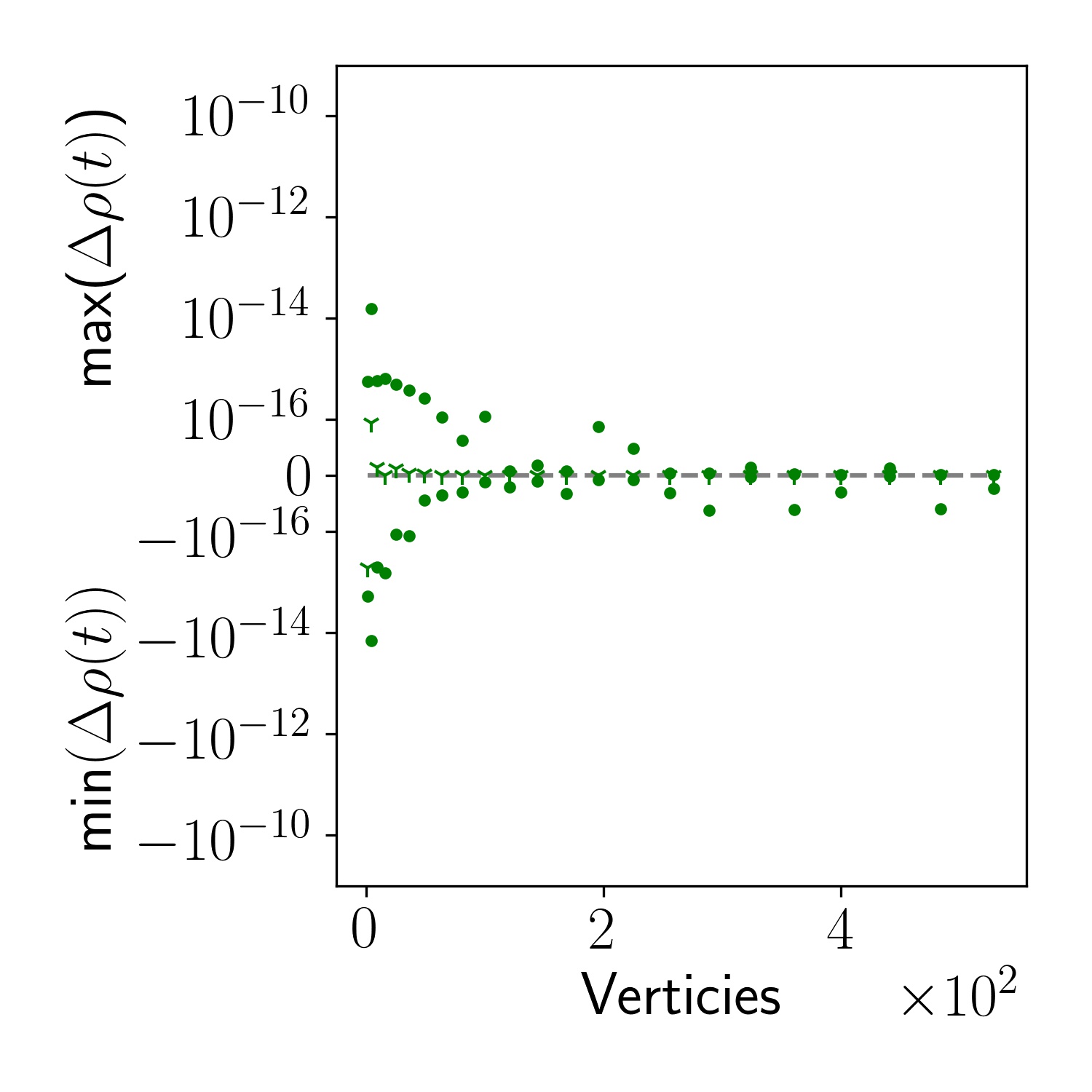} }}%
    \caption*{\footnotesize $\color{C2} \bullet$: $\min(\text{Re}(\Delta\rho(t)))$ or
      $\max(\text{Re}(\Delta\rho(t)))$, $\color{C2} \Ydown$: $\text{mean}(\text{Re}(\Delta\rho(t)))$}
    \subfloat[]{{\includegraphics[width=3.9cm]{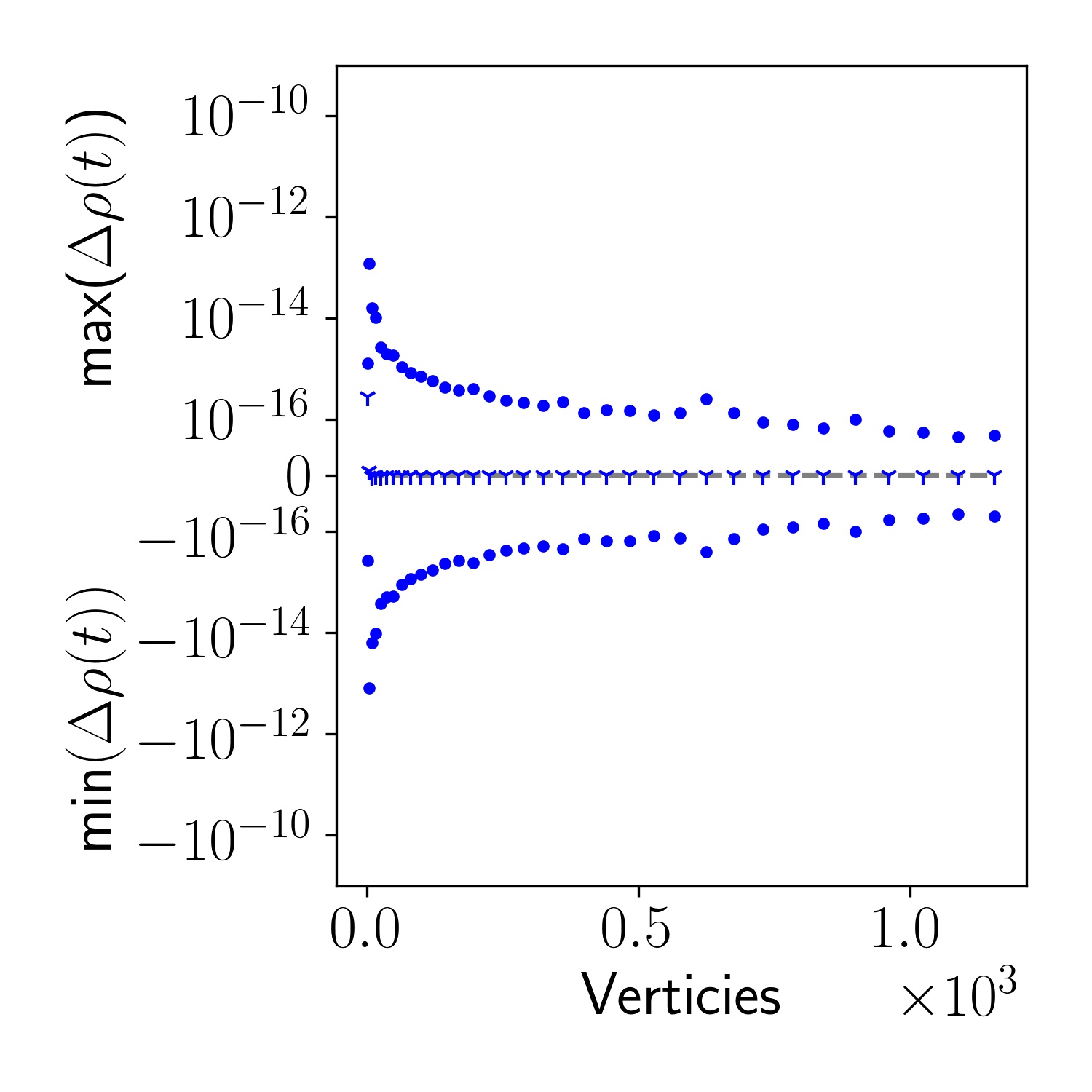} }}%
    \qquad
    \subfloat[]{{\includegraphics[width=3.9cm]{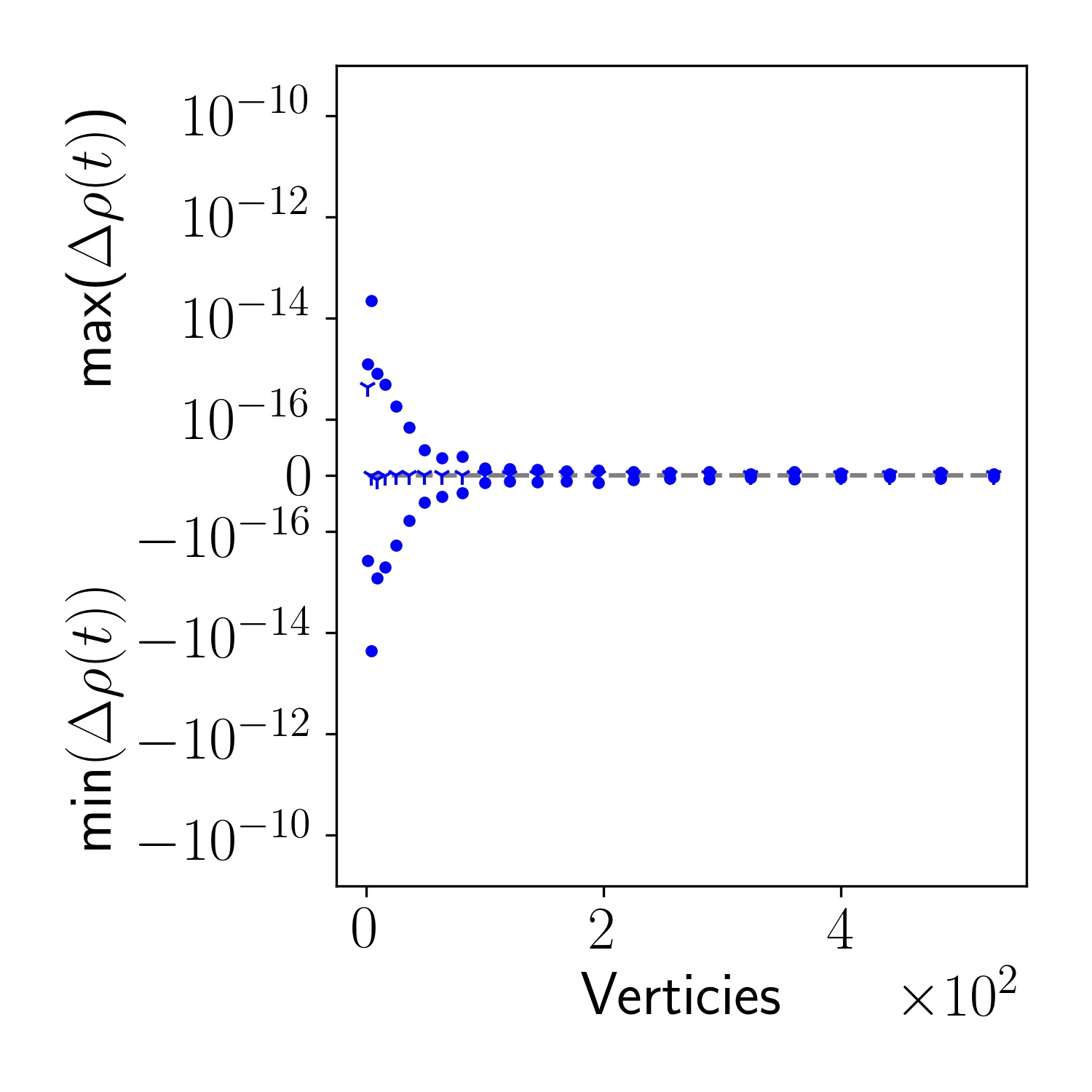} }}%
    \caption*{\footnotesize $\color{blue} \bullet$: $\min(\text{Im}(\Delta\rho(t)))$ or $\max(\text{Im}(\Delta\rho(t)))$, $\color{blue} \Ydown$: $\text{mean}(\text{Im}(\Delta\rho(t)))$}
    \caption{Difference between G-QSW  results calculated to double precision using the QSW\_MPI \texttt{step} method and QSWalk.m for (a) square lattices and (c) Erd\H{o}s-R\'{e}nyi digraphs. $\Delta \rho(t) = \rho_L(t) - \rho_M(t)$ where $\rho_L(t)$ and $\rho_M(t)$ are the results obtained with QSW\_MPI and QSWalk.m.}
    \label{fig:global_QSW_MPI_qswalk_delta}%
\end{figure}

The performance and accuracy of QSW\_MPI was first studied in a workstation-like environment, which allowed for its comparison with preexisting packages. Figure \ref{fig:QSW_MPI_qswalk_delta} examines the difference between L-QSW simulation results obtained using the QSW\_MPI \texttt{step} method as compared to those obtained with the QSWalk.m package. For each of the graph types considered the results generally agree to within an order of $10^{-12}$, with no discrepancies exceeding $10^{-10}$. An equivalent comparison for G-QSWs is shown in Figure \ref{fig:global_QSW_MPI_qswalk_delta} for the square lattice and Erd\H{o}s-R\'{e}nyi digraph sample sets, where the difference does not exceed the order of $10^{-12}$ and is generally below $10^{-14}$. As the mean difference in Figures \ref{fig:QSW_MPI_qswalk_delta} and \ref{fig:global_QSW_MPI_qswalk_delta} is generally centred around $0$, it does to appear that the maximum and minimum differences are indicative of significant systematic error. A minor exception to this does occur in the real components of the Erd\H{o}s-R\'{e}nyi and complete digraphs (see Figure \ref{fig:QSW_MPI_qswalk_delta} (c) and (d)), however, the magnitude of the mean error decreases with digraph size and does not exceed the order of $10^{-12}$. For the real and complex components of all graph types, the maximum and minimum difference decreases with digraph size. This is to be expected with increased dispersion of the density operator. 

\begin{figure}[p]
  \centering
  \subfloat[]{{\includegraphics[width=4.3cm]{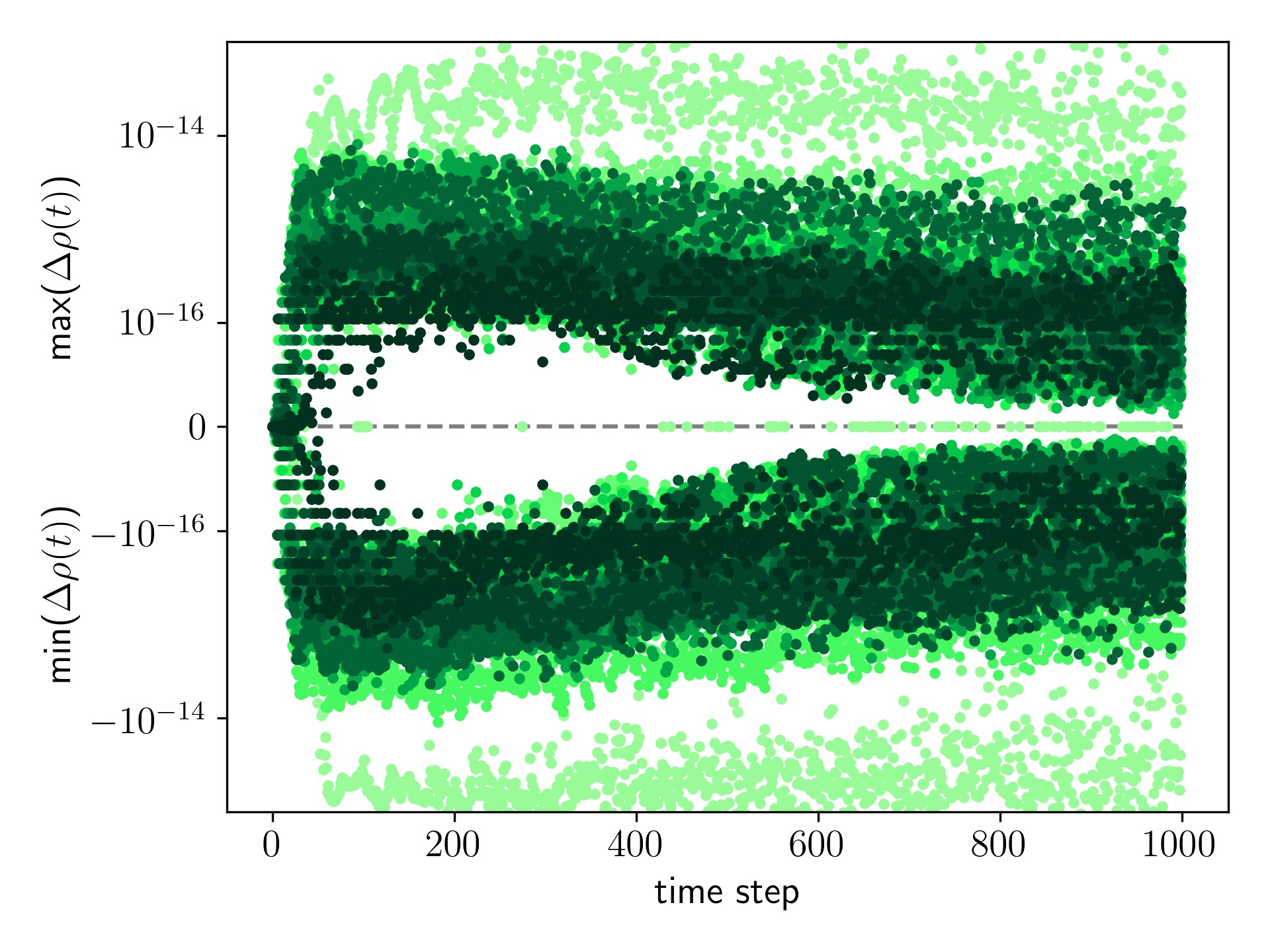}\includegraphics[width=4.3cm]{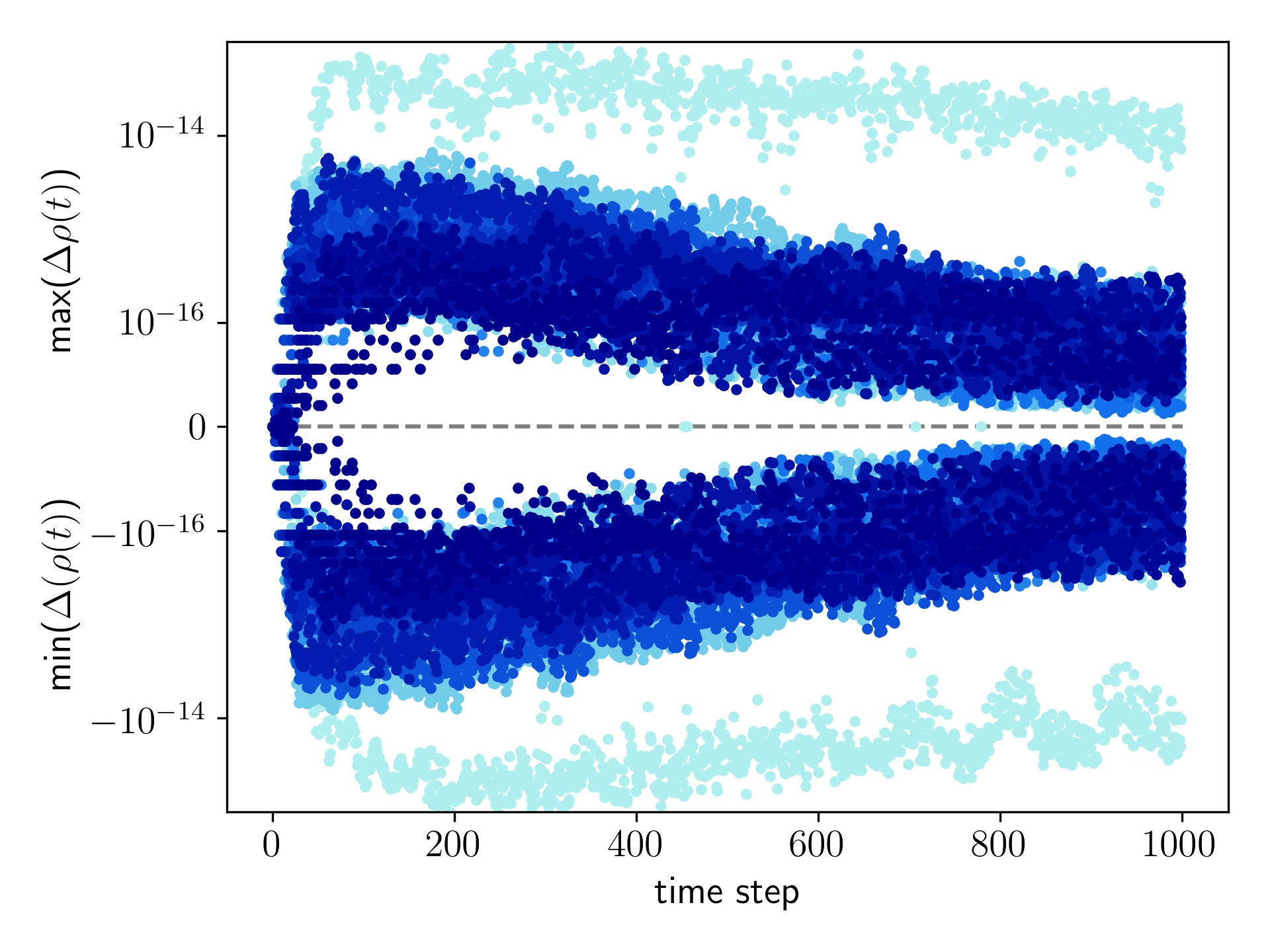}}}
  \newline
  \subfloat[]{{\includegraphics[width=4.3cm]{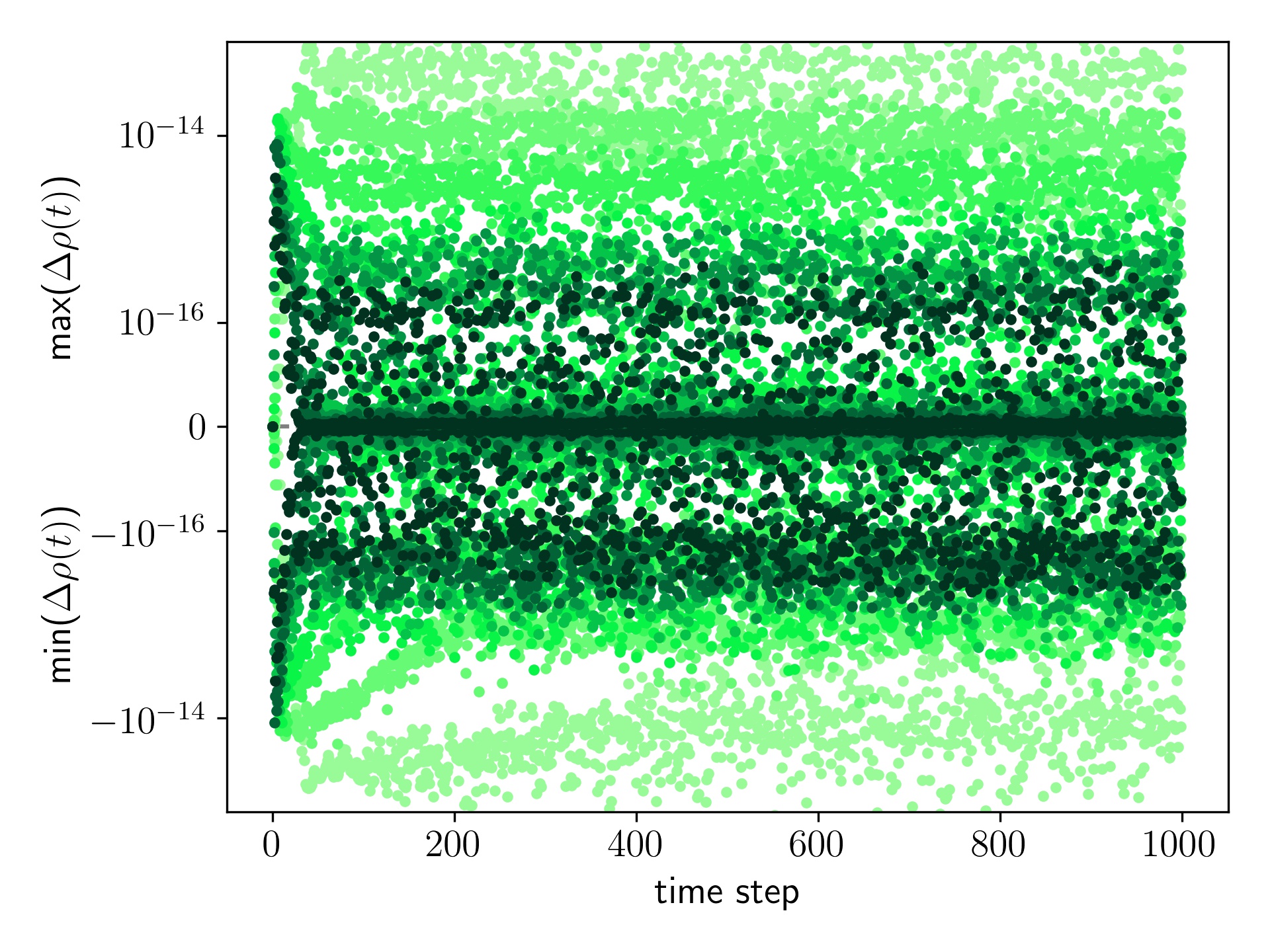}\includegraphics[width=4.3cm]{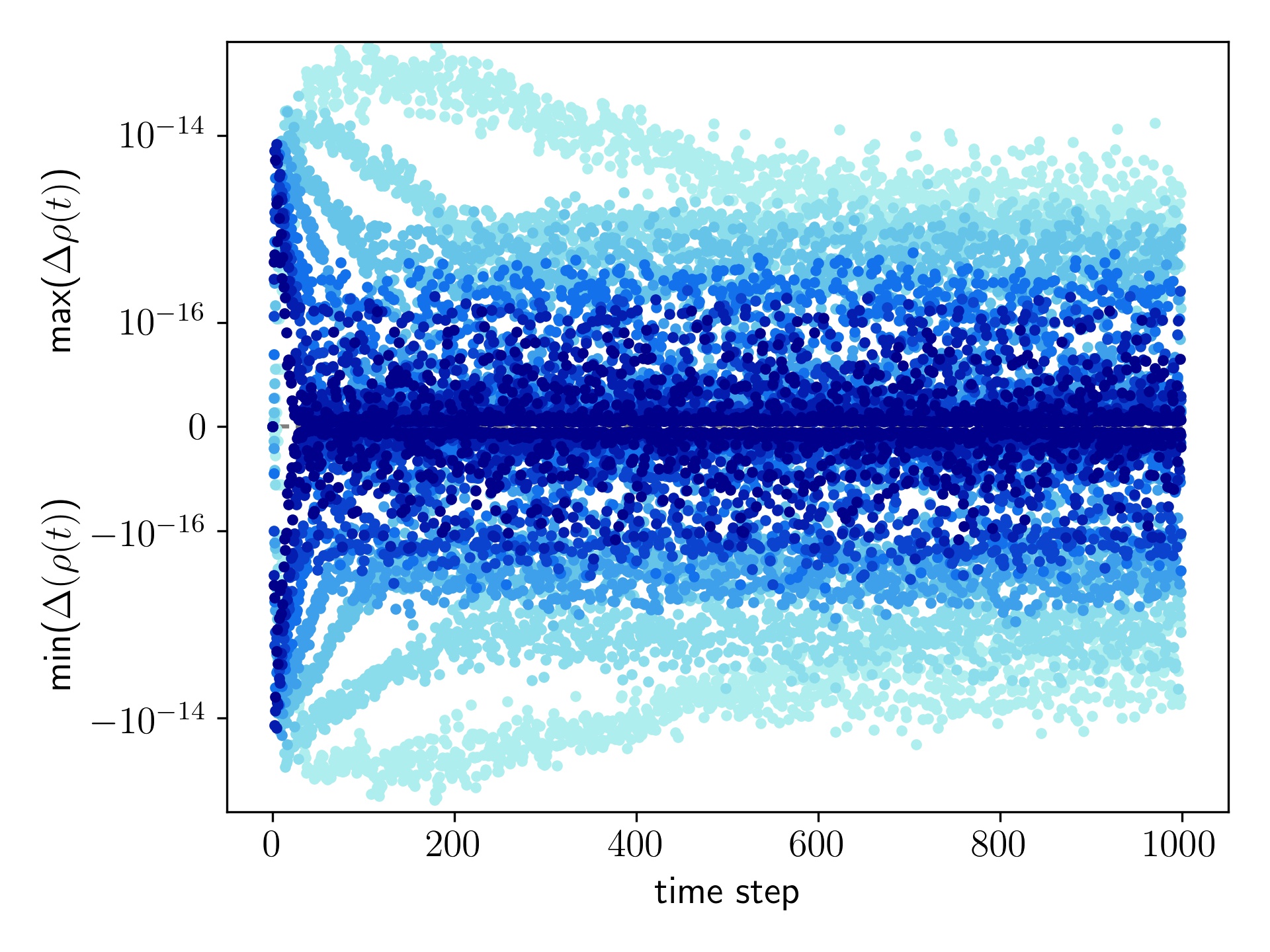}}}
  \newline
  \caption*{\footnotesize $\color{green} \bullet$: real component (left), $\color{blue} \bullet$: imaginary component (right)}
  \caption{Maximum and minimum difference in $\rho(t)$ as calculated for the time-series of an L-QSW by the \texttt{step} and \texttt{series} methods for (a) line digraphs and (b) complete digraphs. For all walks, $t_1 = 0$, $t_2 = 100$ and $\Delta t = 0.5$. The graph sample sets correspond to those shown in Figure \ref{fig:step_vs_series_time}, with the darker data-points corresponding to an increasing number of non-zeros. $\Delta \rho(t) = \rho_\text{s}(t) - \rho_\text{q}(t)$ where $\rho_\text{s}(t)$ and $\rho_\text{q}(t)$ are the results obtained with \texttt{step} and \texttt{series}.}
  \label{fig:step_vs_series_delta}
\end{figure}

Figure \ref{fig:step_vs_series_delta} compares L-QSW time-series calculations obtained via the QSW\_MPI \texttt{step} and \texttt{series} methods on the line digraph and complete digraph sample sets. A consistent concurrence to the order of $10^{-14}$ or below is observed which, in combination with the QSW\_MPI/QSWalk.m comparison results shown in Figure \ref{fig:QSW_MPI_qswalk_delta}, supports the accuracy of the \texttt{series} method.

\begin{figure}[p]
    \centering
    \subfloat[]{{\includegraphics[width=3.9cm]{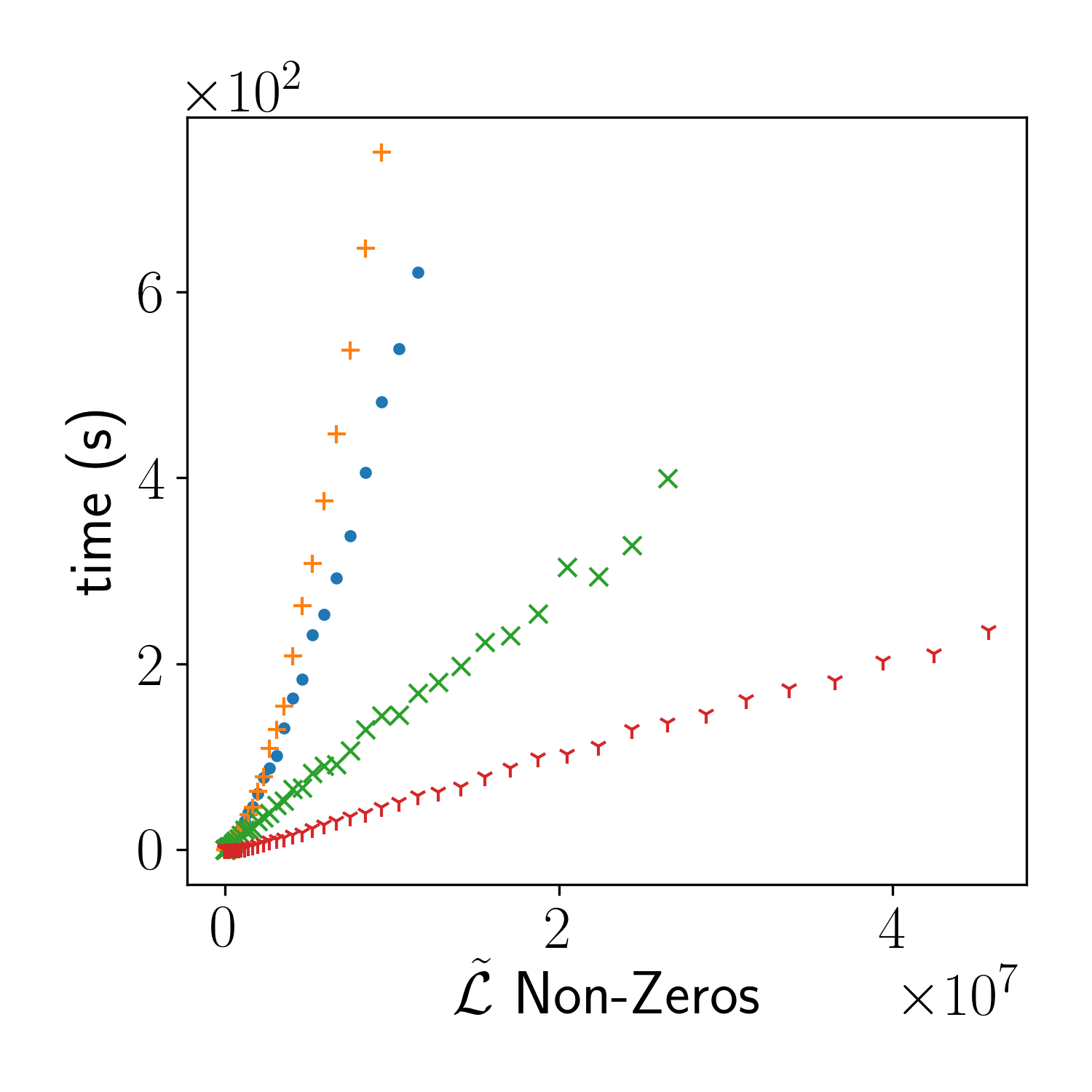} }}%
    \qquad
    \subfloat[]{{\includegraphics[width=3.9cm]{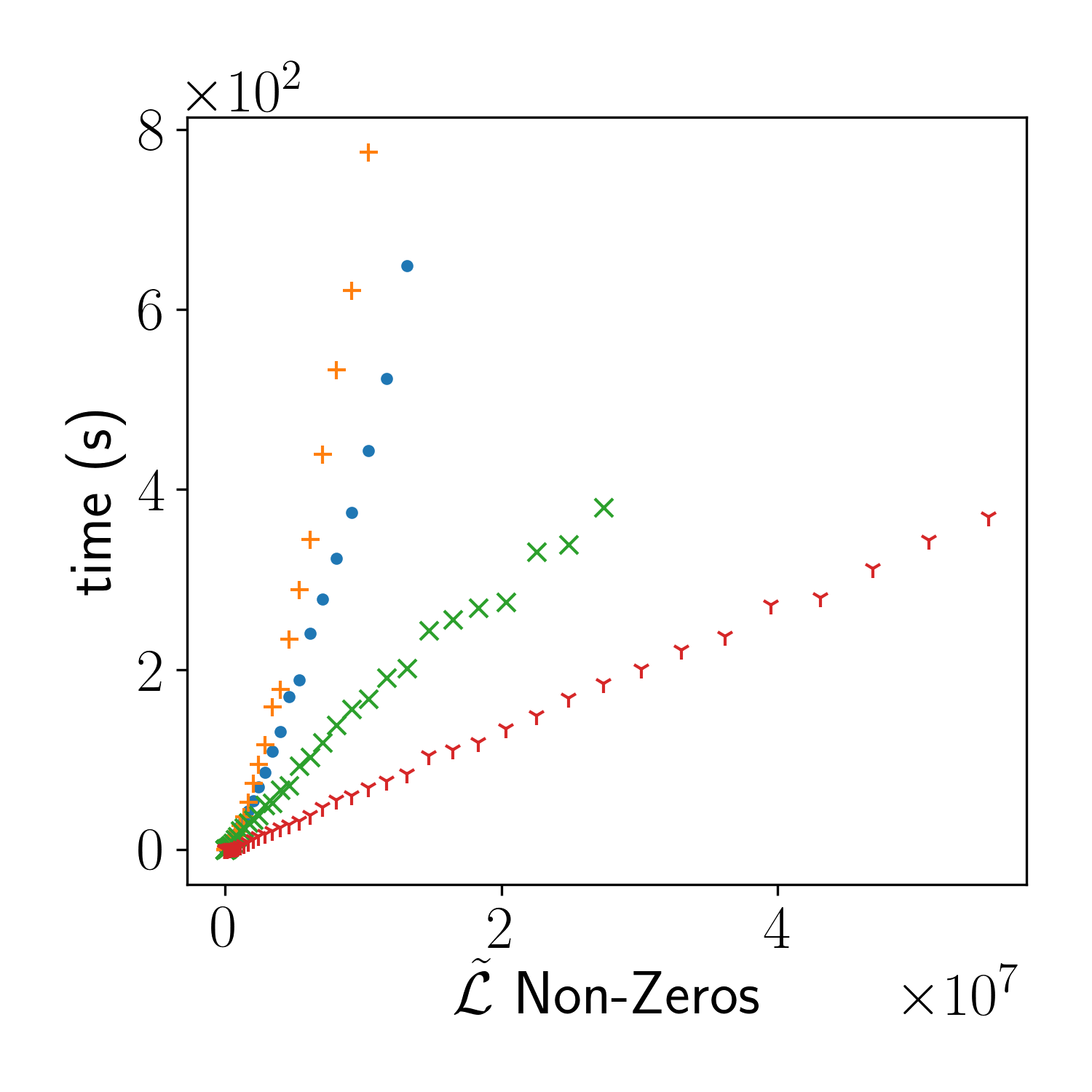} }}%
    \newline
    \subfloat[]{{\includegraphics[width=3.9cm]{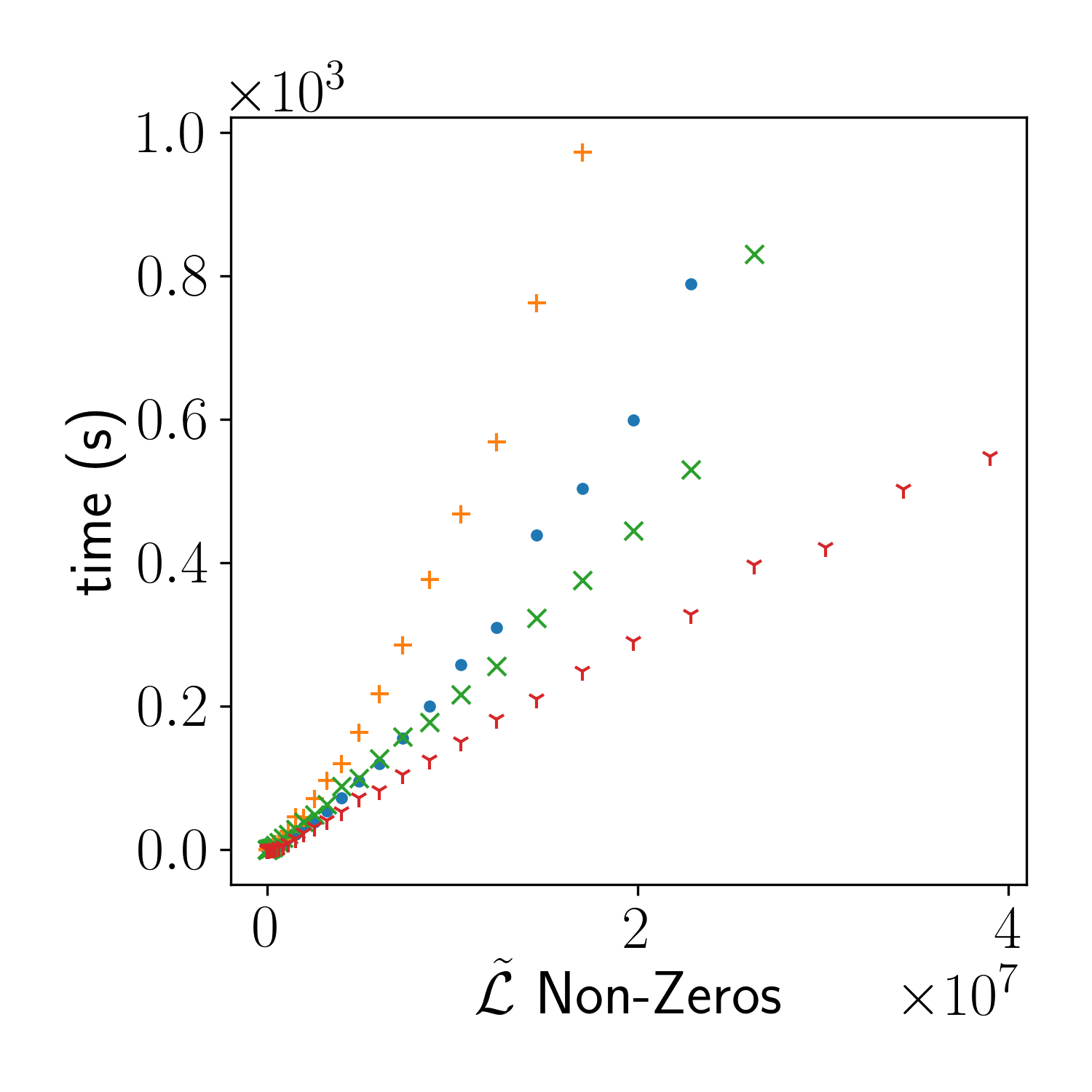} }}%
    \qquad
    \subfloat[]{{\includegraphics[width=3.9cm]{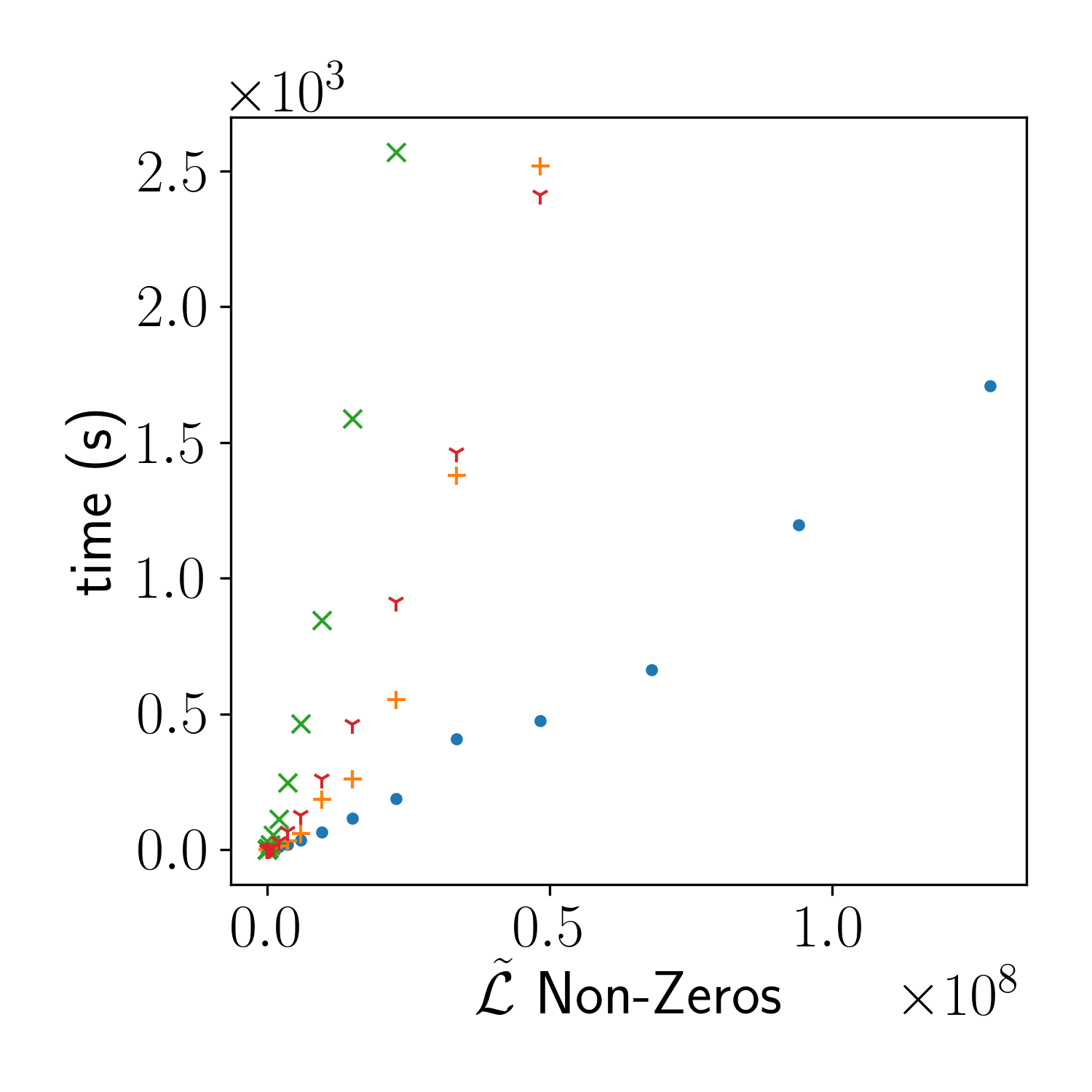}}}%
    \newline
    \caption*{\footnotesize $\color{C0} \bullet$: QSWalk.jl, $\color{C1} \times$: QSWalk.m,}
    \caption*{\footnotesize $\color{C2} \times$: QSW\_MPI (1 MPI process), $\color{C3} \Ydown$: QSW\_MPI (16 MPI processes).}
    \caption{In-program time of QSW packages simulating an L-QSW simulation on a workstation. The QSWalk.jl and QSWalk.m packages consist of single-threaded programs, while is QSW\_MPI shown running with one and sixteen MPI processes.  (a) Shows performance on the set of line digraphs, (b) the square lattices, (c) the Erd\H{o}s-R\'{e}nyi digraphs, and (d) the complete digraphs. System specifications: AMD $2^\text{nd}$ generation Epyc Processor 16 cores at 2.4 GHz and 64 GB RAM.}
    \label{fig:desk}%
\end{figure}

\begin{figure}[p]
  \centering
  \subfloat[]{{\includegraphics[width=4.0cm]{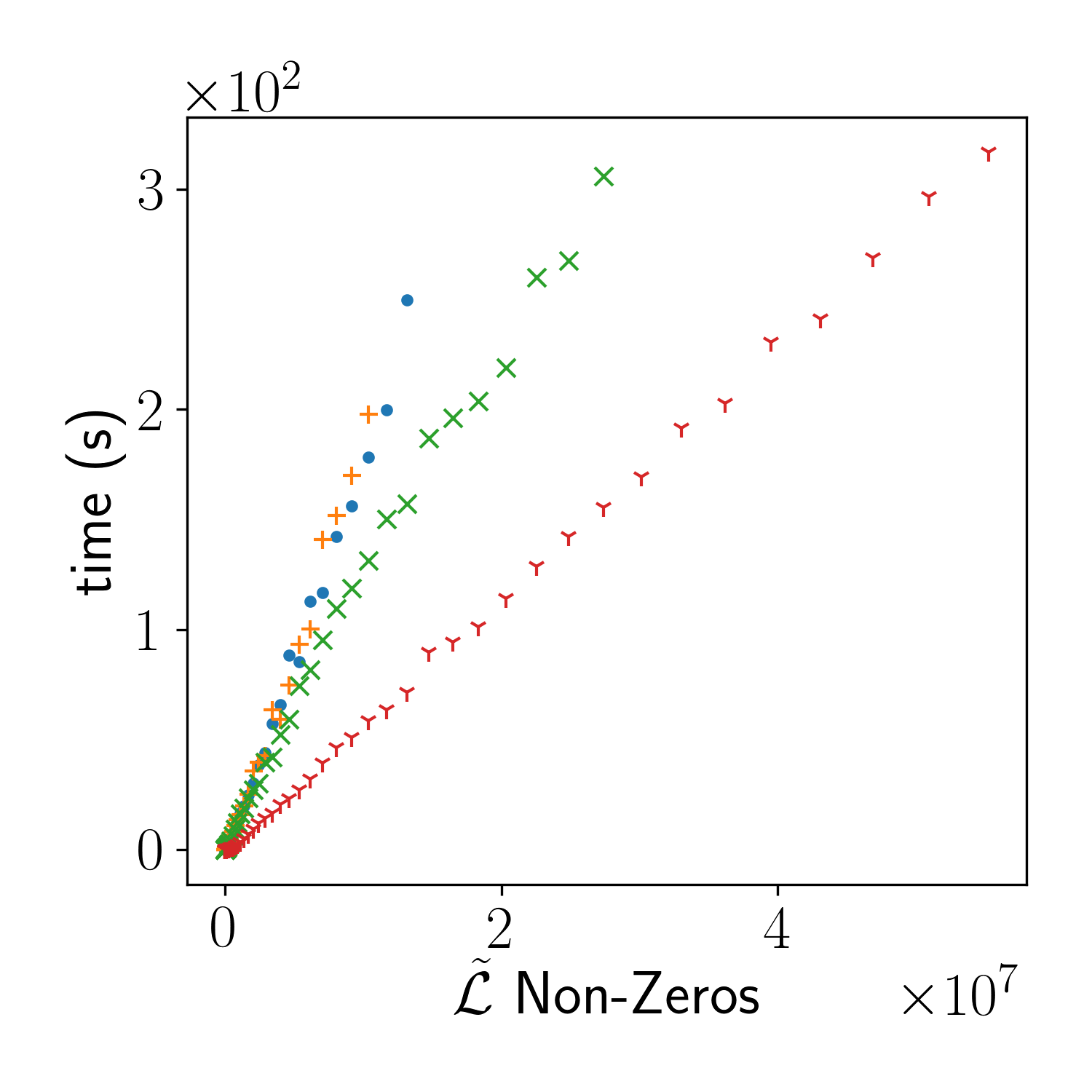}}}
  \qquad
  \subfloat[]{{\includegraphics[width=4.0cm]{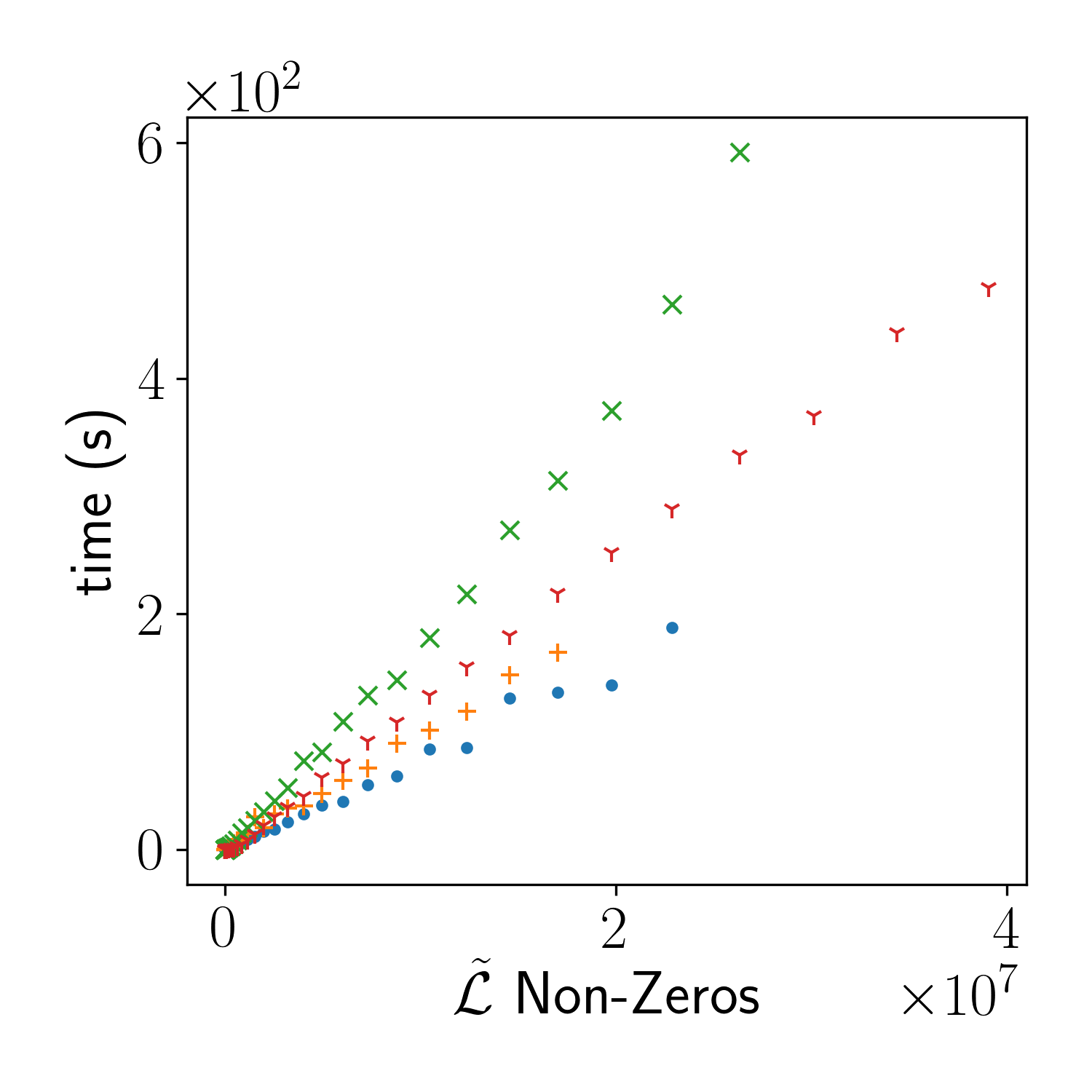}}}
  \caption*{\footnotesize $\color{C0} \bullet$: QSWalk.jl, $\color{C1} \times$: QSWalk.m,}
  \caption*{$\color{C2} \times$: QSW\_MPI (1 MPI process), $\color{C3} \Ydown$: QSW\_MPI (16 MPI processes).}
  \caption{In-program L-QSW exponentiation time of QSWalk.m, QSWalk.jl and the QSW\_MPI \texttt{step} method for the (a) Erd\H{o}s-R\'{e}nyi digraph and (b) complete digraph sample sets shown in Figure \ref{fig:desk}. System specifications match those of Figure \ref{fig:desk}}
  \label{fig:exp_time}
\end{figure}

\begin{figure}[t]
  \centering
  \subfloat[]{{\includegraphics[width=3.9cm]{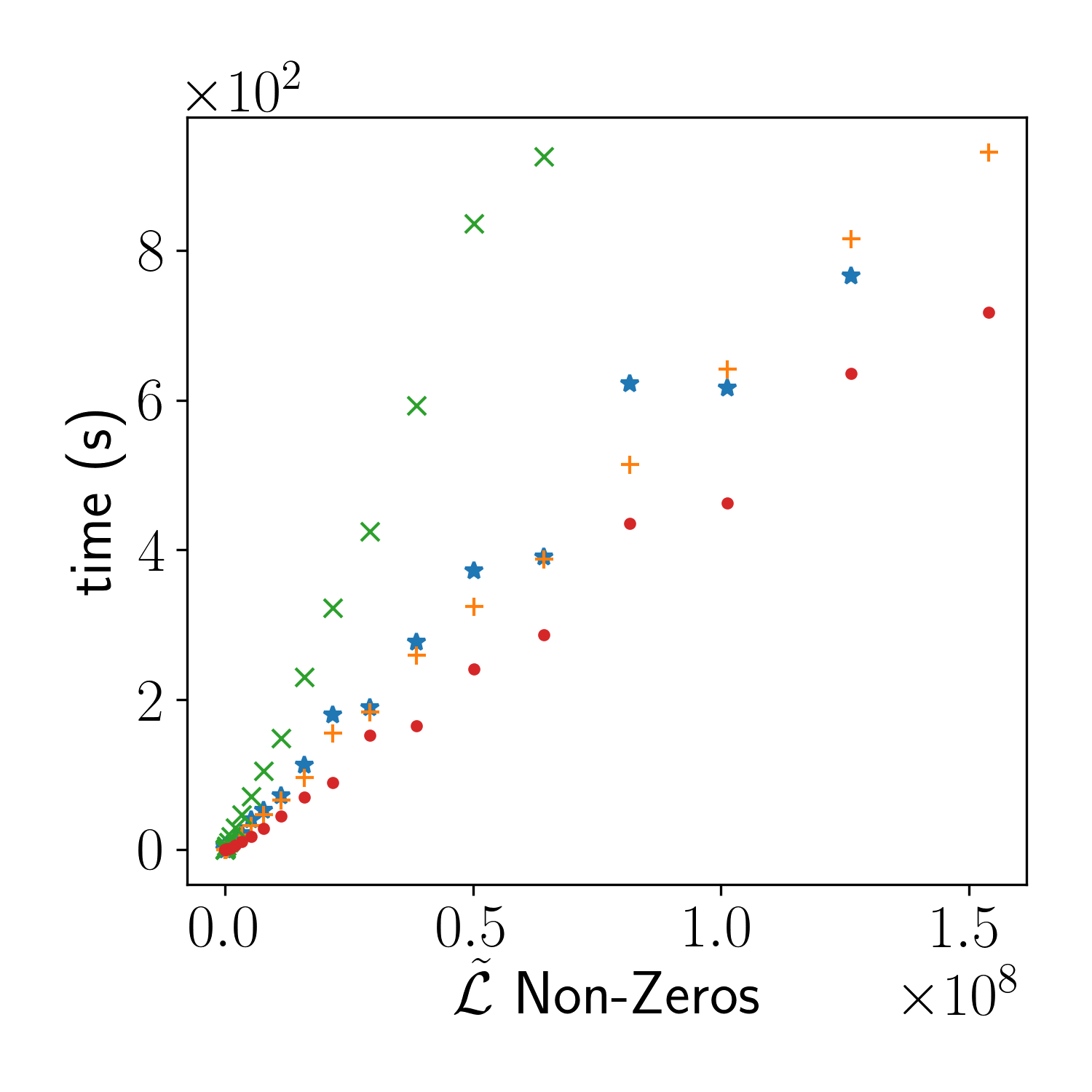} }}%
  \qquad
  \subfloat[]{{\includegraphics[width=3.9cm]{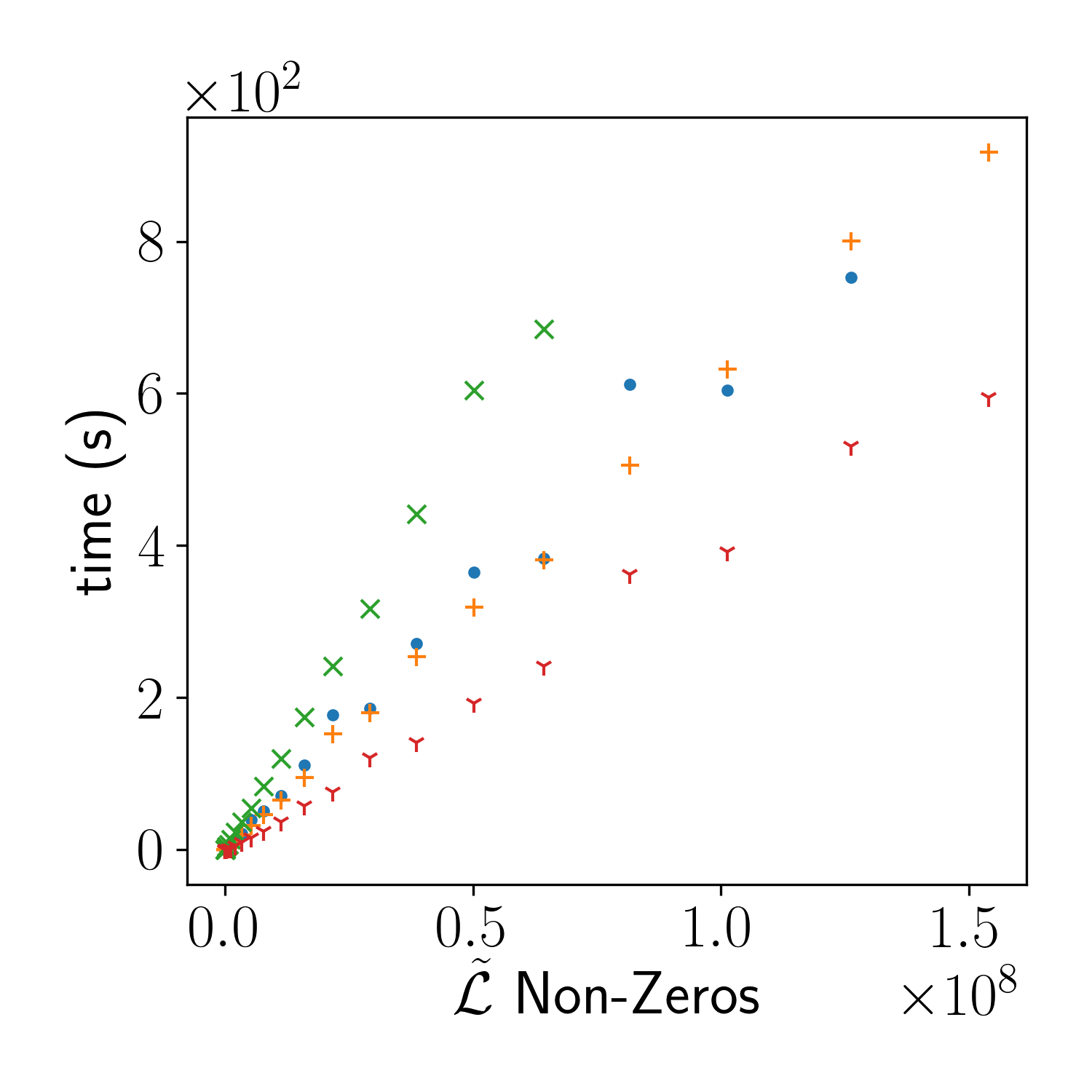} }}%
  \caption*{\footnotesize $\color{C0} \bullet$: QSWalk.jl, $\color{C1}
    \times$: QSWalk.m,}
  \caption*{$\color{C2} \times$: QSW\_MPI (1 MPI process), $\color{C3}
    \Ydown$: QSW\_MPI (16 MPI processes).}
  \caption{G-QSW total (a) in-program time and (b) exponentiation time of QSWalk.m, QSWalk.jl and the QSW\_MPI \texttt{step} method for the Erd\H{o}s-R\'{e}nyi digraph set. System specifications match those of Figure \ref{fig:desk}}.
  \label{fig:g_desk_exp}
\end{figure}

The simulation time of the QSW\_MPI \texttt{step} method is depicted with comparison to QSWalk.m and QSWalk.jl for L-QSWs in Figure \ref{fig:desk}. For all but the case of complete digraphs, QSW\_MPI exhibits superior performance with one MPI process, which is further improved when running QSW\_MPI on all available CPU cores. Examining only the exponentiation phase of the three packages, as shown in Figure \ref{fig:exp_time}, we find that QSW\_MPI outperforms the preexisting methods for the Erd\H{o}s-R\'{e} digraph set, but is significantly slower overall for the complete digraph set. As the similarity observed in exponentiation time between QSW\_MPI running with one MPI process and the preexisting packages in Figure \ref{fig:exp_time} (a) does not reflect the speedup observed in Figure \ref{fig:desk} (c), it is evident that a significant portion of the speedup is attributable to the optimised L-QSW $\tilde{\mathcal{L}}$ construction subroutines. Figure \ref{fig:g_desk_exp} shows the total in-program time and exponentiation time for G-QSWs on the Erd\H{o}s-R\'{e} digraph set. Single process performance of QSW\_MPI is slower than both QSWalk.m and QSWalk.m, but with multiple processes it is again able to provide a speedup.  

\begin{figure}[h!]
   \centering
   \includegraphics[width=6.5cm]{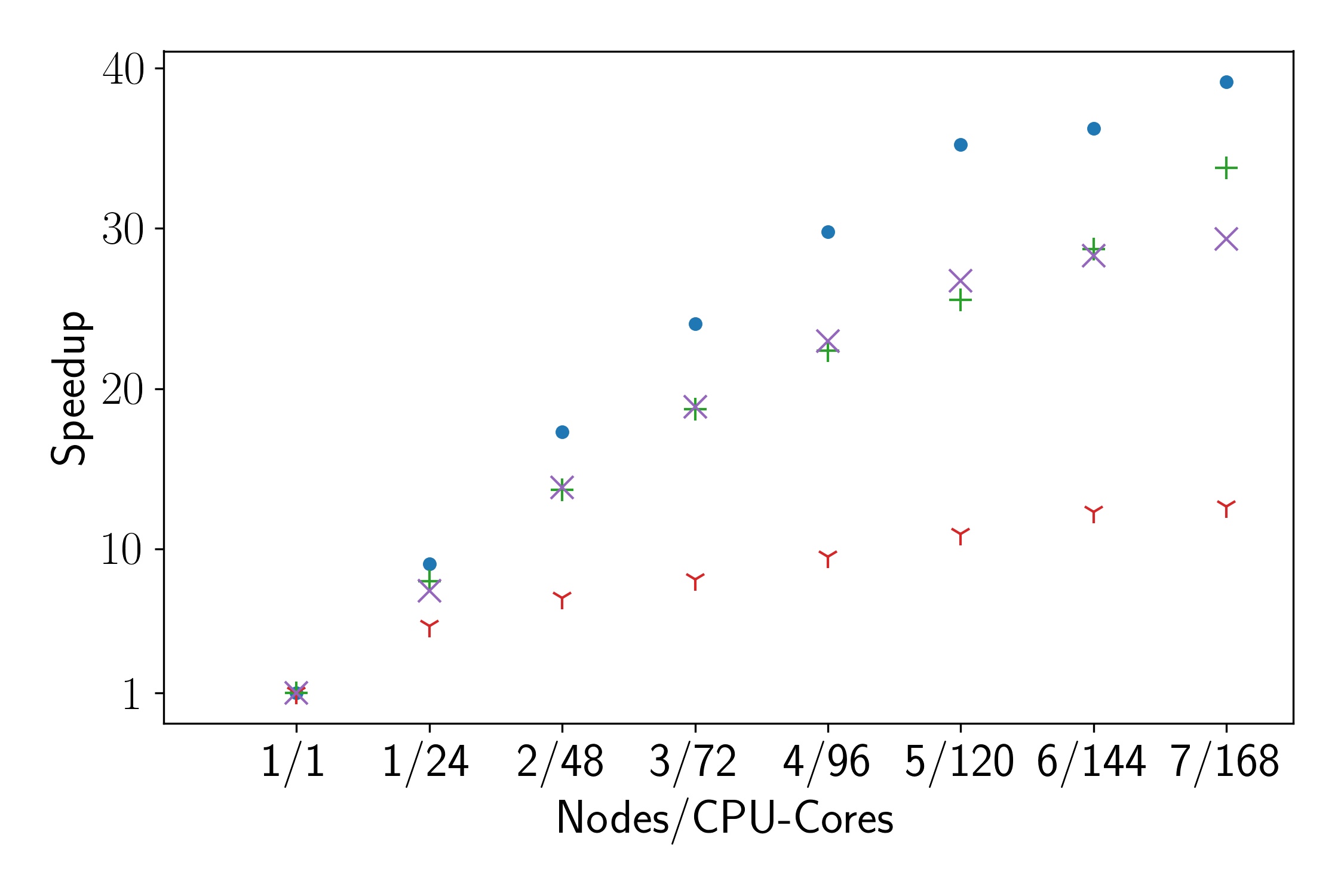}
   \caption*{(a)}
   \includegraphics[width=6.5cm]{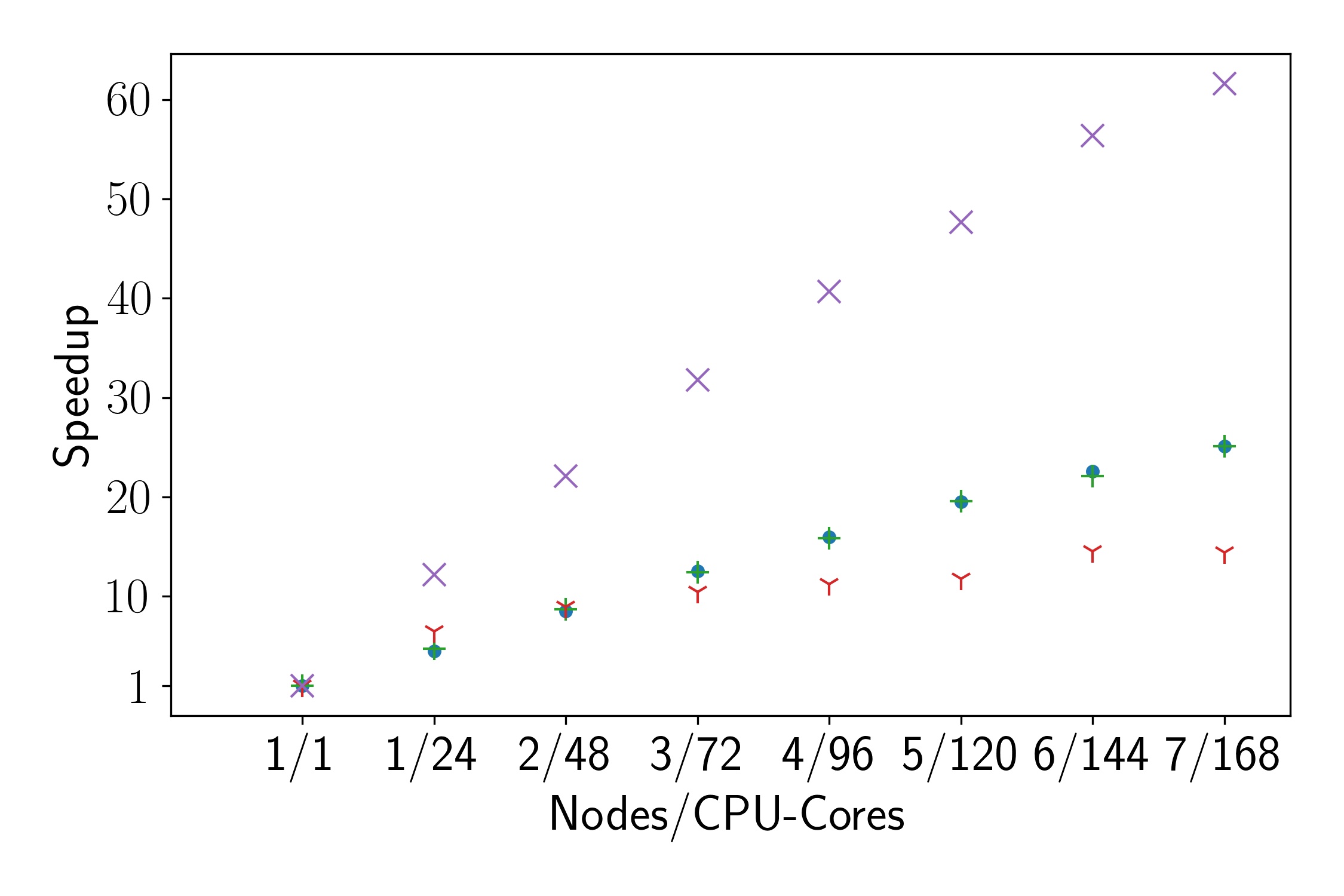}
   \caption*{(b)}
  \caption*{\footnotesize $\color{C0} \bullet$: line graph, $\color{C2} +$: grid
    graph, $\color{C3} \Ydown$: random graph, $\color{C4} \times$: complete graph}
  \caption{Proportional wall time speedup of an L-QSW simulation via the \texttt{step} method with increases to the number of supercomputer nodes for (a) 12 MPI processes per CPU, and (b) 1 MPI process with 12 OpenMP threads per CPU. Single process program wall time was 1973 s for the line digraph (5050 vertices), 2010 s for the square lattice (3844 vertices), 2842 s for the Erdos\H{o}s-R\'{e}nyi digraph (2020 vertices) and 27821 s for the complete digraph (400 vertices). System specifications: XC40 Series Supercomputer, 2 $\times$ Intel Xeon E5-2690V3 with 12 cores at 2.6 GHz and 64 GB RAM per node.}
  \label{fig:scale}
\end{figure}

\begin{figure}[t]
  \centering
  \subfloat[]{{\includegraphics[width=3.9cm]{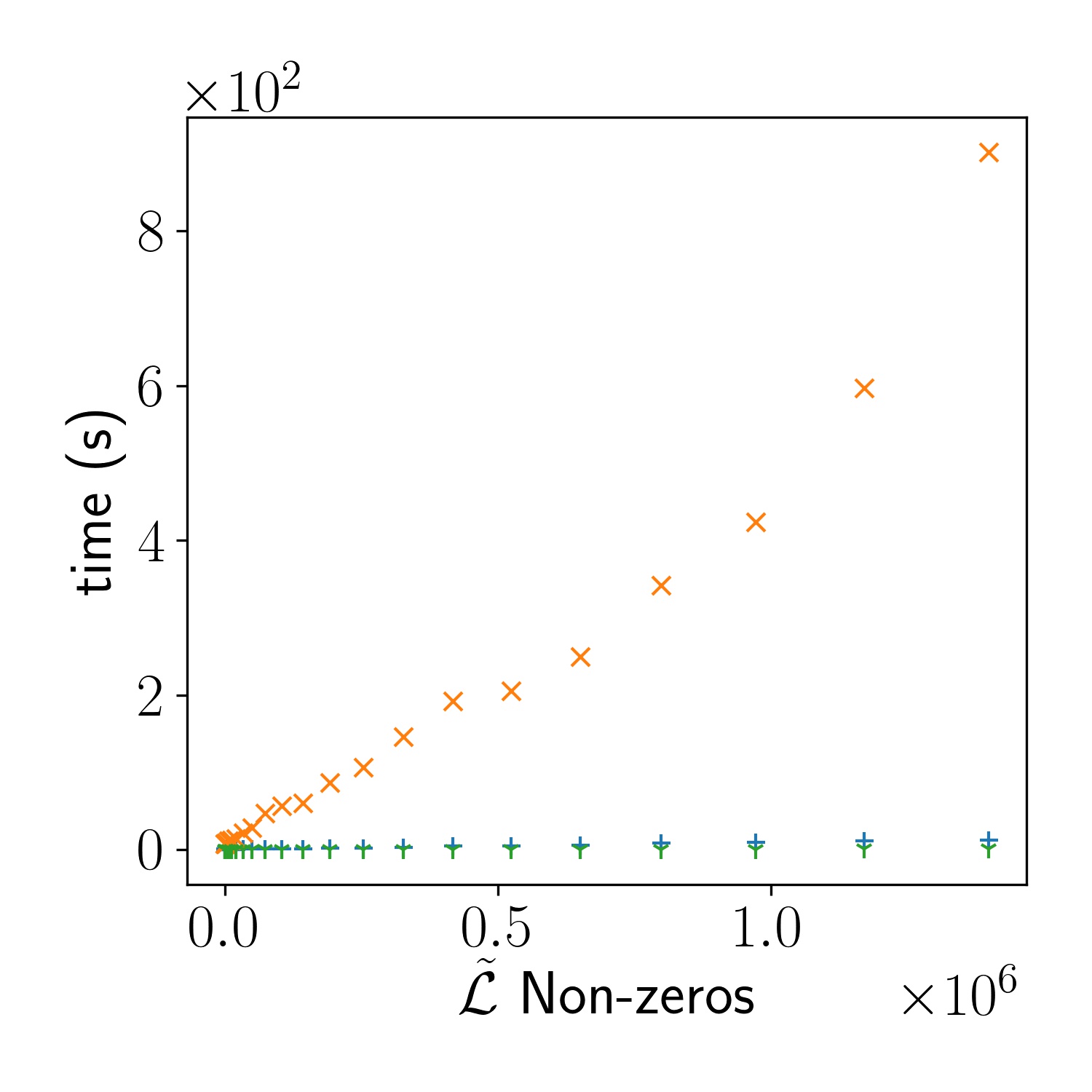}\includegraphics[width=3.9cm]{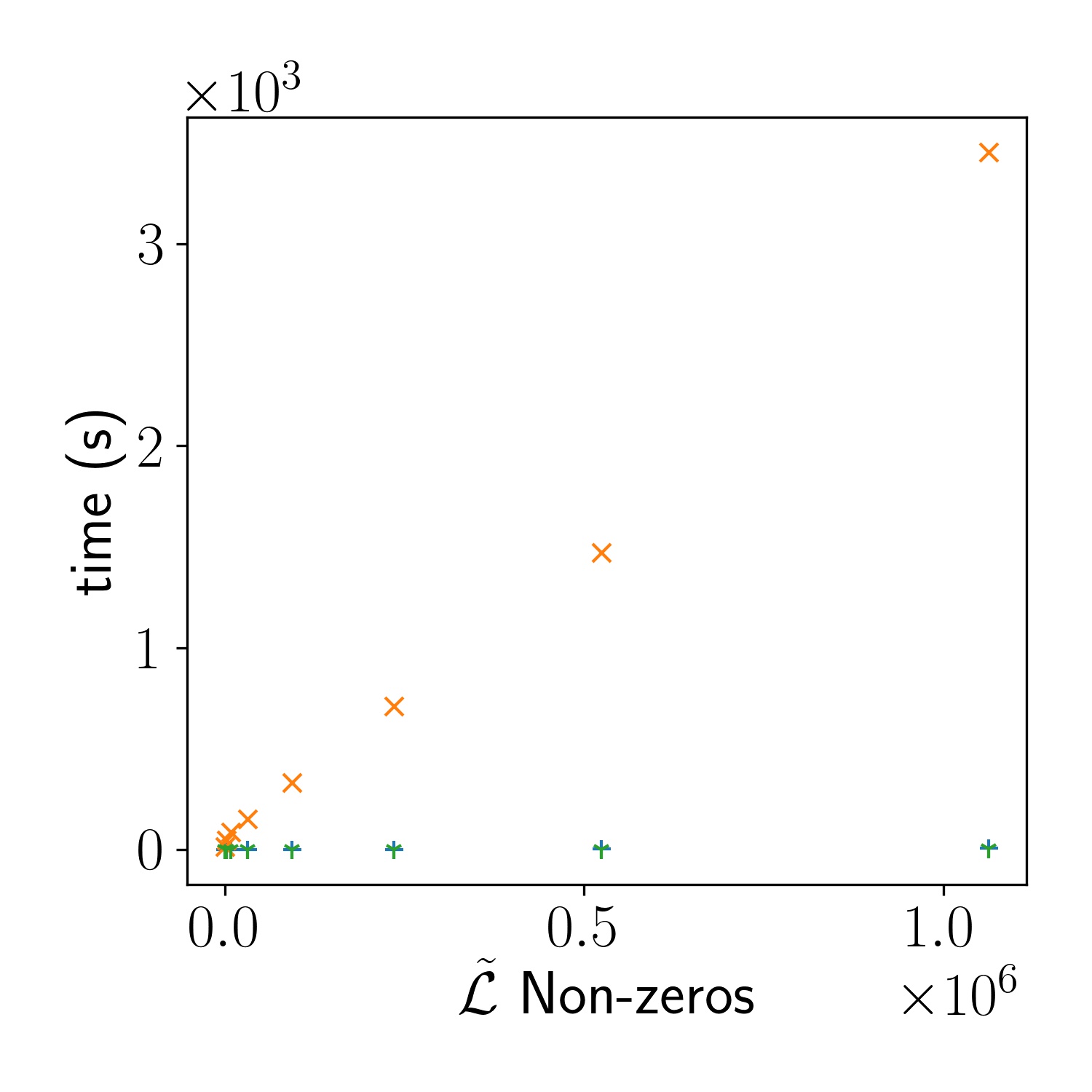}}}
  \newline
  \subfloat{\footnotesize $\color{C0} +$: series method, $\color{C1} \times$: step method, $\color{C2} \Ydown$: step method average per $\Delta t$.}
  \caption{In-program time taken to simulate the time series of a L-QSW from $t_1 = 0$ to $t_q = 100$ with $\Delta t = 0.5$ as a function of superoperator non-zeros on a (a) line digraph and (b) complete digraph.}
  \label{fig:step_vs_series_time}
\end{figure}

A motivating factor in the selection of the scaling and squaring algorithm used by QSW\_MPI is its ability to calculate the time series of the system evolution efficiently \cite{al-mohy_computing_2011} . This is demonstrated in Figure \ref{fig:step_vs_series_time}, which compares the time taken to calculate a time-series with repeated calls to \texttt{step} or a single call to the \texttt{series} method for the line digraph and complete digraph sample sets. With a time-step size of $\Delta t = 0.5$, \texttt{series} executes with a time proportional to a single \texttt{step}. In combination with the results provided in Figure \ref{fig:step_vs_series_delta}, this supports the efficacy of QSW\_MPI for this use case.  

The scalability of QSW\_MPI in a super-computing environment is shown in Figure \ref{fig:scale}. Proportional speed-up of L-QSW simulation via the \texttt{step} method with increases to the number of supercomputer nodes was studied for a line digraph, square lattice, Erd\H{o}s-R\'{e}nyi digraph and complete digraph, each within 1\% of having $1.28 \times 10^8$ arcs. This allowed for the parallelism of QSW\_MPI to be considered in terms of the MPI communication overhead induced by the digraph connectivity. Two scenarios were considered, 12 MPI processes per CPU and 1 MPI process with 12 OpenMP threads per CPU. As before, L-QSWs were propagated from an initial maximally mixed state to $t = 100$ via the \texttt{step} method.

Figure \ref{fig:scale} (a) shows the results for pure MPI parallelism. For each graph, speed-up is observed with an increasing number of processes, with a higher degree of speed-up associated with a lower degree of graph connectivity as expected. Consistent speed-up is also observed when using MPI + OpenMP, as shown in Figure \ref{fig:scale} (b), but to a lesser extent for the line, square lattice and Erd\H{o}s-R\'{e}nyi digraphs. Conversely, the complete digraph exhibits a maximum speed up of $\sim 62$ times using MPI + OpenMP as opposed to a maximum of $\sim 30$ when using only MPI. For the less connected graphs, the overhead induced through the creation of the OpenMP threads is likely disproportionate to the time spent in sparse matrix multiplication, whereas the complete digraph is sufficiently dense so as to benefit from the finer degree of parallelism and less frequent through-network communication. As the advantages offered by OpenMP are limited to this narrower class of simulations, the feature is present QSW\_MPI as a compile-time option.

The Erd\H{o}s-R\'{e}nyi digraph achieves a significantly lower degree of proportional speed-up, the maximum being $\sim 13$ times, as compared to the other digraph types which have maximums falling between $33$ and $62$ times. Irregularity in the digraph structure likely impacts the performance of sparse matrix multiplication, either through poor cache utilisation or poor performance of the collective MPI constructs responsible for the inter-process transfer of $\tilde{\rho}(t)$ elements. QSW\_MPI clearly performs best on sparse graphs and digraphs with a high degree of regularity.

\section{Conclusion} \label{sec:conclusion}

QSW\_MPI provides a means of quantum stochastic walk simulation on graphs and digraphs with many thousands of vertices through the use of MPI parallelism and memory-efficient of sparse data structures. Key to this was the implementation of a relatively recent method of matrix exponentiation, stable for both Hermitian and non-Hermitian matrices. As far as we know, this is the first implementation of this specific algorithm in the context of distributed memory computing.

The package performs particularly well in the context of time-series calculations and walks on sparse graphs and digraphs with a high degree of regularity. Also of note is the speed-up achieved through a detailed analysis of the structure of the L-QSW vectorised superoperator. While somewhat mundane from a theoretical or technical standpoint, the resulting performance highlights the value of `tailor-made' approaches compared to the use of `off the shelf' software libraries when simulating extensive systems. We note that, in a workstation context, that preexisting packages offer superior performance when calculating QSWs states on highly connected graphs for a single time-point. Still, nevertheless, QSW\_MPI significantly increases the scope of possible QSW simulations.     

As QSW\_MPI implements its own basic linear algebra subroutines, its parallel performance may not be optimal as compared to an implementation utilising preexisting parallel libraries such as PetSc. However, the installation of such libraries is not always straightforward; thus QSW\_MPI, dependent only on packages included on most Unix-based operating systems, gains a good deal of portability due to this omission. Nevertheless, it would be beneficial to extend QSW\_MPI to allow for the use of external linear algebra libraries as a user-specified compile-time option. This would be relatively easy to achieve due to the modularity of the package's underlying Fortran libraries.   

\vspace{-0.4cm}

\section*{Ackowlegements}

This work was supported by the Muriel and Colin Ramm Postgraduate Scholarship in Physics, and an Australian Government Research Training Program Scholarship at the University of Western Australia. The Pawsey Supercomputing Centre provided computational resources with funding from the Australian Government and the Government of Western Australia.

\vspace{-0.4cm}





\bibliographystyle{elsarticle-num}
\bibliography{short}







\end{document}